\documentclass[twocolum]{aa}  
\usepackage{amsmath}

\newcommand{\br}{\boldsymbol{r}}

\newcommand{\buu}{\boldsymbol{u}}

\newcommand{\bnabla}{\boldsymbol{\nabla}}

\newcommand{\dd}{{\rm d}}
\newcommand{\ab}{a_{\rm b}}
\newcommand{\eb}{e_{\rm b}}

\newcommand{\bB}{\boldsymbol{B}}

\usepackage{makecell}
\usepackage{graphicx}
\usepackage{xcolor}
\usepackage{txfonts}   
\usepackage{hyperref}
\hypersetup{
  unicode=true,      
  pdftoolbar=true,    
  pdfmenubar=true,    
  pdffitwindow=true,   
  pdftitle={},
  pdfauthor={Gagnier \& Pejcha},
  pdfkeywords={},     
  pdfnewwindow=true,   
  colorlinks=true,    
  linkcolor=blue,     
  citecolor=blue,     
  filecolor=gray,     
  urlcolor=blue      
}
\begin{document} 

\title{Journey to the center of the common envelope evolution}
\subtitle{Inner dynamics of the post-dynamical inspiral}
  \authorrunning{Gagnier \& Pejcha} 


\author{Damien Gagnier\inst{\ref{inst1}, \ref{inst2}, \ref{inst3}} \and
          Ond\v{r}ej Pejcha\inst{\ref{inst3}}}

\institute{Astronomisches Rechen-Institut, Zentrum für Astronomie der Universität Heidelberg, Mönchhofstr. 12-14, 69120 Heidelberg,\\ Germany \email{damien.gagnier@uni-heidelberg.de}\label{inst1} \and Heidelberger Institut für Theoretische Studien, Schloss-Wolfsbrunnenweg 35, 69118 Heidelberg, Germany  \label{inst2} \and Institute of Theoretical Physics, Faculty of Mathematics and Physics, Charles University, V Hole\v{s}ovi\v{c}k\'{a}ch 2, Praha 8, 180 00, Czech Republic  \label{inst3}
}

\date{\today}

\abstract{
Three-dimensional hydrodynamical simulations of common envelope evolution are often terminated soon after the initial dynamical plunge of the companion transitions into a long-lasting post-dynamical inspiral with slowly varying semi-major axis, $a_\text{b}$. This premature termination is often due to insufficient numerical resolution and challenges associated with the softening of the gravitational potential of the two cores. In this work, we use statically-refined 3D hydrodynamical simulations to study binaries orbiting inside a common envelope, exploring the effects of varying numerical resolution, $\delta$, gravitational potential softening prescriptions, and the associated softening lengthscale, $\epsilon$. We find that quantities such as the binary inspiral timescale or the volume-averaged shearing rate typically converge to asymptotic values only for $\epsilon \le 0.1 a_\text{b}$ and $\delta \le 6 \times 10^{-3}a_\text{b}$ with smaller $\epsilon$ requiring correspondingly smaller $\delta$. This suggests that many of the contemporary simulations could be effectively under-resolved. After a few tens of binary orbits, the two cores become surrounded by a corotating, nearly hydrostatic gas structure, resembling the shared envelope of a contact binary. We propose that this structure is responsible for the slowing down of the dynamical inspiral, leading to an asymptotic inspiral timescale of approximately $10^5$ orbital periods for a binary mass ratio $q=1/3$, and approximately $10^6$ orbital periods for a binary mass ratio $q=1$. By investigating kinetic helicity, we argue that the magnetic field is unlikely to organize into large-scale structures via the usual $\alpha$--effect during the post-dynamical phase. Nonetheless, even in the absence of magnetic fields, we observe intermittent polar outflows collimated by partially centrifugally evacuated polar funnels. We discuss implications for the long-term evolution in the post-dynamical inspiral phase and the ultimate emergence of the post-common-envelope binary.

}

\maketitle

\section{Introduction}

Common envelope evolution (CEE) is a phase in the life of binary systems during which one of the components expands and initiates a dynamically unstable mass transfer towards its more compact companion \citep[e.g.,][]{Paczynski1976,Meyer1979,Ivanova2013,Roepke2023}. The companion later becomes engulfed in the now shared envelope and the two stellar objects then rapidly  spiral towards each other, depositing energy and angular momentum into the surrounding gas. The dynamical approach of the two cores may eventually slow down due to the redistribution of the gas and a phase of slow, quasistationary approach ensues. Ultimately, the common envelope will disperse and the two stellar cores will emerge as a post-common envelope binary with properties often significantly different from the original binary. Because of the wide range of temporal and spatial scales that would need to be resolved, \emph{ab initio} three-dimensional (3D) (magneto)hydrodynamical simulations of this post-dynamical inspiral phase are often terminated well before the evolution naturally concludes.

In \citet{Gagnier2023,Gagnier2024}, we studied the post-dynamical inspiral phase of CEE by constructing idealized numerical experiments. In these simulations, a spherical envelope is first spun up to emulate angular momentum deposition by the dynamically inspiralling companion, and then subjected to the evolving gravitational field of the two orbiting cores at the center. The (magneto)hydrodynamical simulations were performed in 3D using a spherical mesh with both adaptive and static mesh refinement. To reduce the computational cost of the simulations, we excised a spherical region around the two orbiting cores, removing them from the simulation domain. This also allowed us to control accretion onto the central regions by applying either an inflow or a reflecting inner boundary condition. With these simulations, we addressed issues such as angular momentum transfer between the binary and the envelope, the amplification of magnetic fields, and the effects of these fields on the envelope structure and dynamics, and on the binary's orbital evolution. We also showed that the structure and dynamics of the common envelope is remarkably similar to those of circumbinary disks. This similarity was also implicated in other recent \emph{ab initio} 3D simulations \citep{Ondratschek2022,wei2024,landri2024,Vetter2024} and used in 1D long-term models of CEE \citep{Tuna2023,valli2024}.

Recent simulations of circumbinary disks in a more general context \citep[e.g.][]{Moody2019,Munoz2019,Munoz2020,Duffell2020,Paschalidis2021,Li2021,Dittmann2021,Combi2022,Duffel2024} have highlighted the crucial importance of adequate resolution of the cavity containing the binary in numerical simulations. This is essential in order to precisely quantify the torques that dictate the orbital evolution \citep[e.g.,][]{Dittmann2021,Duffel2024}, to analyze the potential mass transfer between the two binary components \citep[e.g.,][]{Bowen2017,Avara2023}, to study the evolution of their rotational velocity, or to understand the process of self-consistent launching of jets \citep[e.g.,][]{Paschalidis2021,Combi2022}, for example. 

In the context of the late phases of CEE, sufficient numerical resolution may become even more crucial than in circumbinary disks because of the potential absence of a low-density cavity encompassing the binary due to the more spherical geometry of the system. Consequently, our earlier simulations with the extracted central sphere presented in \citet{Gagnier2023,Gagnier2024} provide limited information about processes occurring in the immediate vicinity of the central binary. In other CEE simulations, both the core of the primary star and the companion are generally represented as point masses, requiring the use of artificial softening of the gravitational potentials to avoid singularities. However, this approach inevitably  results in reduced injection of orbital energy and angular momentum, as well as in nonphysical flow structures near the cores. This can potentially lead to erroneous envelope ejection rates and gravitational torques, and diminished magnetic energy amplification. This problem is further exacerbated during the late phases of the CEE, when the stellar cores are close and the two softened regions constitute a nonnegligible fraction of the intraorbital volume. Furthermore, the intraorbital region is often significantly underresolved during this late phase.

To provide several examples of gravitational softening in CEE, in the MHD simulations of \citet{Ondratschek2022} the softening radius is fixed at $\sim 40 \%$ of the minimum orbital separation. In the hydrodynamical simulations of \cite{Kramer2020}, \citet{Moreno2022}, and \citet{MoranFraile2024}, the softening radius is adaptively reduced to avoid the softened regions to overlap. In \citet{Chamandy2020}, the softening radius reaches up to $\sim 34\%$ of the orbital separation. This quantity reaches up to   $62 \%$ of the final separation in \citet{Prust2019}. \citet{Ohlmann2016,Ohlmann2016b},  \citet{Chamandy2019b}, \citet{Sand2020}, \citet{Chamandy2024}, and \citet{Vetter2024} maintain the softening radius below $20 \%$ of the binary separation. All of these works used \citet{Hernquist1989}'s spline function to soften the gravitational potentials. \citet{Sandquist1998,Sandquist2000}, \citet{Passy2012}, \citet{Staff2016a}, \citet{Iaconi2018}, and \citet{Shiber2019} instead use \cite{Ruffert1993}'s softened potentials with softening radii often of the order of, or larger than, the final orbital separation. Perhaps even more alarming is the fact that, in some of these simulations, the final orbital separation only amounts to the width of a couple of mesh's cells.
In the context of circumbinary and protoplanetary disks, a Plummer-softened potential is almost exclusively used. However, \citet{Dong2011} showed that torques can be highly sensitive to potential shapes, and that low-order gravitational potentials such as Plummer's, are particularly inaccurate. Such a wide spectrum of approaches and parameters raises the fundamental question of the potential impact of gravitational softening and intraorbital resolution on the ejection of the common envelope, on the amplification of magnetic energy within it, on the morphology of emerging planetary nebulae and the final separation of orbits, as well as on the reliability of current predictions on these issues. 

In this paper, we investigate the gas dynamics in the binary's immediate vicinity and its effects on the orbital evolution with a special emphasis on the role of gravitational softening and intraorbital numerical resolution. To this end, we construct a post-dynamical inspiral model similar to that of \citet{Gagnier2023,Gagnier2024}, emulating the preceding dynamical stage. In this work, however, we include both the binary and the intraorbital region in the numerical domain, using a statically refined Cartesian mesh to focus on the intraorbital region at the expense of the outer envelope.
This paper is organised as follows. In Sect.~\ref{sec:setup} we present our physical model and describe our numerical setup. In Sect.~\ref{sec:results}, we present our simulations and our analysis of gravitational softening formulations, gravitational torques, envelope ejection and morphology, kinetic helicity, and shear rates. In Sect.~\ref{sec:discussion}, we summarize and discuss our results in the context of the entire post-dynamical inspiral phase of CEE, as well as potential similarities with ordinary contact binaries.

\section{Physical model and numerical setup}\label{sec:setup}
We construct our post-dynamical inspiral model within the inertial frame centered on the binary's center of mass. $M_1$ is the mass of the primary's core located at $\{x_1,y_1,z_1\}$ and $M_2$ is the mass of the secondary object, whether a main-sequence star or a compact object, located at $\{x_2,y_2,z_2\}$. Both objects are not resolved and are treated as point masses. We set the gravitational constant $G$, the total binary mass $M = M_1 + M_2$, the orbital separation $a_{\rm b}^i$ at the onset of our simulations, and the associated orbital angular velocity $\left({GM/\ab^{i3}}\right)^{1/2}$ to unity. As a choice, we set the radius of the primary star $R_\ast=50$. To simplify our model and establish continuity with our prior works, we solely consider circular orbits ($\eb=\dot{e}_\text{b}=0$) and neglect mass accretion onto the individual cores. The mass of the envelope beyond the binary's orbit is set to $M_\text{env} = 2$ in our units.

To convert the non-dimensional units used in the simulations to dimensional cgs units, the reader should multiply the non-dimensional values by the appropriate reference quantities. As an example, take a non-dimensional density $\rho$, a reference mass of the binary $M= M_\odot$, and a reference length $a_{\rm b}^i =  R_\ast/50$ with $R_\ast = 50 R_\odot$. The dimensional density is then calculated as
\begin{equation}
\rho_\text{dim} =  5.9\,\text{g\,cm}^{-3}\, \rho\, \left(\frac{M}{M_\odot} \right) \left(\frac{a_\text{b}^i}{\,R_\odot} \right)^{-3} \ .
\label{eq:units}
\end{equation}

Instead of tracking the prior evolution of the inspiraling binary, we replicate its outcome through a method  described by \citet{Morris2006,Morris2007,Morris2009}, \citet{Hirai2021}, and \citet{Gagnier2023,Gagnier2024}, where the envelope is spun-up until a satisfactory amount of total angular momentum,
\begin{equation}
J_z = ( 1+\beta)\frac{M_2(M_1+M_{\rm env})}{M+M_{\rm env}}\sqrt{G(M + M_{\rm env})m\ab^{i}}  - \mu\sqrt{GM\ab^{i}},
\end{equation}
is imparted. Here, $\mu = M_1M_2/M$ is the binary's reduced mass and, in this work, we set $\beta=0.1$ and $m=100$.  Our selection of initial parameters for the binary and envelope broadly mirrors the outcomes of \emph{ab initio} simulations across various progenitor types. Table~\ref{tab:runs} summarizes the properties of our models. 
In the subsequent sections,  we describe the equations used for solving the problem (Sect.~\ref{sec:equations}), the initial conditions (Sect.~\ref{sec:IC}), and the mesh structure (Sect.~ \ref{sec:mesh}).

\subsection{Equations of hydrodynamics}\label{sec:equations}

We use the Eulerian code Athena++ \citep[][]{Stone2020} to solve the compressible equations of hydrodynamics
    \begin{align} \label{eq:mass}
        \frac{\partial \rho}{\partial t} + \bnabla\cdot \rho \buu &= 0\ , \\
        \frac{\partial \rho \buu}{\partial t} + \bnabla \cdot (\rho \buu \otimes \buu + P \boldsymbol{I}) &= - \rho \bnabla \Phi\ ,\label{eq:mom} \\
        \frac{\partial E}{\partial t} + \bnabla \cdot \left((E+P \boldsymbol{I})\buu \right) &= - \rho \bnabla \Phi \cdot \buu\  , \label{eq:etot}
    \end{align}
where $E = e +\rho u^2/2$ is the energy density , $e$ is the internal energy density, $P = (\Gamma - 1)e $ is the pressure, $\Gamma = 5/3$ is the adiabatic index, and $\Phi(\br)$ is the gravitational potential of the binary%
\begin{equation}\label{eq:fullphi}
    \Phi(\br) = \sum_{i=1}^2 \Phi_i(\br) = -\sum_{i=1}^2 GM_i f(\left.\lVert \br - \br_i \right.\rVert ,\epsilon) \ ,
 \end{equation}
where $\epsilon$ is the gravitational softening length, $r_i = \sqrt{x_i^2 + y_i^2 + z_i^2}$ is the distance to the binary's center of mass of either the primary's core, or of the secondary object, and $f(\left.\lVert \br - \br_i \right.\rVert ,\epsilon)$ is the shape of the softened potential. In the Plummer model we have 
\begin{equation}\label{eq:plum}
    f(\left.\lVert \br - \br_i \right.\rVert ,\epsilon)  =  \frac{1}{\sqrt{\left.\lVert \br - \br_i \right.\rVert ^2 + \epsilon^2 }} \ .
\end{equation}
In the \cite{Ruffert1993} model we have
\begin{equation}\label{eq:ruf}
    f(\left.\lVert \br - \br_i \right.\rVert ,\epsilon)  =  \frac{1}{\sqrt{\left.\lVert \br - \br_i \right.\rVert ^2 + \epsilon^2 \exp\left(-\left.\lVert \br - \br_i \right.\rVert ^2/\epsilon^2 \right) }} \ .
\end{equation}
In the \citet{Hernquist1989} spline model we have
\begin{align}\label{eq:spline}
  &f(\left.\lVert \br - \br_i \right.\rVert ,\epsilon) =\nonumber\\ &\frac{1}{h}\begin{cases}
    \left( -\frac{16}{3}u^2 + \frac{48}{5}u^4 -\frac{32}{5}u^5  + \frac{14}{5}\right), & \text{if}\ 0 \leq u < \frac{1}{2}.\\[2mm]
     \left( \frac{-1}{15u} - \frac{32}{3}u^2 +16u^3 - \frac{48}{5}u^4 + \frac{32}{15}u^5 + \frac{16}{5}\right), &  \text{if}\ \frac{1}{2} \leq u < 1.\\[2mm]
   \frac{1}{u}, & \text{if}\ u \ge 1.  
  \end{cases}
  \end{align}
Here, $u = \left.\lVert \br - \br_i \right.\rVert /h$ and $h \equiv 14\epsilon/5$ so that the minimum of the potential has the same depth with the three formulations. We note that this definition of $h$ is arbitrary, it has no physical implication, and it does not establish equivalence between the different formulations (see Appendix~\ref{app:A}). 
In our simulations, we gradually decrease the value of $\epsilon$ from 0.5 to its target value over a span of 20 orbital periods after the initial spinup of the envelope.


\subsection{Initial conditions}\label{sec:IC}

\begin{table*}
\caption{Simulation parameters and results. \label{tab:runs}}
\begin{center}
\begin{tabular}{lcccccccccc}
\hline\hline
{Model} &  {max level}  & {$\delta\ [10^{-3}\ab^i]$} &  {softening} & {$\epsilon\ [\ab^i]$} & {q} & {$\tau_{\ab}\ [10^4 P_{\rm orb}^i]$} & {$\langle |\dot{\gamma_{xy}|}\rangle\ [\Omega_{\rm orb}^i]$} & {orbit} & {$\ab^i$} & {$\ab^f$}\\
\hline
E05.S.f.q1 & 7 & 6.104 & Spline & 0.5 & 1 & 10.68 & 1.738 & fixed & 1 & 1 \\
E02.S.f.q1.vlr & 3 & 97.66 & Spline & 0.2 & 1 & 22.26 & 0.6657 & fixed & 1 & 1\\
E02.S.f.q1.lr & 4 & 48.83 & Spline & 0.2 & 1 & 17.99 & 0.7626 & fixed & 1 & 1\\
E02.S.f.q1 & 7 & 6.104 & Spline & 0.2 & 1 & 12.85 & 1.822 & fixed & 1 & 1\\
E01.S.f.q1 & 7 & 6.104 & Spline & 0.1 & 1 & 16.40& 2.368 & fixed & 1 & 1\\
E005.S.f.q1 & 7 & 6.104 & Spline & 0.05 & 1 & 10.23 & 2.395 & fixed & 1 & 1\\
E005.S.f.q1.hr & 8 & 3.052 & Spline & 0.05 & 1 & 15.84 & 2.730 & fixed & 1 & 1 \\
\hline 
E05.S.l.q1 & 6 & 12.21 & Spline & 0.5 & 1 & 138.6 & 0.4942 & live & 1 & 0.7014 \\
E02.S.l.q1 & 6 & 12.21 & Spline & 0.2 & 1 & 154.3 & 0.5764 & live & 1 & 0.6943 \\
E005.S.l.q1 & 8 & 3.052 & Spline & 0.05 & 1 & 136.2 & 1.374 & live & 1 & 0.6930 \\
\Xhline{0.2\arrayrulewidth}
E05.S.l.q03 & 6 & 12.21 & Spline & 0.5 & 1/3 & 5.397 & 0.4963 & live & 1 & 0.4989 \\
E03.S.l.q03 & 6 & 12.21 & Spline & 0.3 & 1/3 & 7.655 & 0.6429 & live & 1 & 0.4717 \\
E02.S.l.q03 & 6 & 12.21 & Spline & 0.2 & 1/3 & 8.518 & 0.6792 & live & 1 &  0.4721 \\
R.l.q03.lr & 3 & 97.66 & Ruffert & 3$\delta$ & 1/3 &141.8& 0.5592 & live & 1 & 0.6605 \\
E02.S.l.q03.lr& 4 & 48.83 & Spline & 0.2 & 1/3 & 32.14 & 0.7852 & live & 1 & 0.5745 \\
E01.S.l.q03 & 7 & 6.104 & Spline & 0.1 & 1/3 & 9.370 & 0.8464 & live & 1 & 0.4754 \\
E01.R.l.q03 & 7 & 6.104 & Ruffert & 0.1 & 1/3 & 8.654 & 0.9967 & live & 1 & 0.4825 \\
E005.S.l.q03 & 8 & 3.052 & Spline & 0.05 & 1/3 & 9.491 & 1.393 & live & 1 & 0.4842\\
\hline
\end{tabular}
\tablefoot{The orbital separation evolution timescale, $\tau_{\ab} = \langle \dot{\ab}/ \ab \rangle^{-1}$ and the horizontal shear rate, $\langle |\dot{\gamma_{xy}|}\rangle$ are averaged over the last 500 orbits for live simulations, and the last 50 orbits for simulations with fixed orbits. 
$\ab^f$ is the orbital separation measured after 2125 binary orbits.}
\end{center}
\end{table*}

As in \citet{Gagnier2023,Gagnier2024}, we assume that the gas in the envelope is initially in hydrostatic equilibrium and that it can be described by a polytropic equation of state ignoring the gas self-gravity and considering purely radial initial profiles. Because the binary's orbit is now included within the numerical domain, we also initialize the gas within the orbit as a polytrope in hydrostatic equilibrium. Since the equilibrium within and near the orbit is immediately perturbed by the binary motion, the specific details of the initial conditions in the inner regions are not important and we thus provide the corresponding analytical expressions for initial density and pressure in Appendix~\ref{app:ic}.

\subsection{Mesh structure and boundary conditions}\label{sec:mesh}
At root level, we set $256~\times~256~\times~256$ active cells in $\{x,y,z\}$. The domain extends from $-100$ to $100$ in the three directions. On top of the root level, we add multiple levels of static mesh refinement up to level 8 (see Table~\ref{tab:runs}), which corresponds to a grid spacing $\delta \simeq 0.003 \ab^i$. The highest refinement level is concentrated within a box of dimensions $1.6 \ab^i \times 1.6 \ab^i \times 0.2 \ab^i$, centered on the binary's center of mass. We use no inflow boundary conditions in the three directions (``diode'' boundary conditions).

\section{Results}\label{sec:results}
\begin{figure*} 
\centering
\includegraphics[ width=\textwidth]{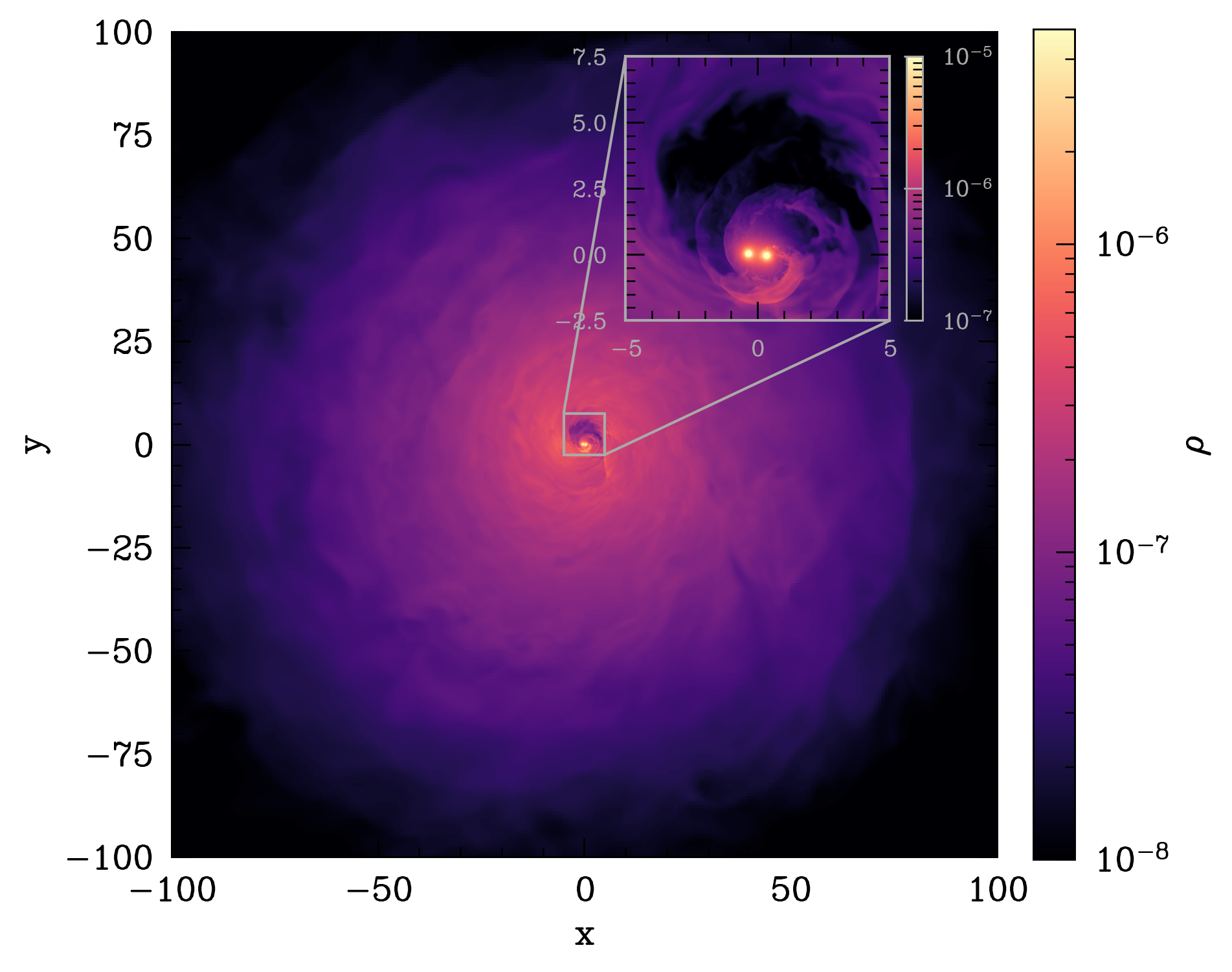}  
   \caption{Snapshot of the gas density cross section in the $xy$ plane for simulation E005.S.l.q1  after $\sim 2550$ orbits.}
\label{fig:rho}
\end{figure*}

Here, we describe results of our simulations. In Fig.~\ref{fig:rho}, we show a density cross-section of one of our simulations, which illustrates a typical output of our runs. We then address different gravitational softening formulations (Sect.~\ref{sec:softening}), gravitational torques (Sect.~\ref{sec:torque}), envelope ejection and its morphology (Sect.~\ref{sec:env_ej}), 
kinetic helicity (Sect.~\ref{sec:helicity}), and shear rate (Sect.~\ref{sec:shear}).

\subsection{Overview of gravitational softening formulations}
\label{sec:softening}

In Appendix~\ref{app:A}, we provide a thorough comparison of different gravitational softening formulations. Here, we simply summarize the most important findings. We show that the Ruffert and spline formulations (Eqs.~(\ref{eq:ruf}) and~(\ref{eq:spline})) yield a more accurate representation of the true gravitational potential of the binary than the Plummer formulation (Eq.~(\ref{eq:plum})), at a distance $\left.\lVert \br - \br_i \right.\rVert \gtrsim  \epsilon/5$ from the core. All three formulations underestimate the gravitational potential by more than a factor 10 at a distance $\left.\lVert \br - \br_i \right.\rVert \lesssim  \epsilon/10$.

The accuracy of the softened potential itself is however not relevant. Instead, it is the gradient of the potential, which appears in the equations of hydrodynamics, that is crucial for meaningful comparison. In Fig.~\ref{fig:grad}, we show that, as expected, the three considered softening formulations effectively dampen the gravitational potential gradient within the softening spheres. While both the Plummer and spline methods achieve effective gradient reduction, the Ruffert method results in a more dramatic dampening without necessarily further relaxing the time step constraint. This variation in the degree of gradient suppression within softening spheres among the different considered formulations may have important implications for the outcome of our simulations. 

In Appendix~\ref{app:A}, we further show that while the spline-softened potential gradient becomes exactly Newtonian at $\left.\lVert \br - \br_i \right.\rVert \geq h$, the Plummer softening formulation still significantly underestimates the potential gradient at distance a $\left.\lVert \br - \br_i \right.\rVert = O(10\epsilon)$ from the point mass. Hence the adopted definition $h \equiv 14\epsilon/5$ is far from implying any form of relevant equivalence between the Plummer and spline formulations
, and we suggest an alternative definition of $h$ that implies conditional equivalence in Appendix~\ref{app:A}. For the sake of consistency with previous studies however, we still adopt $h \equiv 14\epsilon/5$ in this work, while acknowledging its limitations. 
Finally, we show that  an asymptotic regime of small $\epsilon$ may be achieved for the total torque and gravitational body force, provided any nonaxisymmetric structures (e.g. minidisks in circumbinary disks) significantly contributing to  the total torque and body force, are properly resolved outside of the softened regions, that is provided the necessary but insufficient condition (Eq.~(\ref{eq:condition})) is satisfied.   

\subsection{Gravitational torque}
\label{sec:torque}

Here, we address the impact of  the gravitational softening formulation and radius and of the intraorbital resolution on the gravitational torque exerted by the binary on the surrounding envelope.
The $z$-directed gravitational torque exerted by the binary on the envelope reads 
\begin{equation}
\begin{aligned}
        \dot{J}_{z,\rm grav} &= - \int s \rho \frac{\partial \Phi}{\partial \varphi} \dd V \\
         & = \int s  \rho \sum_{i=1}^2 GM_i \frac{\partial f(\left.\lVert \br - \br_i \right.\rVert ,\epsilon)}{\partial \varphi}  \dd V  \ ,
\end{aligned}
\end{equation}
where $s$  
is the radial cylindrical coordinate, and
\begin{equation}
    \frac{\partial f(\left.\lVert \br - \br_i \right.\rVert ,\epsilon)}{\partial \varphi} = -\frac{(x_iy -y_ix)}{\left(\left.\lVert \br - \br_i \right.\rVert ^2 + \epsilon^2\right)^{3/2}}  \ ,
\end{equation}
using a Plummer model,
\begin{equation}
    \frac{\partial f(\left.\lVert \br - \br_i \right.\rVert ,\epsilon)}{\partial \varphi} = -\frac{(x_iy -y_ix) \left( 1 - \exp \left(-\left.\lVert \br - \br_i \right.\rVert ^2/\epsilon^2 \right) \right)}{\left(\left.\lVert \br - \br_i \right.\rVert ^2 + \epsilon^2 \exp\left(-\left.\lVert \br - \br_i \right.\rVert ^2/\epsilon^2 \right)\right)^{3/2}}  \ ,
\end{equation}
using \cite{Ruffert1993}'s softening method, or, using the spline defined in Eq.~(\ref{eq:spline})
\begin{equation}
\begin{aligned}\label{eq:df}
  &  \frac{\partial f(\left.\lVert \br - \br_i \right.\rVert ,\epsilon)}{\partial \varphi} =\\
 &\frac{(x_iy -y_ix) }{h^3}\begin{cases}
    -\frac{32}{3} + \frac{192}{5}u^2 - 32u^3, & \text{if}\ 0 \leq u < \frac{1}{2}.\\[2mm]
      \frac{1}{15u^3}  -\frac{64}{3} + 48u - \frac{192}{5}u^2 +\frac{32}{3}u^3, &  \text{if}\ \frac{1}{2} \leq u < 1.\\[2mm]
    -\frac{1}{u^3}, & \text{if}\ u \ge 1.
  \end{cases}
  \end{aligned}
\end{equation}

In the absence of core accretion, the time derivative of the binary's $z$-directed angular momentum is $\dot{J}_{z,\rm b}  = - \dot{J}_{z,\rm grav}$. Accurately quantifying the gravitational torque is thus crucial for understanding the long-term evolution of the binary system. Achieving this requires careful selection of the gravitational softening formulation and radius, along with sufficient numerical resolution in the binary's vicinity where the torque density is maximum. 

\subsubsection{Gas distribution near the two cores}\label{sec:3.2.1}
\begin{figure*} 
\centering
      \includegraphics[width=\textwidth]{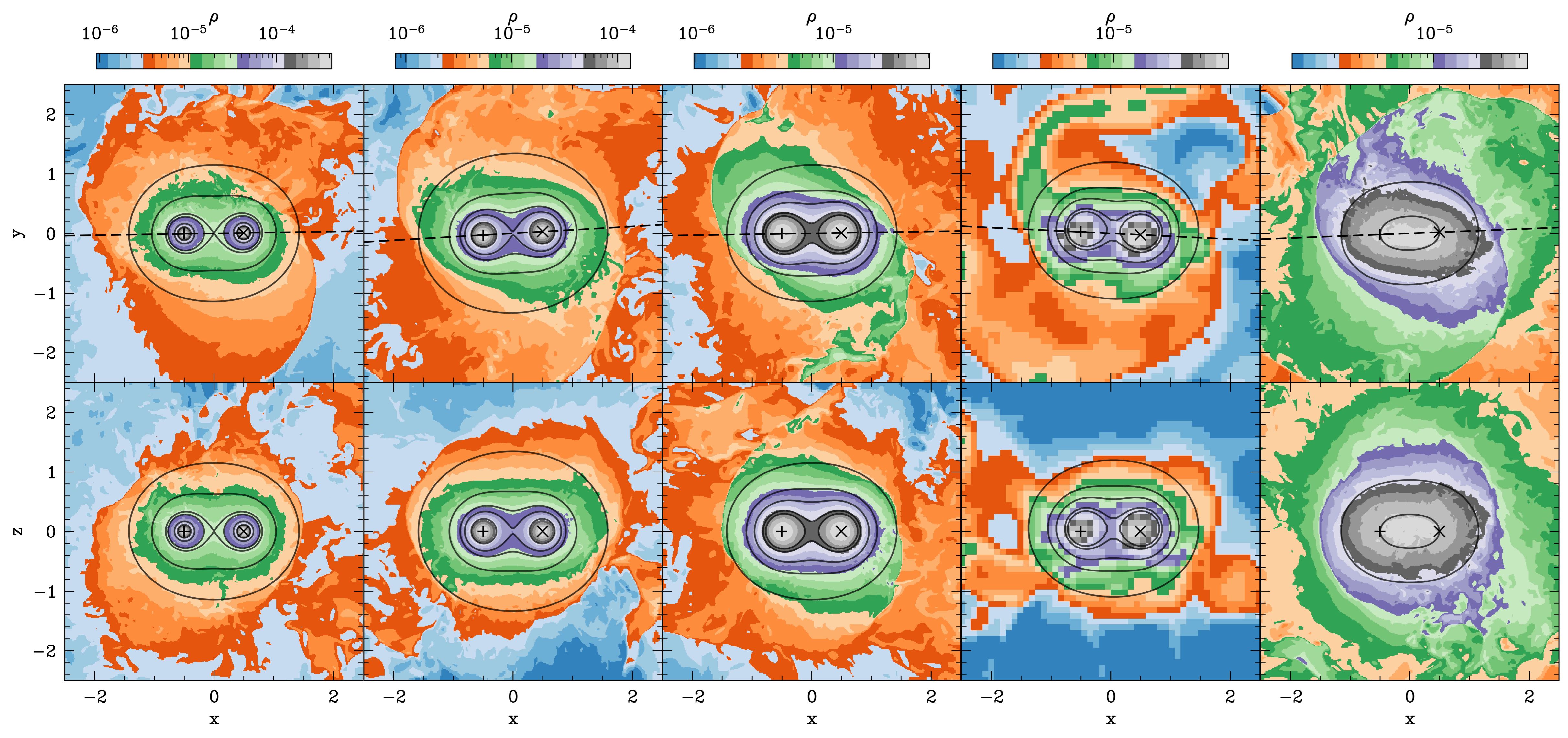}
   \caption{Zoomed-in snapshot of the gas density cross section in the $xy$ (top row) and $xz$ planes (at $y=0$, bottom row) after $\sim 300$ orbits, for $\epsilon = 0.05$, $0.1$, $0.2$, and $0.5$ from left to right, for $q= 1$ with fixed orbital separation and using the spline softening method. The fourth snapshot correspond to the low resolution simulation run E02.S.f.q1.vlr. Solid black lines indicate equipotentials of the smoothed binary potential. The dashed black line marks the intersection between the orbital plane and the plane orthogonal to the orbital plane and parallel to the $\br_i - \br_j$ vector.}
\label{fig:rhoPhiq1}
\end{figure*}

\begin{figure*} 
\centering
      \includegraphics[width=\textwidth]{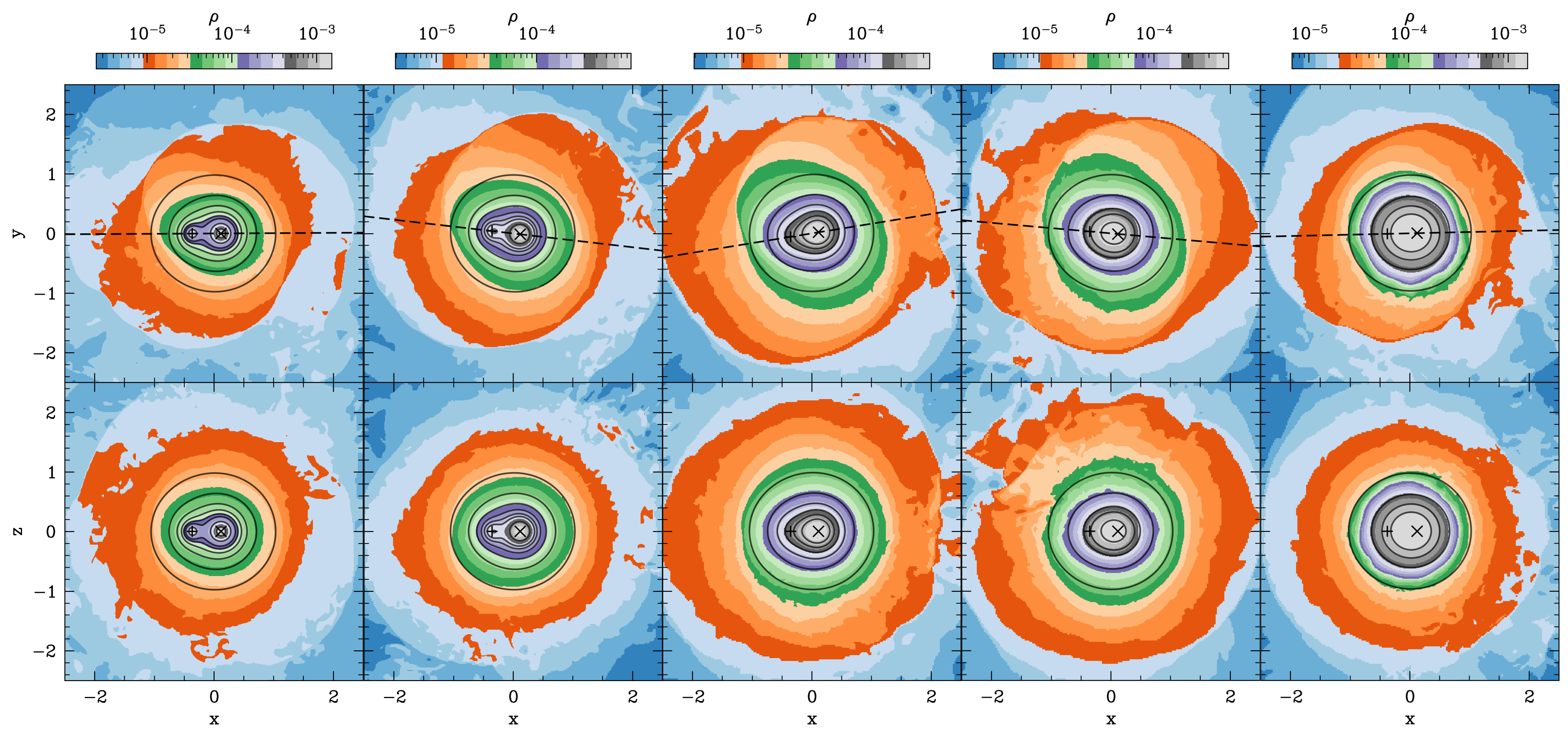}
   \caption{Zoomed-in snapshot of the gas  density cross section in the $xy$ (top row) and $xz$ planes (at $y=0$, bottom row) after $\sim 2100$ orbits, for $\epsilon = 0.05$, $0.1$, $0.2$, $0.3$, and $0.5$ from left to right, for $q= 1/3$ with live orbital separation and using the spline softening method. The black dashed line illustrates the intersection between the orbital plane and the plane orthogonal to the orbital plane and parallel to the $\br_i - \br_j$ vector.}
\label{fig:rhoPhi}
\end{figure*}

\begin{figure} 
\centering
      \includegraphics[width=0.5\textwidth]{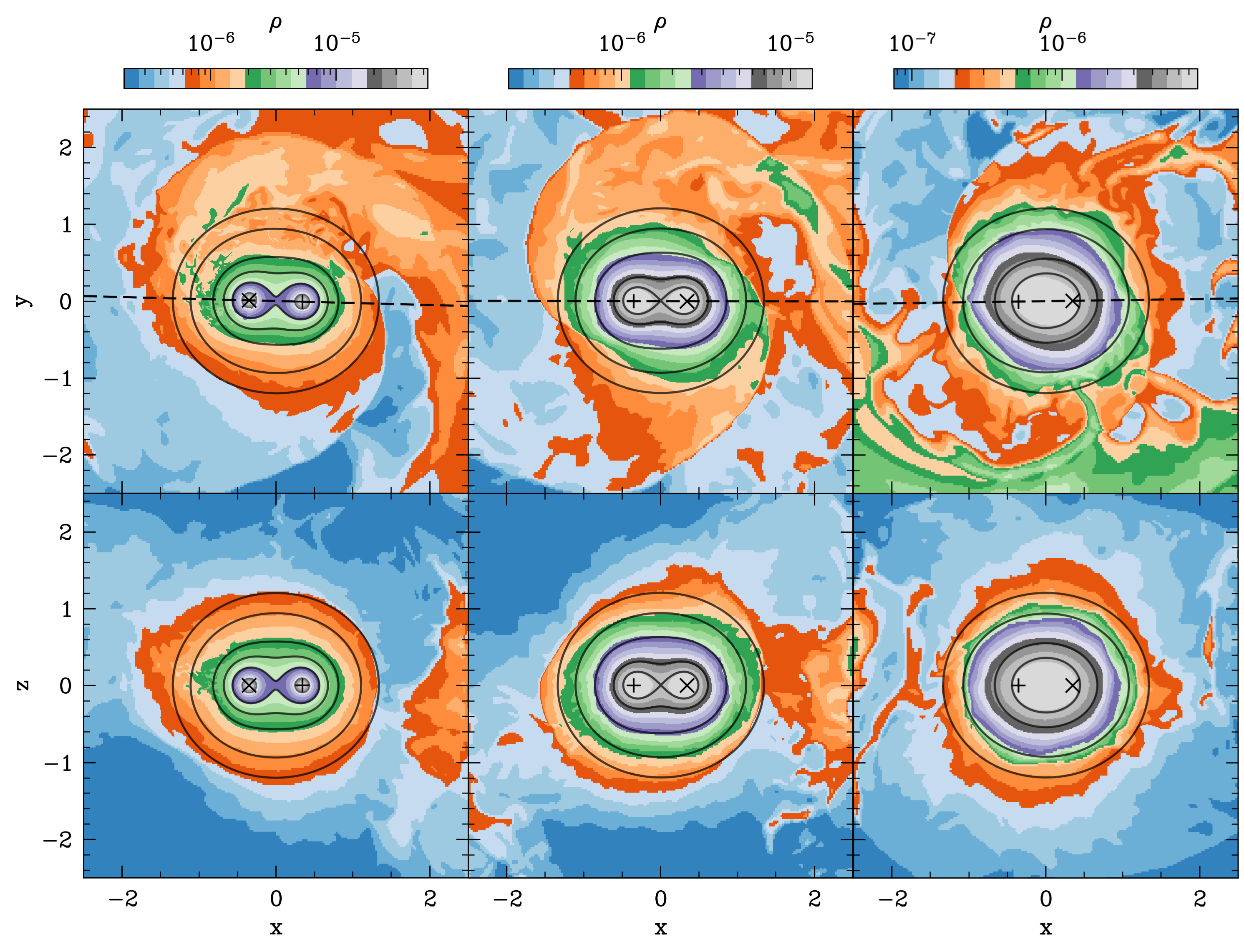}
   \caption{Zoomed-in snapshot of the gas  density cross section in the $xy$ (top row) and $xz$ planes (at $y=0$, bottom row) after $\sim 2100$ orbits, for $\epsilon = 0.05$, $0.2$, and $0.5$ from left to right, for $q= 1$ with live orbital separation and using the spline softening method. The black dashed line illustrates the intersection between the orbital plane and the plane orthogonal to the orbital plane and parallel to the $\br_i - \br_j$ vector.}
\label{fig:rhoPhiq1live}
\end{figure}

\begin{figure*} 
\centering
      \includegraphics[width=\textwidth]{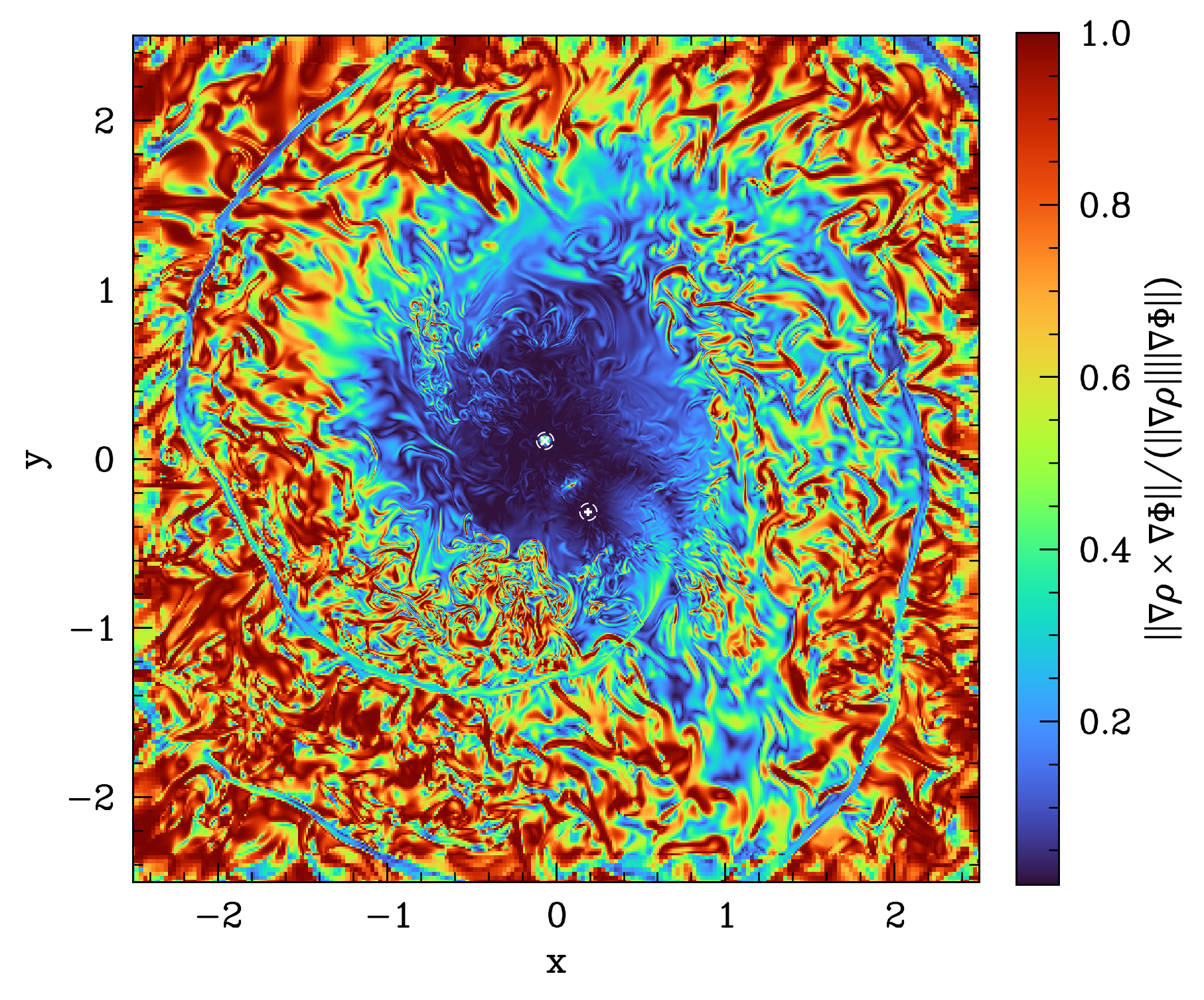}
   \caption{Close-up view of the normalized deviation from hydrostatic equilibrium in the $xy$ plane, after 1500 orbits for simulation run E005.S.l.q03. The cross and plus signs respectively indicate the position of the primary's core and of the companion. The white circles mark the extent of the softened spheres of radius $\epsilon=0.05$.}
\label{fig:HE}
\end{figure*}

In Figs.~\ref{fig:rhoPhiq1}, \ref{fig:rhoPhi}, and \ref{fig:rhoPhiq1live},  we show snapshots of the gas density on the $xy$ and $xz$ planes, as well as gravitational equipotentials, for our $q=1$ and $q=1/3$ runs using the spline softening method. We see that isopycnic surfaces tend to coincide with gravitational equipotentials in the binary's vicinity, in other words, $\bnabla \rho \times \bnabla \Phi \simeq \boldsymbol{0}$. Consequently, the structure around the two cores visually resembles contact binary stars, which share a hydrostatic envelope \citep[e.g.,][]{Lucy1968,Lombardi2011}. To further illustrate the quasihydrostatic structure corotating with the two cores, in Fig.~\ref{fig:HE} we show a snapshot of the normalized deviation from hydrostatic equilibrium, $ \left.\lVert  \bnabla \rho \times \bnabla \Phi \right.\rVert / (\left.\lVert\bnabla \rho \right.\rVert \left.\lVert\bnabla\Phi \right.\rVert)$. We see that the equilibrium condition is satisfied in the immediate vicinity of the two cores.

Hydrostatic equilibrium implies $\rho = \rho(\Phi)$, that is, $\rho$ has the same symmetry properties as $\Phi$. In particular, 
\begin{equation}
    \rho(x_i + x, y_i - y, z) = \rho(x_i - x, y_i + y, z) \ .
\end{equation}
Conversely, $\partial \Phi / \partial \varphi$ is antisymmetric with respect to the plane orthogonal to the orbital plane and parallel to the $\br_i - \br_j$ vector, that is
\begin{equation}
 \left. \frac{\partial \Phi}{\partial \varphi} \right|_{(x_i + x, y_i - y, z)} = - \left. \frac{\partial \Phi}{\partial \varphi} \right|_{(x_i - x, y_i + y, z)} \ .
\end{equation}
Consequently, the gravitational torque density, $s \rho \partial \Phi / \partial \varphi$, is also antisymmetric with respect to such plane, yielding near zero gravitational torque contribution from the quasihydrostatic regions. We emphasize that the reduction in gravitational torque arises from the formation of a corotating quasi-hydrostatic equilibrium structure, rather than from simply gas corotation, as is often assumed.

Maintaining the hydrostatic equilibrium near the cores largely depends on the gravitational softening formulation and on the grid resolution. For instance, while the spline and Ruffert softening methods behave very similarly for $\left.\lVert \br - \br_i \right.\rVert > \epsilon$, the Ruffert softened potential gradient is much weaker than the spline-softened potential gradient within the softening spheres (see Fig.~\ref{fig:grad}). Such a shallower potential can be resolved with fewer cells. However, the potential gradients for both the Ruffert and spline methods reach their maximum values outside the softening radius, at a distance of about $\sim 1.2\epsilon$ from the core (see Fig.~\ref{fig:grad}). Hence, the supposed advantage of the less steep Ruffert potential inside the softening sphere may be rendered ineffective, as the region just outside the softening radius, where the potential gradient is at its maximum, likely dictates the grid resolution requirements. In fact, the use of Ruffert's softening method may be problematic because the associated shallower potential within the softening spheres implies weaker hydrostatic support with less concentrated mass, potentially compromising the stability and mass distribution of the gas in the surrounding regions.

The hydrostatic equilibrium of the gas surrounding the binary is a result of mass accumulation about the point masses. 
Such mass accumulation and equilibrium prevent the formation of non-axisymmetric structures such as circumstellar disks, and is permitted by the absence of core mass accretion in our simulations that may be circumvented by the use of numerical mass sinks, which we discuss in Sect.~\ref{sec:disc}.
However, the ability of the primary's core and/or of the secondary star to accrete matter is likely strongly dependent on their nature and evolutionary stage as well as on the thermodynamical and radiative properties of the gas at the core-envelope interfaces \citep[e.g.,][]{Hjellming1991,Webbink2008}.

\subsubsection{Torques with fixed orbital separation}

\begin{figure} 
\centering
      \includegraphics[width=0.5\textwidth]{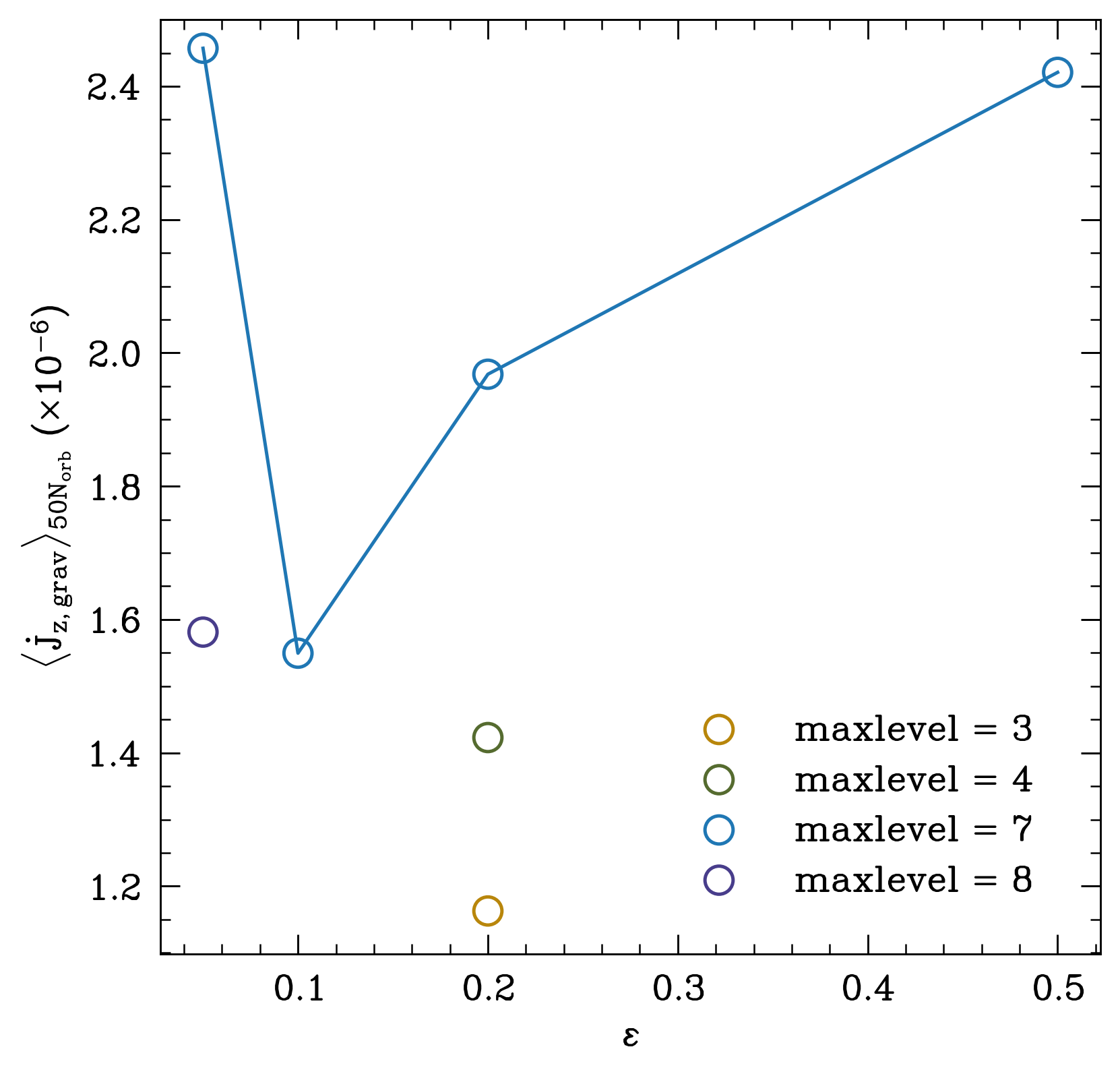}
   \caption{
   Gravitational torque averaged over the last 50 orbital periods as a function of the softening radius $\epsilon$, for all of our $q=1$ simulations with fixed orbital separation.}
\label{fig:jgrav_fix}
\end{figure}

In Fig.~\ref{fig:jgrav_fix}, we show the gravitational torque exerted by the binary on the envelope, averaged over the last 50 orbits, as a function of $\epsilon$ for our simulation runs with $q=1$ and fixed orbital separations (see Table~\ref{tab:runs}). This allows us to isolate the effect of the softening radius using a spline softening method. We observe that smaller softening radii generally lead to reduced gravitational torque at quasisteady state.

This trend occurs because increasing $\epsilon$ increases the critical distance to the cores, beyond which the gravitational potential gradient is significantly underestimated (Eq.~\ref{eq:spline_rel}). As a result, the hydrostatic support of the gas surrounding the binary weakens, making it more vulnerable to perturbations from the surrounding dynamics. These perturbations disrupt the symmetry of the gas distribution, resulting in a net gravitational torque. Therefore, although a larger $\epsilon$ reduces the contribution from existing density asymmetries to the gravitational torque (Fig.~\ref{fig:ratio}), it also promotes gas distribution asymmetries by artificially weakening hydrostatic equilibrium. This effect can be seen in the slight misalignment of isopycnic lines and equipotentials near the cores in Fig.~\ref{fig:rhoPhiq1} for $\epsilon = 0.5$.

However, our simulation run E005.S.f.q1 deviates from this trend, indicating that an asymptotic regime of small $\epsilon$ is not reached. This is because the maximum refinement level of 7 is insufficient to resolve the pressure gradient within the softening sphere of radius 0.05 at all times. As a result, hydrostatic equilibrium is not consistently maintained, and the gas distribution fails to match the symmetry of the gravitational potential, leading to artificial gravitational torques and impacting the large-scale density structure (see Sect.~\ref{sec:env_ej}). By increasing the maximum refinement level to 8 (run E005.S.f.q1.hr), the simulation produces a quasisteady gravitational torque and a mass distribution similar to that observed in run E01.S.f.q1, where the softening radius $\epsilon$ is set to 0.1.

\subsubsection{Torques with ``live'' orbital separation}\label{sec:live}

\begin{figure} 
\centering
      \includegraphics[ width=0.5\textwidth]{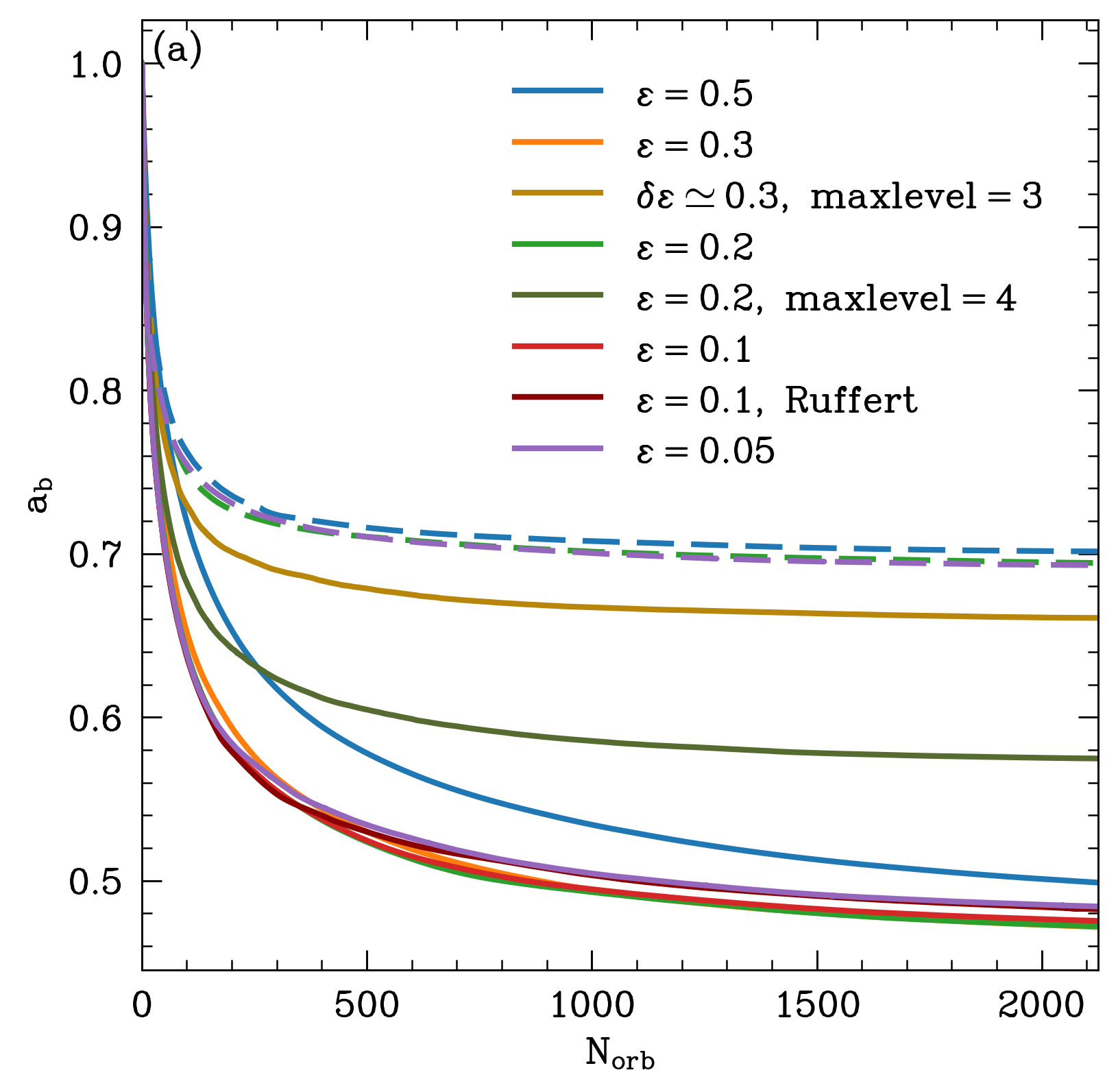}\\
      \includegraphics[ width=0.5\textwidth]{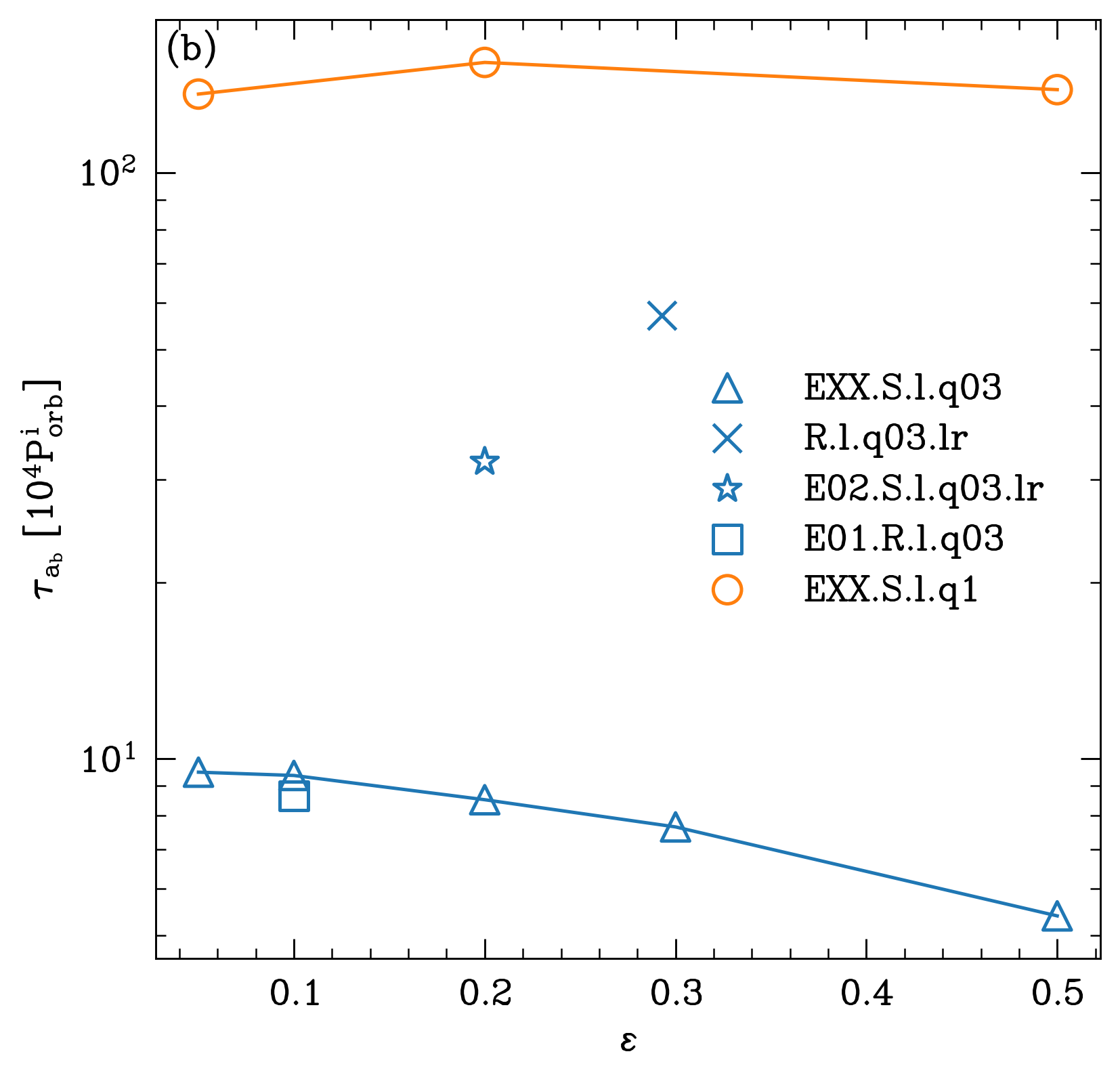}\\
   \caption{Panel (a): Orbital separation evolution for all ``live'' binary simulations (see Table~\ref{tab:runs}). Panel (b): Orbital separation evolution timescale, $\tau_{\ab} = \langle \dot{\ab}/ \ab \rangle_{500N_{\rm orb}}^{-1}$ averaged between orbits 1625 and 2125, as a function of the softening radius $\epsilon$.}
\label{fig:liveb}
\end{figure}

We now aim to understand the role of the gravitational softening formulation and radius, as well as of the intraorbital resolution on the secular evolution of binaries' orbits.  In this work, we fix the orbital eccentricity $\eb$ to zero, and we neglect core accretion. In this simplified setup, the time derivative of the binary's angular momentum can be written as the 
orbital separation evolution equation
\begin{equation}\label{eq:ab}
    \frac{\dot{\ab}}{\ab} = 2\frac{\dot{J}_{z,\rm b} }{J_{z,\rm b} } \ .
 \end{equation}
We use this equation to evolve the binary semi-major axis directly in the simulation.
 
In Fig.~\ref{fig:liveb}, we show the orbital separation evolution  and  the orbital contraction timescale averaged over the last 500 orbital periods as a function of the softening radius, for all of our ``live'' binary simulations (see Table~\ref{tab:runs}). We find that simulations E03.S.l.q03, E02.S.l.q03, E01.S.l.q03, and E005.S.l.q03 yield relatively similar outcome, and that simulations with $\epsilon \leq 0.1$ result in a final orbital evolution timescale of $\sim 10^5\,P_{\rm orb}^i$. For $q=1/3$, we find that the orbital evolution of our simulation with the largest softening radius (simulation E05.S.l.q03) stands out from the other well-resolved simulations.  That is because this simulation does not satisfy  criterion (\ref{eq:condition}). In particular, we find that the contribution from the (extended) softening regions to the total gravitational torque exceeds the total gravitational torque measured in the simulation (see Fig.~\ref{fig:condition}). This simulation may thus not be used to obtain any physical insight nor to make predictions.

Simulations E02.S.l.q03 and E02.S.l.q03.lr are initialized identically, except that the binary's vicinity and the intraorbital region are less well resolved in the latter (see Table~\ref{tab:runs}). We find that, for the same softening radius and method, the less well resolved run results in a final orbital separation approximately 20\% larger than its more resolved counterpart, and an orbital contraction timescale about four times larger. We note that despite having lower resolution in the binary's vicinity than simulation E02.S.l.q03, simulation E02.S.l.q03.lr is still likely to be better resolved than most CEE simulations in the literature. Simulation R.l.q03.lr uses a Ruffert softening method with $\epsilon = 3 \delta$ with a minimum grid cell width $\delta \simeq \ab^f/5$. These parameters broadly represent the late phases of CEE simulations of \cite{Sandquist1998,Sandquist2000}, \cite{Passy2012}, \cite{Staff2016a}, and \cite{Iaconi2018}, for example. Here, we find a final orbital separation approximately 30\% larger and, crucially, an orbital contraction timescale more than ten times larger than in our well-resolved simulations. These results suggest that high spatial resolution is crucial for accurate binary separation evolution, as it is required to capture the relevant spatial scales associated with the physical processes involved in mass redistribution. Achieving a sufficiently small softening radius also requires high resolution to resolve the pressure gradients within and around the softening spheres. 
Conversely, simulations with $q=1$ using three different softening radii, where pressure gradients within the softening spheres are well-resolved, yield similar final orbital separations and contraction timescales. The final separation is 50\% larger compared to binaries with $q=1/3$, while the orbital contraction timescale is approximately ten times longer, around $10^6\ P_{\rm orb}^i$.

We further note that the final separation increases for increasing mass ratio. This has been previously noted by \cite{Passy2012} and \cite{Nandez2016}. However, unlike our work where the total binary mass $M_1 + M_2$ is conserved as the mass ratio varies, their results were based on a constant primary mass $M_1$.
 
Finally, we note that the orbital separation evolution is likely significantly impacted by the absence of core accretion in our simulations. Indeed, mass and angular momentum accretion may have significant impact on the orbital evolution by adding an extra term to Eq.~(\ref{eq:ab}), which would then read
\begin{equation}\label{eq:ab2}
 \frac{\dot{a_{\rm b}}}{a_{\rm b}} = \frac{\dot{M}}{M} \left( 2 \frac{M \dot{J}_{z,\rm b}}{\dot{M}J_{z, \rm b}} - 3\right) \ ,
 \end{equation}
 where in this case $\dot{J}_{z,\rm b} = - \dot{J}_{z,\rm grav} - \dot{J}_{z,\rm accr}$ and $-\dot{J}_{z,\rm accr}$ is the total $z$-directed angular momentum accretion rate onto the binary.
Accretion processes are typically implemented as numerical mass sinks  \citep[][]{Dittmann2021,Dempsey2022}, which may also significantly affect the gravitational torque by hampering the hydrostatic equilibrium of the gas surrounding the binary resulting from mass accumulation (see discussion in Sect.~\ref{sec:disc}). 

\subsection{Envelope ejection and morphology}\label{sec:env_ej}

\begin{figure} 
\centering
      \includegraphics[ width=0.5\textwidth]{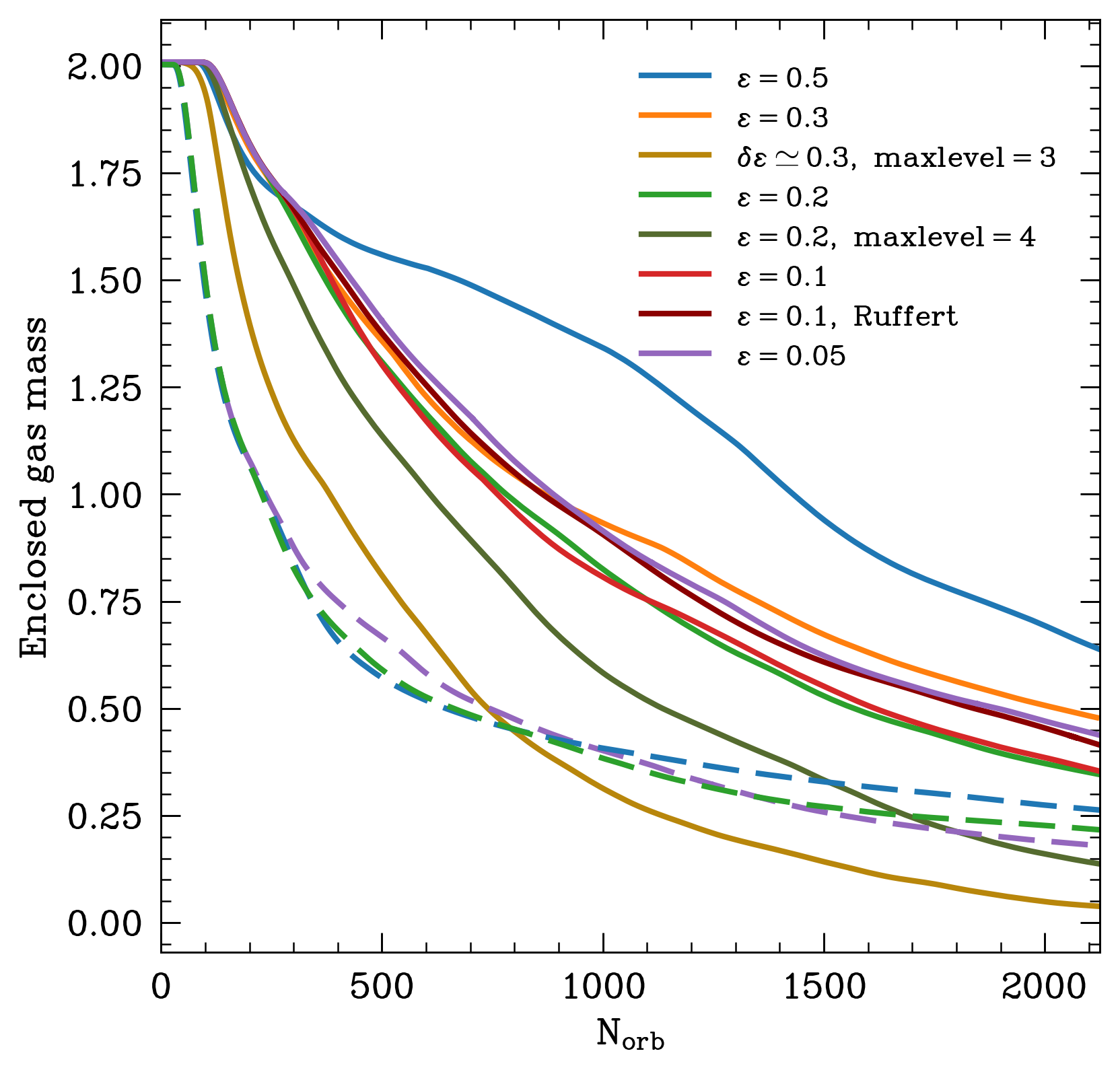}
   \caption{Enclosed gas mass for various values of $\epsilon$ using the spline-softening formulation.}
\label{fig:massspline}
\end{figure}

\begin{figure} 
\centering
      \includegraphics[ width=0.49\textwidth]
      {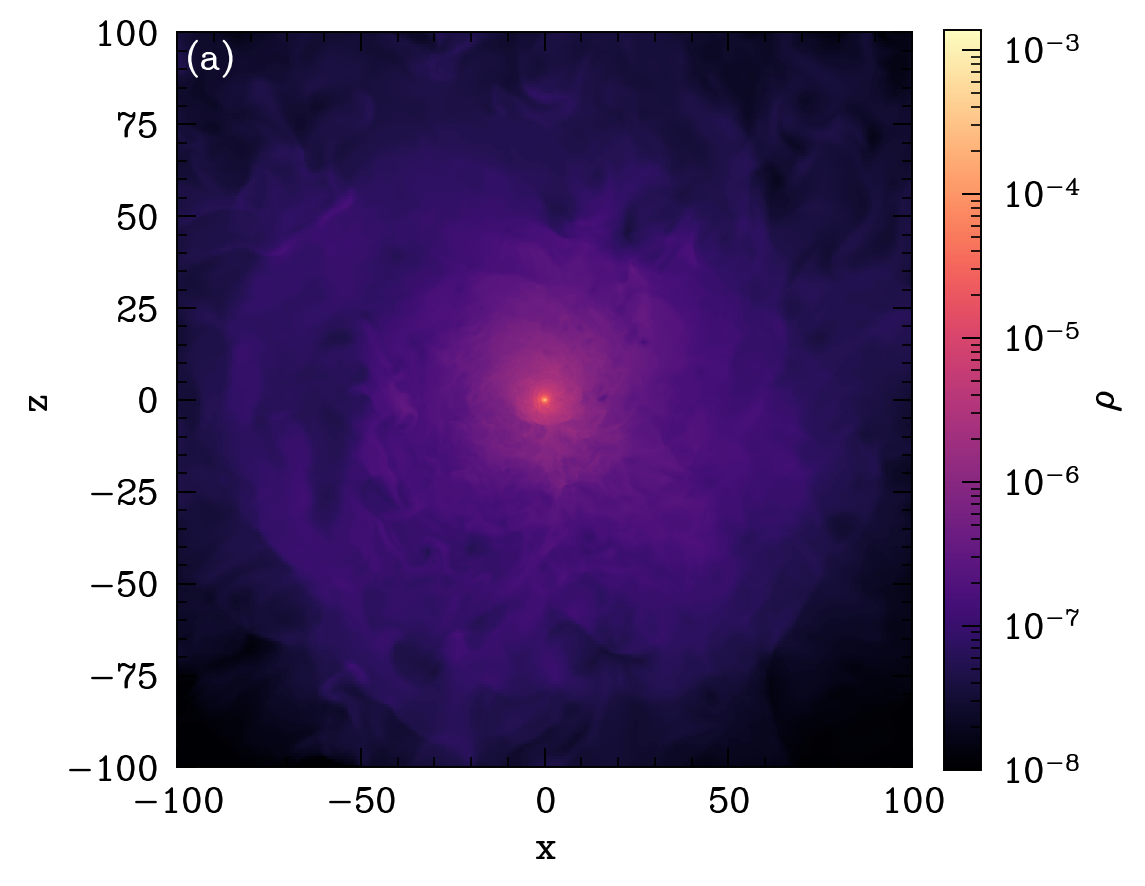}\\
      \includegraphics[ width=0.49\textwidth]
      {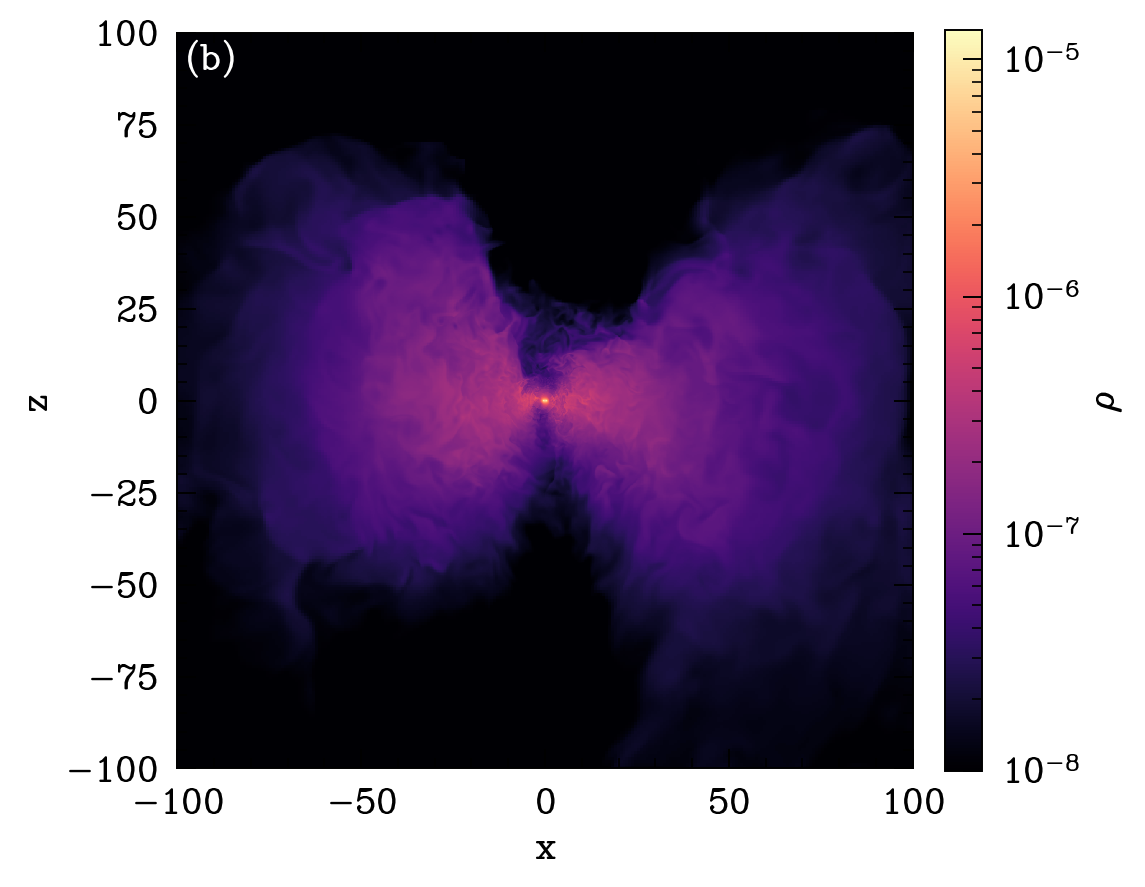}
   \caption{Snapshot of the gas density cross-section in the $xz$-plane after 2150 orbits, for simulations E005.S.l.q033 (panel a) and E005.S.l.q1 (panel b).}
\label{fig:iso_aniso}
\end{figure}

We show the evolution of gas mass within the numerical domain for all our ``live'' simulations in Fig.~\ref{fig:massspline}. The softening radius introduces significant nonlinearity in envelope ejection outcomes. A larger softening radius reduces the available gravitational energy that can be transferred from the binary to the shared envelope. Specifically, a larger $\epsilon$ decreases the gravitational energy available for envelope ejection by reducing the depth of the binary potential. Conversely, a larger $\epsilon$ decreases the binding energy density of the gas near the binary, potentially making the remaining envelope easier to expel.
Additionally, changes in orbital separation influence the amount of gravitational energy injected into the envelope. However, the orbital evolution is also directly affected by the mass of the remaining envelope and its spatial distribution. This interplay between the softening radius $\epsilon$ and the orbital separation makes predicting the precise impact on envelope ejection challenging.

However, extreme cases are more straightforward to interpret. A large softening radius, such as $\epsilon = 0.5$ (e.g., run E05.S.l.q03), significantly limits the total gravitational energy available for envelope ejection, resulting in reduced mass ejection. Similarly, low grid resolution runs (e.g., runs R.l.q03.lr and E02.S.l.q03.lr) fail to capture the relevant spatial scales involved in mass redistribution and do not accurately resolve pressure gradients near the cores. This leads to an inaccurate representation of gravitational forces and energy transfer and, in runs R.l.q03.lr and E02.S.l.q03.lr, in the ejection of (almost) the entire envelope.

\begin{figure*} 
\centering
      \includegraphics[ width=\textwidth]{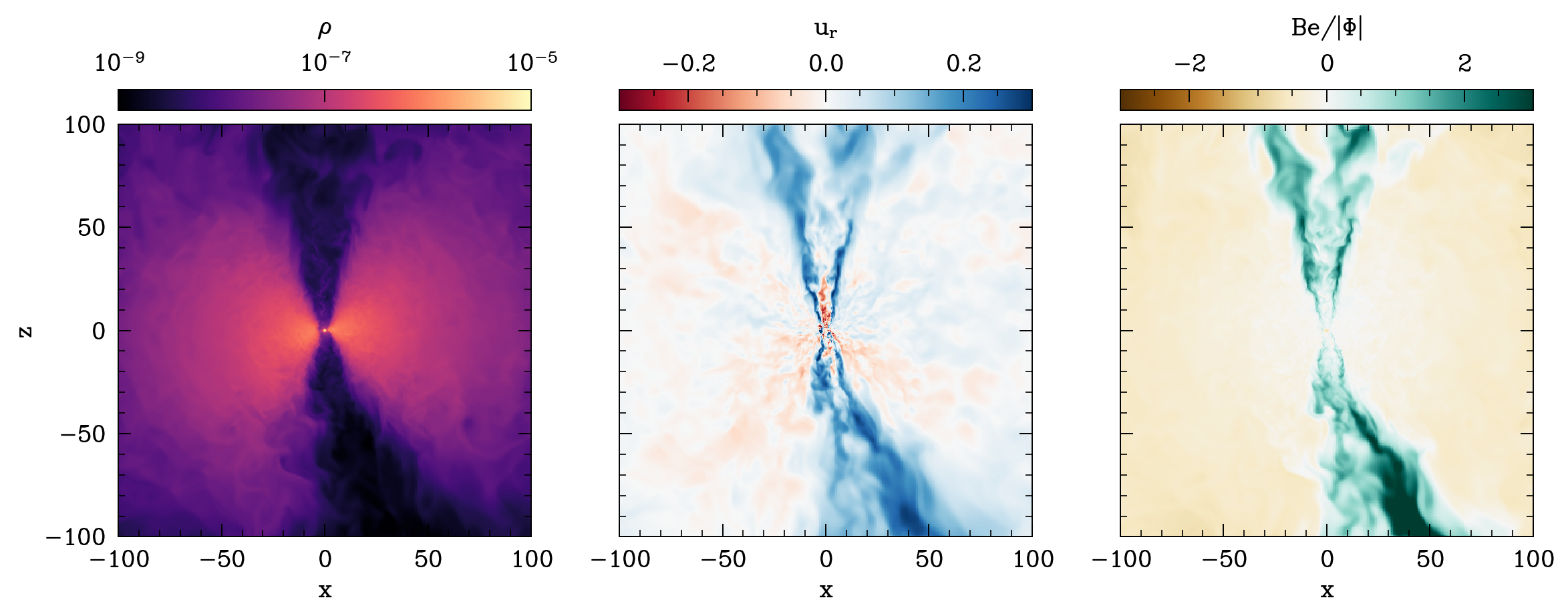}   
            \includegraphics[ width=\textwidth]{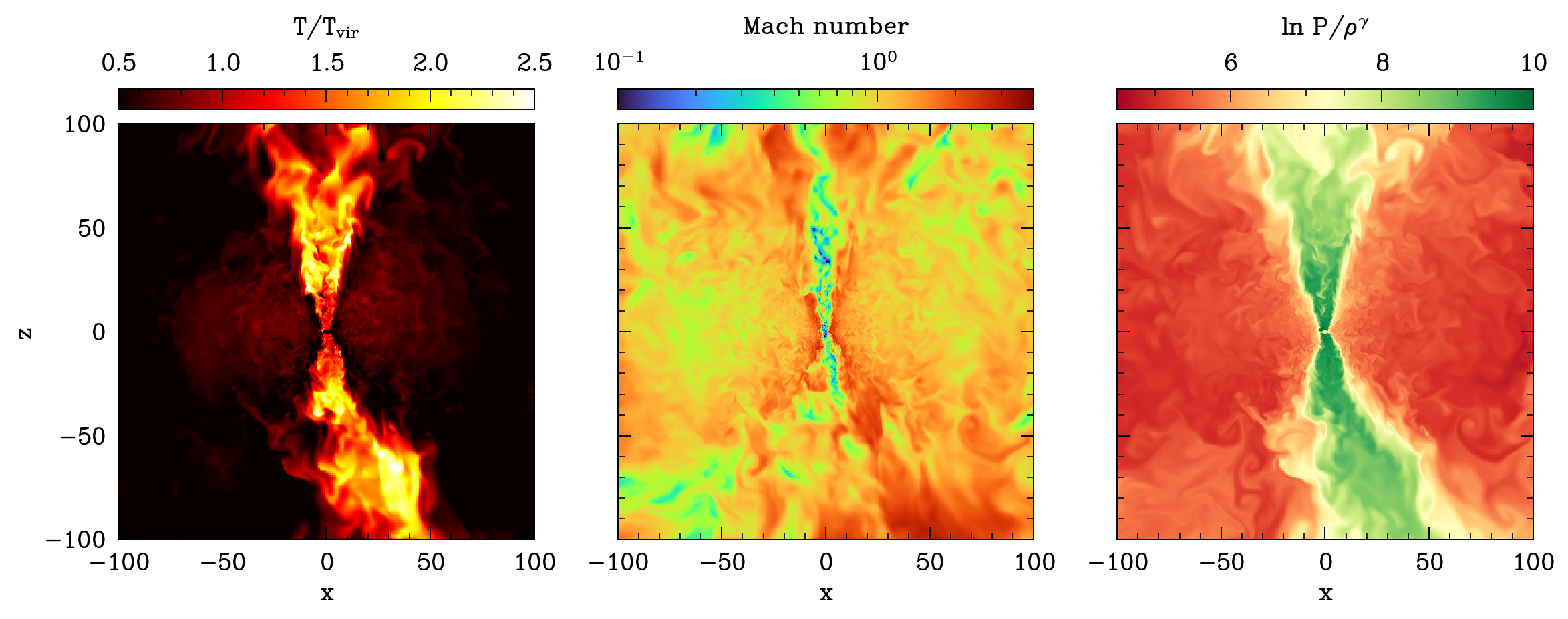}    
   \caption{Snapshots of the gas density, radial velocity, normalized Bernoulli parameter, temperature ratio $T/T_{\rm vir}$, specific entropy, and Mach number in the $xz$ plan for simulation E05.S.l.q1 after $\sim  2250$ orbits.}
\label{fig:opt12e05}
\end{figure*}

In addition to the significant differences in the amount of mass ejected from the computational domain, the large-scale morphology of the remaining envelope strongly depends on the binary's mass ratio, softening radius, and intraorbital numerical resolution. Notably, in all simulations with a mass ratio $q=1/3$, the gas distribution remains nearly isotropic, regardless of resolution or softening radius. In contrast, simulations with $q=1$ result in the formation of a circumbinary disk, driven by the higher amount of angular momentum injected into the envelope. This is illustrated in Fig.~\ref{fig:iso_aniso}, where we show snapshots of the gas density cross-section in the $xz$-plane at the end of simulations E005.S.l.q033 and E005.S.l.q1, as well as in Figs.~\ref{fig:opt12e05}, \ref{fig:opt12e02}, and \ref{fig:opt12e005}.

Finally, despite the absence of magnetic fields in our simulations, we observe jet-like polar outflows in simulations with $q=1$, where sufficient angular momentum has been injected into the envelope to clear, at least partially, a low-density polar funnel and form a circumbinary disk. These polar outflows are present for all three simulations with live orbits and $q=1$. The outflows can alternate between unipolar and bipolar configurations and sometimes take the form of rising hot bubbles when the polar funnel is not yet cleared. We illustrate this in Figs.~\ref{fig:opt12e05}, \ref{fig:opt12e02}, and \ref{fig:opt12e005}, by showing snapshots of the gas density, the radial velocity, and the normalized Bernoulli parameter
\begin{equation}
\begin{aligned}
    \frac{Be}{ |\Phi|} &= \frac{1}{2} \frac{|\buu|^2}{|\Phi|} + \frac{\gamma P }{(\gamma - 1)\rho|\Phi|}  -1 \\
    & = \frac{1}{2} \frac{|\buu|^2}{|\Phi|} + \frac{H}{|\Phi|} -1 \ ,
\end{aligned}
\end{equation}
where $H$ is the enthalpy. The Bernoulli parameter $Be$ represents the total specific energy of a gas parcel. A negative value of $Be$ indicates that the gas parcel is gravitationally bound to the binary, while a positive value of $Be$ indicates that it is unbound. In these figures, we also show the gas temperature ratio $T/T_{\rm vir} = 3P/(\rho|\Phi|)$, the Mach number, and the specific entropy of the gas $\ln P/\rho^\gamma$.

\begin{figure*} 
\centering
      \includegraphics[ width=\textwidth]{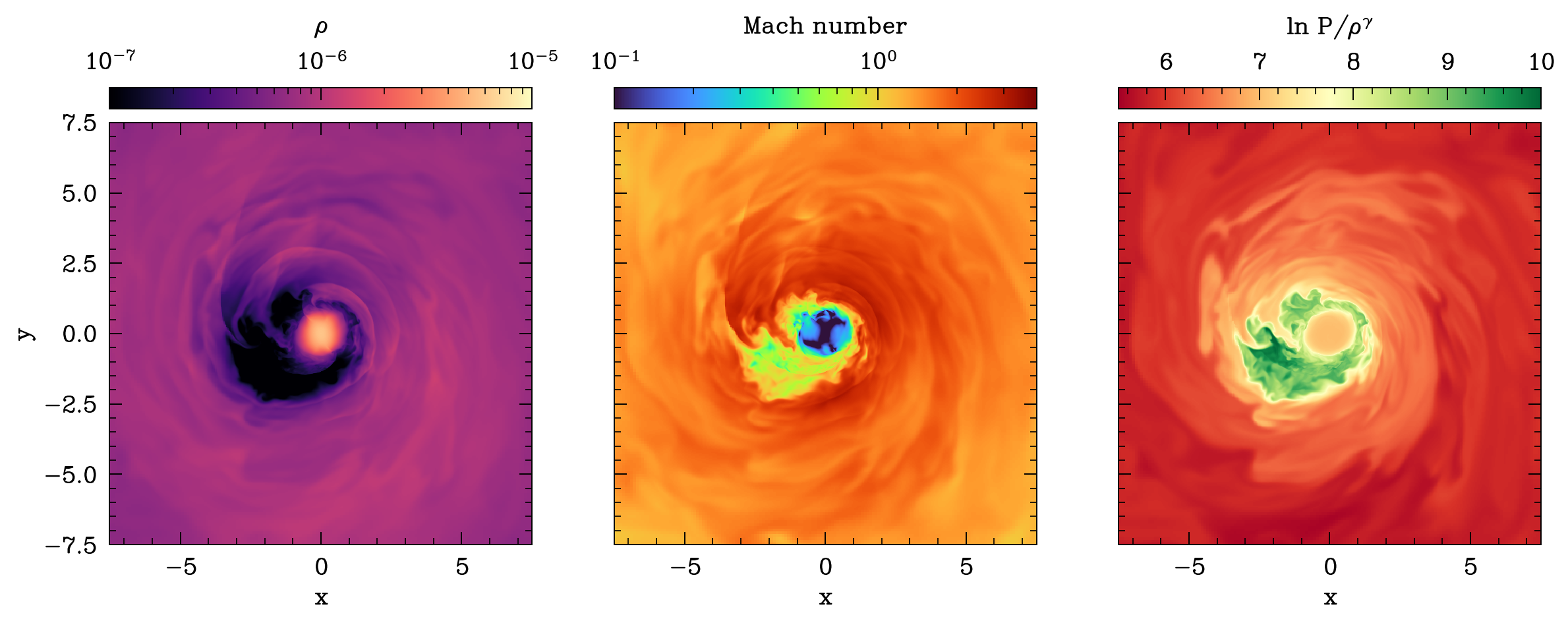} 
   \caption{Snapshots of the gas density, Mach number, and specific entropy in the $xy$ plan for simulation E02.S.l.q1 after $\sim  2250$ orbits.}
\label{fig:opt4e05}
\end{figure*}

\begin{figure*} 
\centering
      \includegraphics[ width=\textwidth]{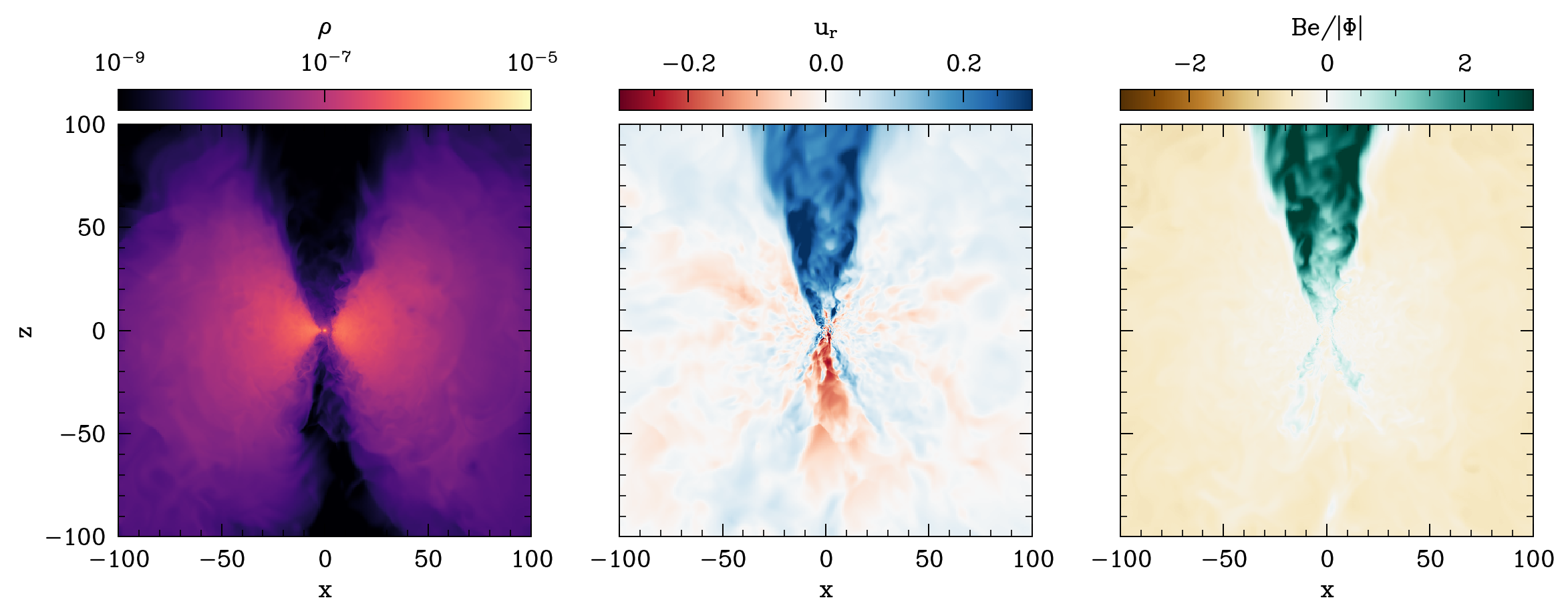}  
    \includegraphics[ width=\textwidth]{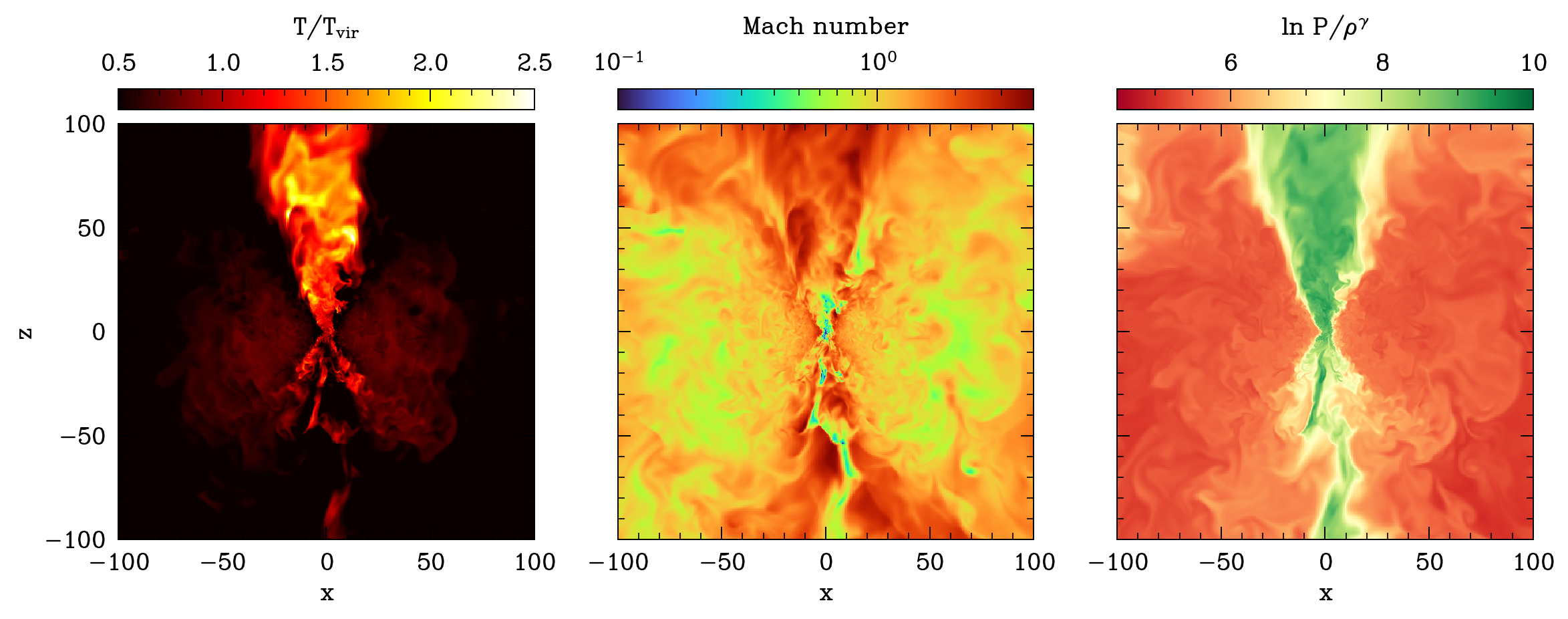}    
   \caption{Same as Fig.~\ref{fig:opt12e05}, but for simulation run E02.S.l.q1.}
\label{fig:opt12e02}
\end{figure*}

\begin{figure*} 
\centering
      \includegraphics[ width=\textwidth]{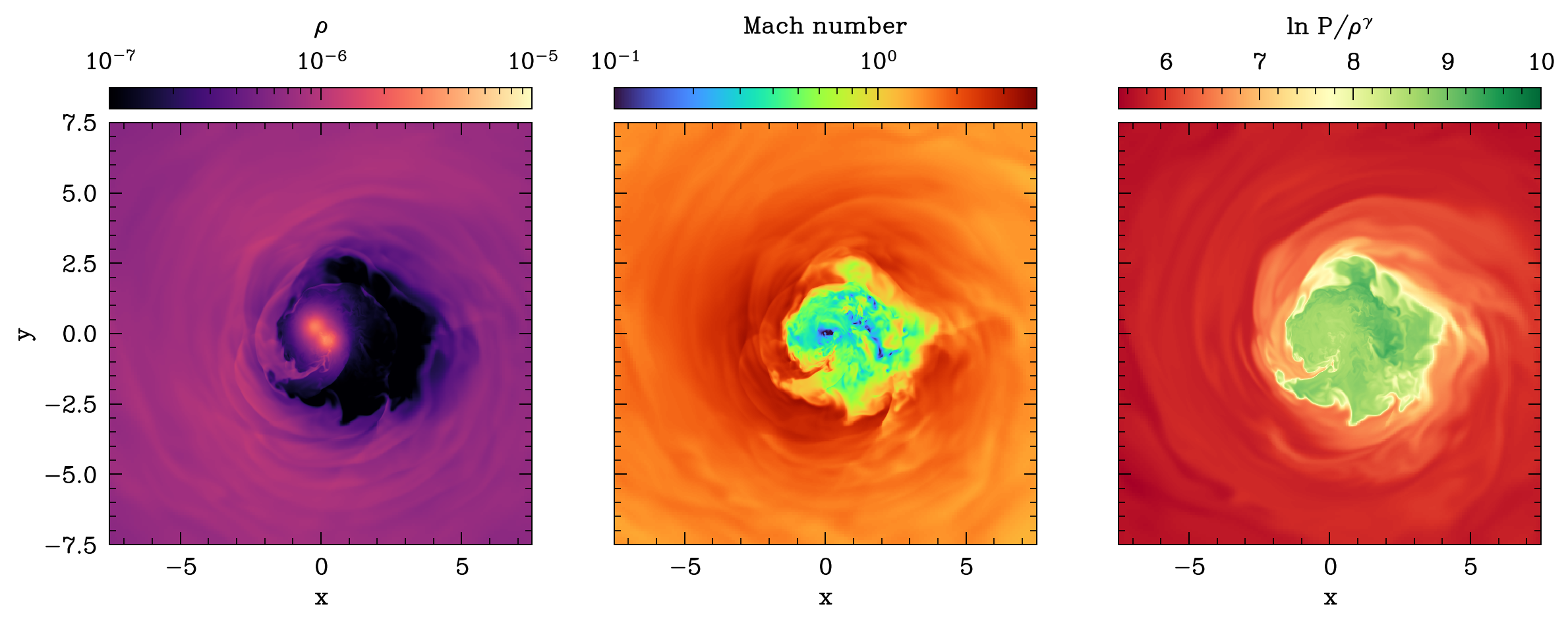} 
   \caption{Same as Fig.~\ref{fig:opt4e05}, but for simulation run E02.S.l.q1.}
\label{fig:opt4e02}
\end{figure*}

\begin{figure*} 
\centering
      \includegraphics[ width=\textwidth]{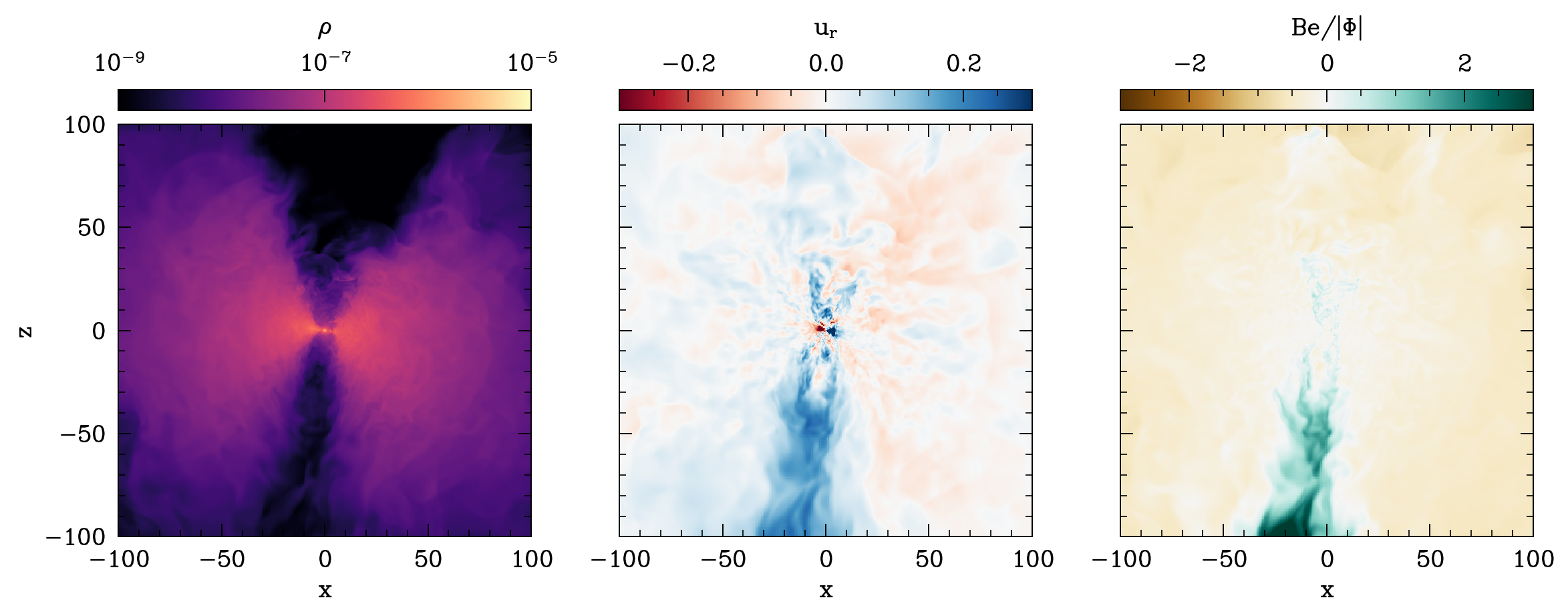}  
    \includegraphics[ width=\textwidth]{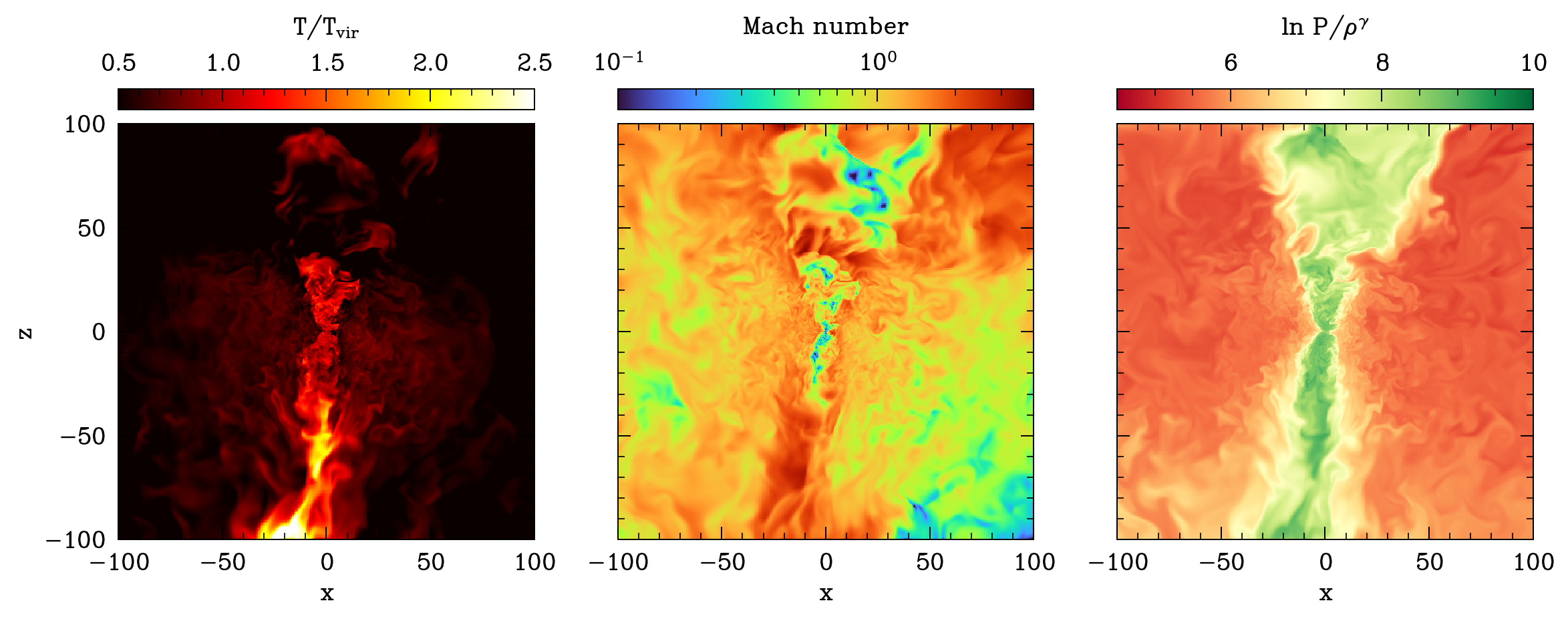}    
   \caption{Same as Fig.~\ref{fig:opt12e05}, but for simulation run E02.S.l.q1.}
\label{fig:opt12e005}
\end{figure*}

\begin{figure*} 
\centering
      \includegraphics[ width=\textwidth]{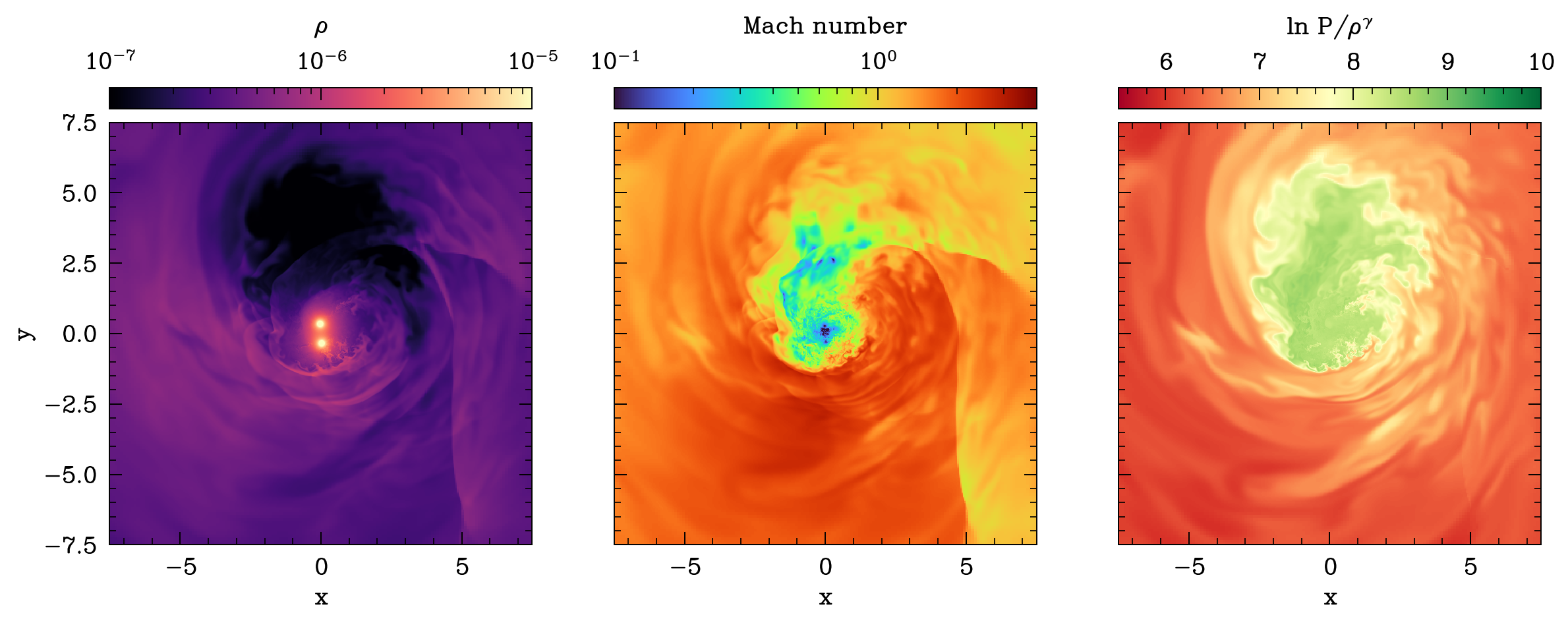} 
   \caption{Same as Fig.~\ref{fig:opt4e05}, but for simulation run E005.S.l.q1.}
\label{fig:opt4e005}
\end{figure*}

We find that these polar outflows are characterized by positive Bernoulli parameters, suggesting that they have sufficient kinetic and thermal energy to overcome the binary's gravitational potential, and therefore be effectively unbound.
We further find that these polar outflows are very hot compared to the circumbinary disk and have high specific entropy.  This suggests that the accreting disk material is strongly heated in the binary's vicinity, and that part of this heating is irreversible. The heated material then may expand along the polar axis, either freely as a pressure-driven jet-like outflow \citep[e.g.][]{Nobuta1999}, or as buoyant hot bubbles. In our in simulations, the main source of gas heating in the binary's vicinity is shock-heating, which can be decomposed into two main contributions, adiabatic compression $- P \bnabla \cdot \buu$, and irreversible heating from viscous dissipation at the shock front, the latter being the source of specific entropy in our simulations. 

In our three live binary simulations with a mass ratio of $q=1$, we observe a predominantly unipolar outflow that persists for most of the simulations' duration. This outflow arises as part of the kinetic energy from the cold inflowing material, typically coming from the opposite pole, is converted to thermal energy through shock heating as it collides with the hydrostatic structure surrounding the cores. This results in a fast, hot, high-entropy jet-like structure propagating outward along the polar axis (Fig.~\ref{fig:opt12e02}). The strong vertical pressure gradient resulting from the shock heating in the binary's vicinity can eventually quench polar accretion, and, doing so, limit the polar outflow in the opposite hemisphere (Fig.~\ref{fig:opt12e005}).

In the absence of strong polar accretion, the existence of a polar outflow depends on the binary's ability to clear a centrifugally supported low-density cavity around itself. This process inhibits mass accretion onto the binary and leads to the generation of additional shocks. These shocks occur where accretion streams, drawn from the inner edge of the disk, collide with the hydrostatic structure surrounding the two cores (e.g., Fig.~\ref{fig:opt4e02}), as well as at the cavity edge where the streams collide \citep[e.g.,][]{Farris2015a}. The increase in temperature and entropy associated with these shocks is dependent on the pre-shock Mach number of the accretion streams. This Mach number is directly related to the size of the cavity, the boundary of which corresponds to the location of the centrifugal barrier. In our simulations, the cavity size is linked to the prescribed gravitational softening length around the cores, with larger softening radii resulting in a more underestimated gravitational torque in the vicinity of the binary (see Appendix~\ref{app:A}). This can be seen by comparing Figs.~\ref{fig:opt4e05}, ~\ref{fig:opt4e02}, and ~\ref{fig:opt4e005}, for instance. These figures also show that, regardless of the softening radius, the disk is off-centered and the cavity is elliptical. This has already been observed in a large number of hydrodynamical simulations of circumbinary disks around equal mass ratio binaries in a circular orbit \citep[e.g.,][]{Shi2012,Penzlin2022,Penzlin2024} and is associated with the disk's eccentricity \citep[see also][]{Gagnier2023}. Such eccentricity is thought to be excited by stream impacts at the cavity edge \citep[e.g.,][]{Shi2012} and/or by resonant Lindblad excitation \citep[e.g.,][]{Lubow1991a, Ogilvie2007}, and to be a long-lived mode trapped by the steep density gradient at the cavity edge \citep[][]{Munoz2020}. The disk eccentricity may in turn excite the orbital eccentricity of the binary \citep[e.g.,][]{Papaloizou2001}. Understanding the excitation mechanisms and long-term evolution of orbital eccentricity is crucial, as eccentricity in post-CEE binaries strongly affects both the merger rates and the properties of their gravitational wave signals. Investigating this in more detail should be the focus of future studies.

Simulation run E05.S.l.q1 represents a special case. The large softening radius of 0.5 leads to the formation of a nearly spherical, quasihydrostatic, and gravitationally bound gas bubble surrounding the binary (Sect.~\ref{sec:3.2.1}). Within this bubble, the gravitational potential gradient is significantly smoothed, which inhibits strong shocks or compression. Consequently, the bubble maintains low entropy and low Mach number. As the binary begins to clear a low-density cavity, the bubble becomes isolated from the disk. Direct collisions between supersonic accretion streams and such isolated bubble results in the formation of strong shocks at the bubble's boundary, dramatically increasing local heat and specific entropy production. This process generates a hot bipolar outflow that efficiently clears the polar funnel in both hemispheres. The bubble eventually erodes due to mixing with hotter, higher-entropy material in the cavity and destabilization caused by compression from the accretion streams, leading to its dissolution. The collapse of the bubble results in reduced entropy production, weakening the polar outflow and leading the funnel and cavity to partially refill. In simulation run E05.S.l.q1, this bipolar outflow persists from approximately the 1900th to the 2400th orbit. The bubble may eventually reform, potentially leading to the recurrence of jet-like outflows.

To conclude, non-magnetic jet are an interesting phenomenon, but a more detailed study of jet-launching mechanisms is beyond the scope of this work and will be addressed in future studies.

\subsection{Kinetic helicity}
\label{sec:helicity}

Large-scale open magnetic fields are crucial for the acceleration and collimation of magnetized jet-like outflows in accretion disks \citep[e.g.,][]{konigl2000,spruit2010}, however, their origin remains unclear. Such fields threading the disk could either result from large-scale dynamo within the disk \citep[e.g.,][]{Branden05,armitage1998}, or be advected from an external weak field \citep[e.g.,][]{spruit2005,Salvesen2016,Li2019}. In the context of CEE, the launching of such bipolar jet-like outflows may give rise to the characteristic bipolar shapes often observed in planetary nebulae  \citep[e.g.,][]{GarciaSegura1999,GarciaSegura2018,GarciaSegura2020,Zou2020,Ondratschek2022, Vetter2024}.

\begin{figure*} 
\centering
      \includegraphics[ width=0.49\textwidth]{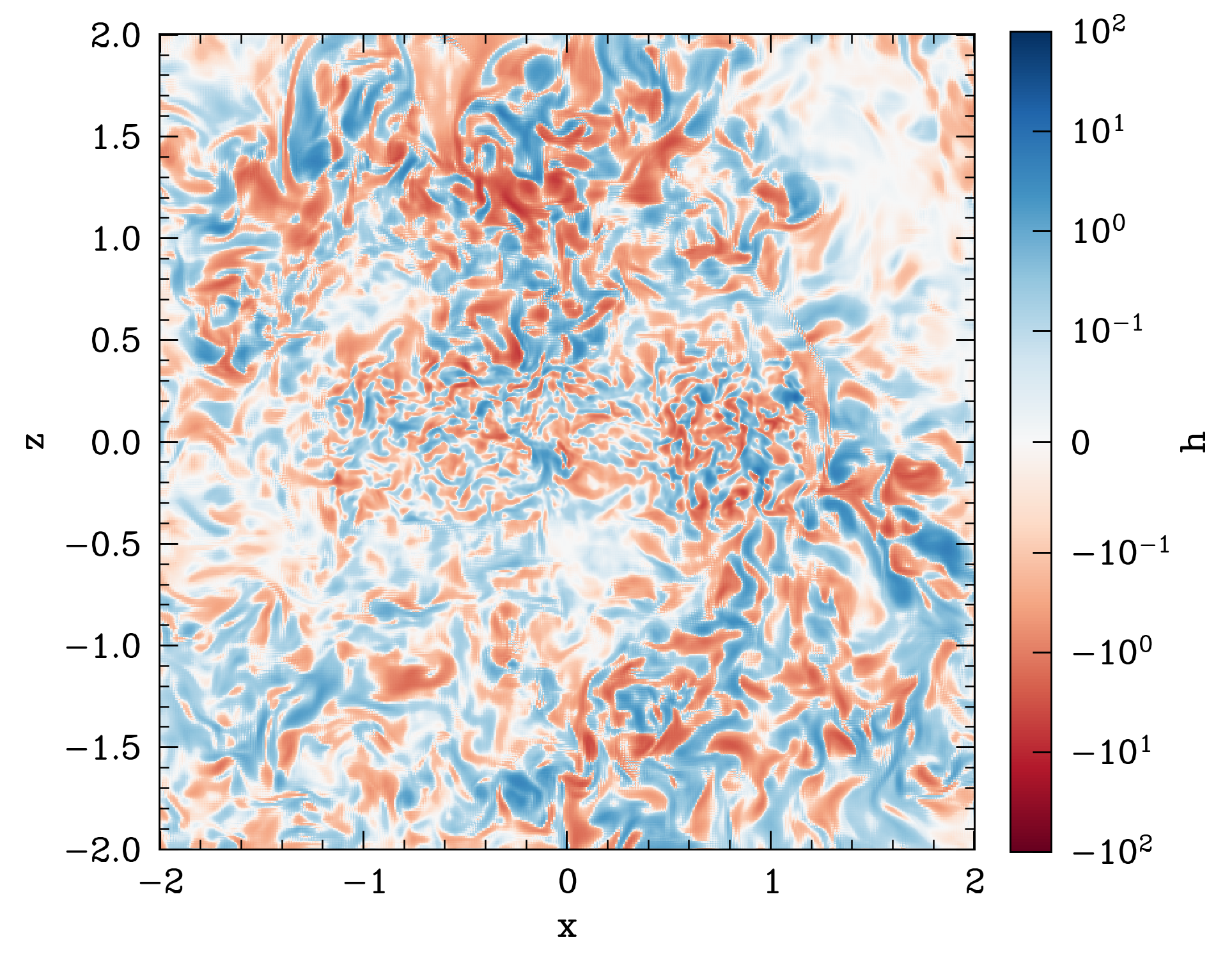}  
      \includegraphics[ width=0.49\textwidth]{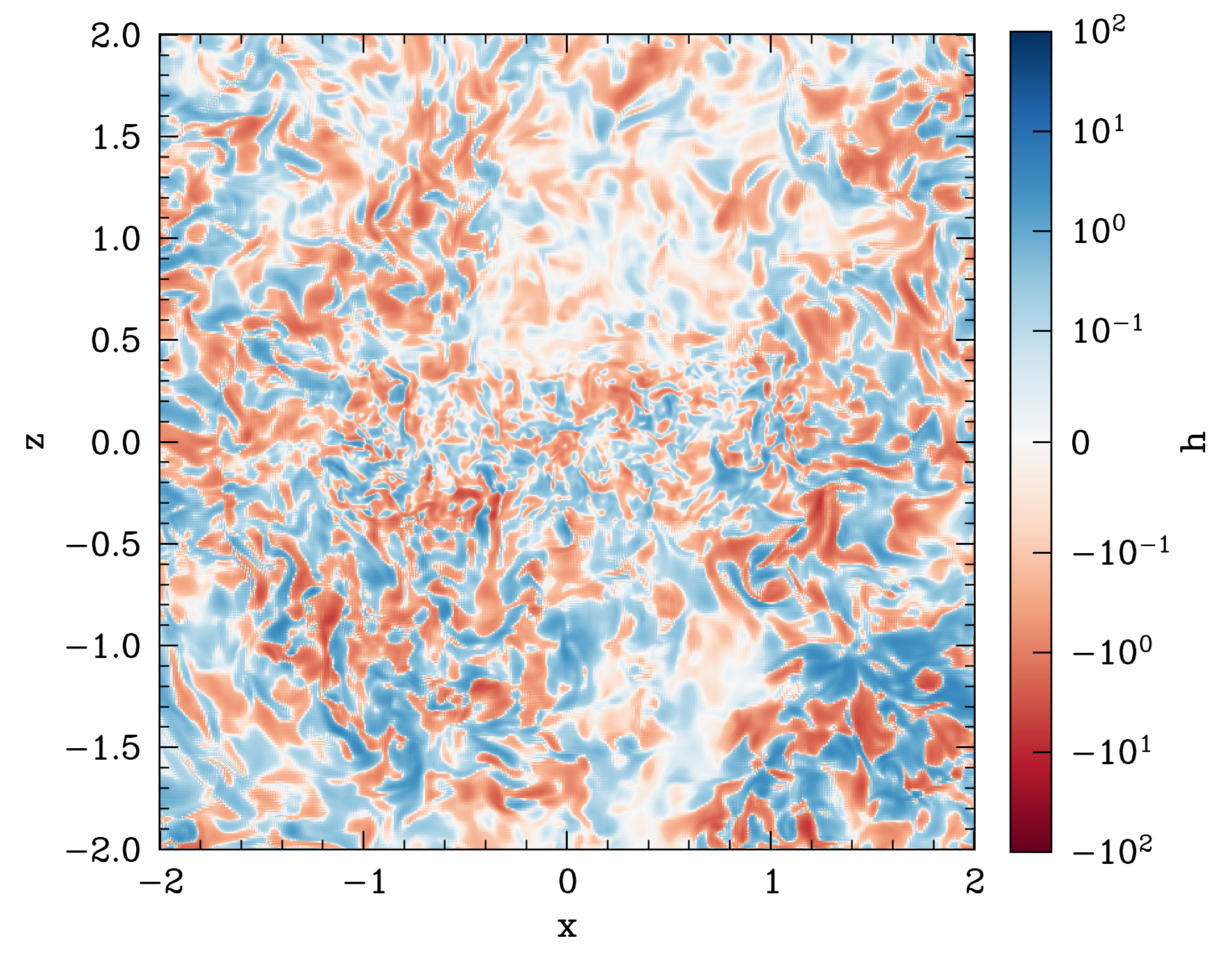}  
   \caption{Zoomed-in snapshot of the kinetic helicity $h = \buu^{\prime\prime} \cdot \bnabla \times  \buu^{\prime\prime}$ cross section at $t = 41~P_{\rm orb}^i$ (left) and   $t = 2000~P_{\rm orb}^i$ (right), for simulation E01.S.l.q03 (see Table~\ref{tab:runs}). }
\label{fig:h}
\end{figure*}

Helicity, a measure of the chirality of the flow, that is of its lack of mirror symmetry, has long been known to be of crucial importance in turbulent dynamo theory by being responsible for the $\alpha$-effect \citep[][]{SKR66}, which is key to the growth and the maintaining of large-scale magnetic fields \citep[e.g.,][]{Moffatt2014}. In the context of planetary and solar dynamo, if the helicity density is uniformly of one sign in the northern hemisphere, and of the other sign in the southern hemisphere, the local electromotive forces will add up in each hemisphere, giving rise to global currents which will support a large scale (dipolar) magnetic field \citep[e.g.][]{brandenburg_etal12}.  In other applications where there is no spatial segregation of the helicity density, the magnetic field usually ends up being disorganized, and of a scale that is not greater than the turbulent motion that generated it. In the context of common envelopes, \cite{Gagnier2024} suggest the absence of large-scale magnetic field production, with a maximum of magnetic energy amplification on the scale of the binary-driven spiral density waves wavelength. Unlike \cite{Gagnier2024} however, here we also include the region inside the orbit to our numerical domain. It is therefore of great interest to see whether linking intra- and extra-orbital dynamics may affect the development of large scale magnetic fields. Simulations in this work are purely hydrodynamical, we therefore restrict ourselves to simple predictions that we will verify in future magnetohydrodynamical studies. 
 The kinetic helicity density reads
\begin{equation}
    h = \buu^{\prime\prime} \cdot \bnabla \times  \buu^{\prime\prime} \ ,
\end{equation}
where 
\begin{equation}
    \buu^{\prime\prime} = \buu - \widetilde{\buu} \ ,
\end{equation}
is the velocity perturbation, $\widetilde{\buu} = \overline{\rho \buu}/\overline{\rho}$ is the Favre averaged velocity  \citep[e.g.,][]{Gagnier2024}, and
\begin{equation}
    \overline{X} = \int_{t^\prime-P_{\rm orb}}^{t^\prime+P_{\rm orb}} X  \dd t^\prime \ .
\end{equation}
We show a zoomed-in snapshot of the kinetic helicity density cross section at $t = 40\,P_{\rm orb}^i$ and  $t = 2000\,P_{\rm orb}^i$, for simulation E01.S.l.q03  in Fig.~\ref{fig:h}. We find no distinguishable kinetic helicity density segregation in our simulations, as well as near zero kinetic helicity $H = \int h \dd V$. This suggests the absence of large-scale magnetic field growth (from the usual $\alpha$-effect), but does not rule it out. Indeed, the inhomogeneous kinetic helicity density spatial fluctuations may still produce large scale magnetic fields by interacting with the binary-driven large scale differential rotation, analogously to the $\alpha\Omega$ dynamo \citep[e.g.][]{Silantev2000,Proctor2007,Kleeorin2008,Sridhar2014}. Similarly, the time fluctuations of the kinetic helicity $H$ may lead the growth of large scale magnetic fields,  both  in the presence and absence of large scale shear \citep[e.g.][]{Mitra2012}. This question will be studied in more details in future works.



\subsection{Shear rate}
\label{sec:shear}

\begin{figure*} 
\centering
      \includegraphics[ width=\textwidth]{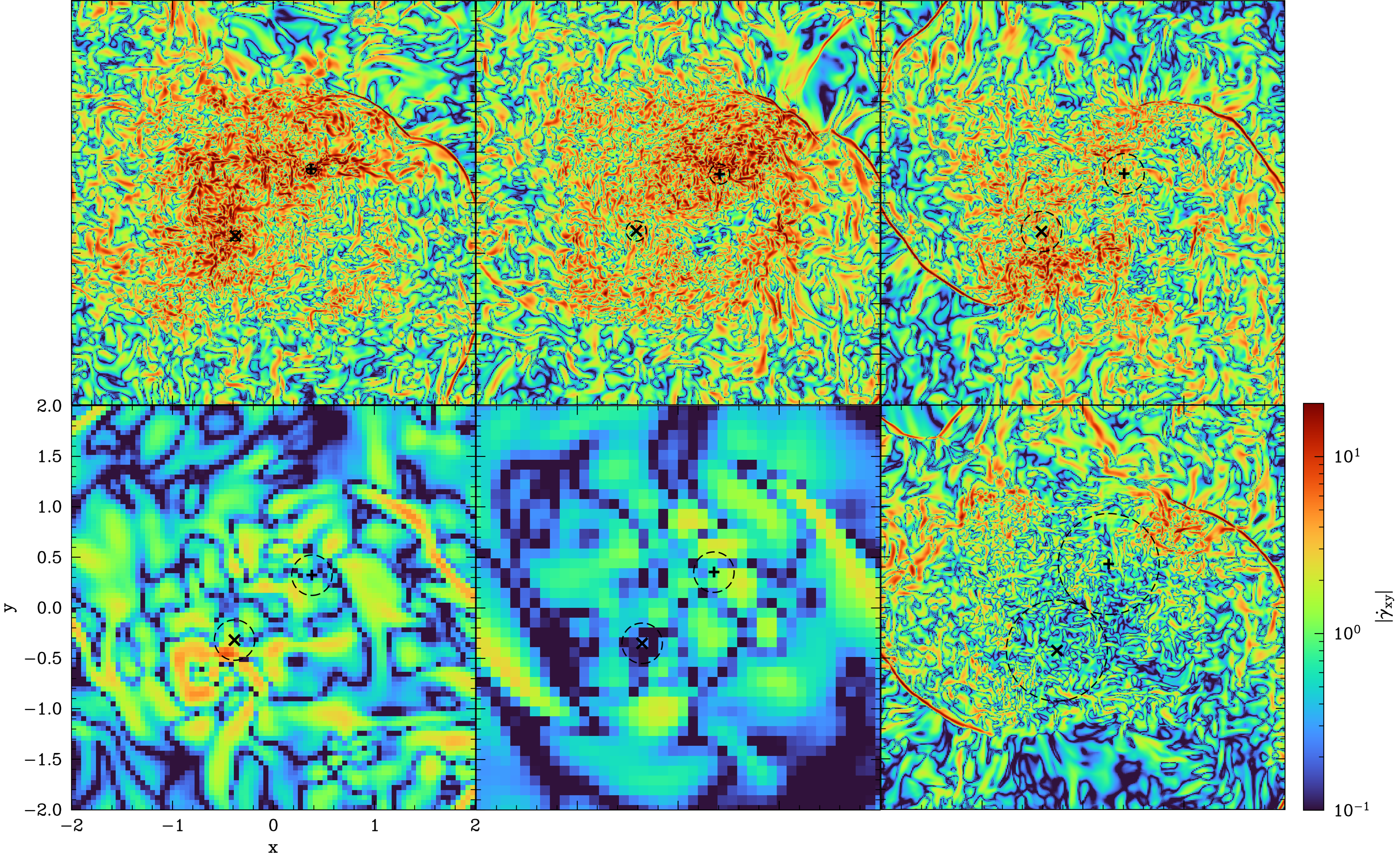}
   \caption{Absolute horizontal shear rate in the orbital plane after $\sim 300$ orbits with fixed separation, for $q=1$ with a spline softening formulation, and for $\epsilon= 0.5$, $0.2$, $0.1$, and $0.05$, increasing from left to right and top to bottom in a clockwise direction. The bottom left and bottom center snapshots respectively correspond to simulation runs E02.S.f.q1.lr and E02.S.f.q1.vlr  (see Table~\ref{tab:runs}). The cross and plus signs respectively indicate the position of the primary's core and of the companion. The black circles mark the extent of the softened spheres of radius $\epsilon$. }
   \label{fig:Shearq1}
\end{figure*}

\begin{figure*} 
\centering
      \includegraphics[ width=\textwidth]{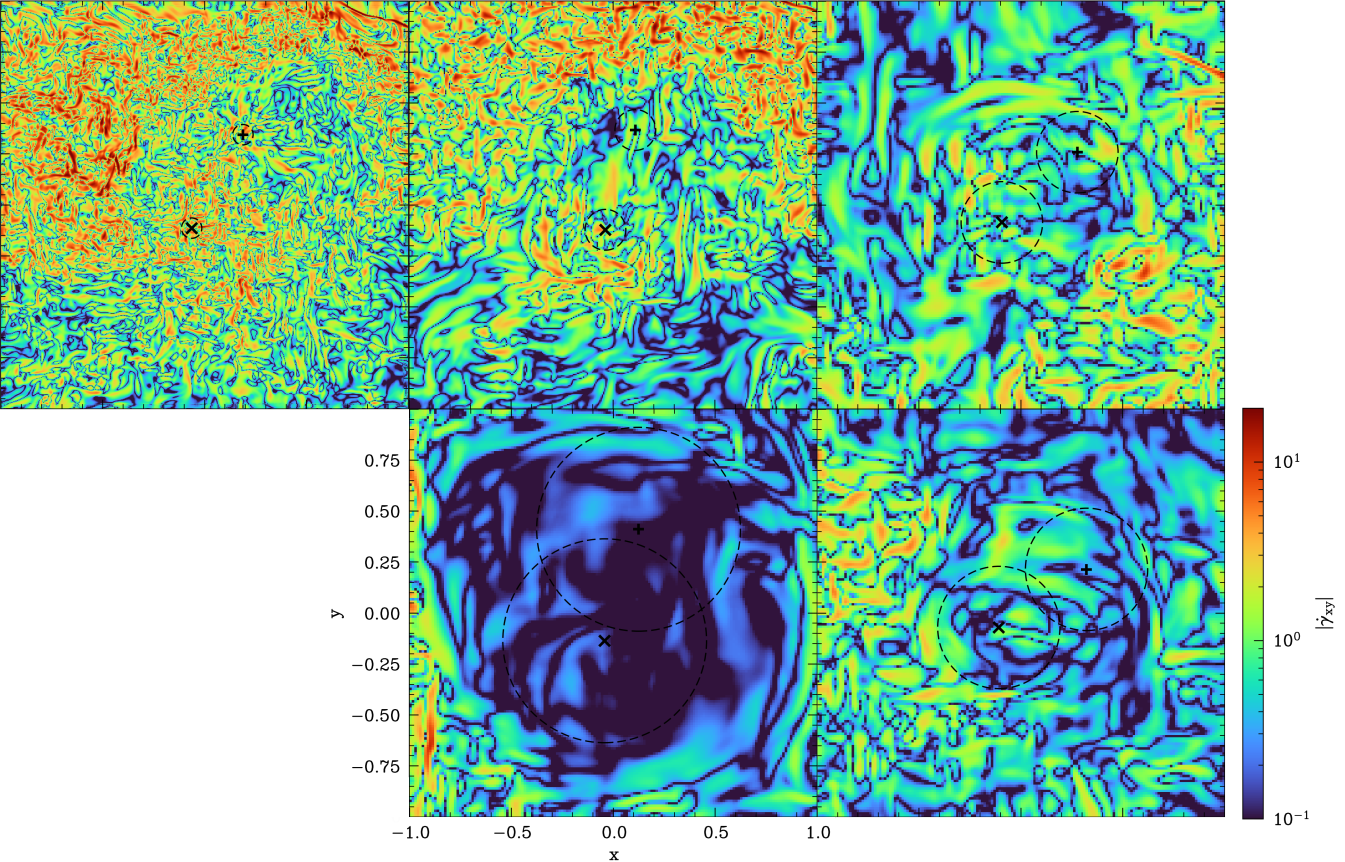}
   \caption{Absolute horizontal shear rate on the orbital plane after $\sim 650$ orbits, for $q=1/3$ with a spline softening formulation, and for $\epsilon= 0.5$, $0.3$, $0.2$, $0.1$, and $0.05$, increasing from left to right and top to bottom in a clockwise direction. The cross and plus signs respectively indicate the position of the primary's core and of the companion. The black circles mark the extent of the softened spheres of radius $\epsilon$. }
   \label{fig:Shearq033}
\end{figure*}

\begin{figure} 
\centering
      \includegraphics[ width=0.5\textwidth]{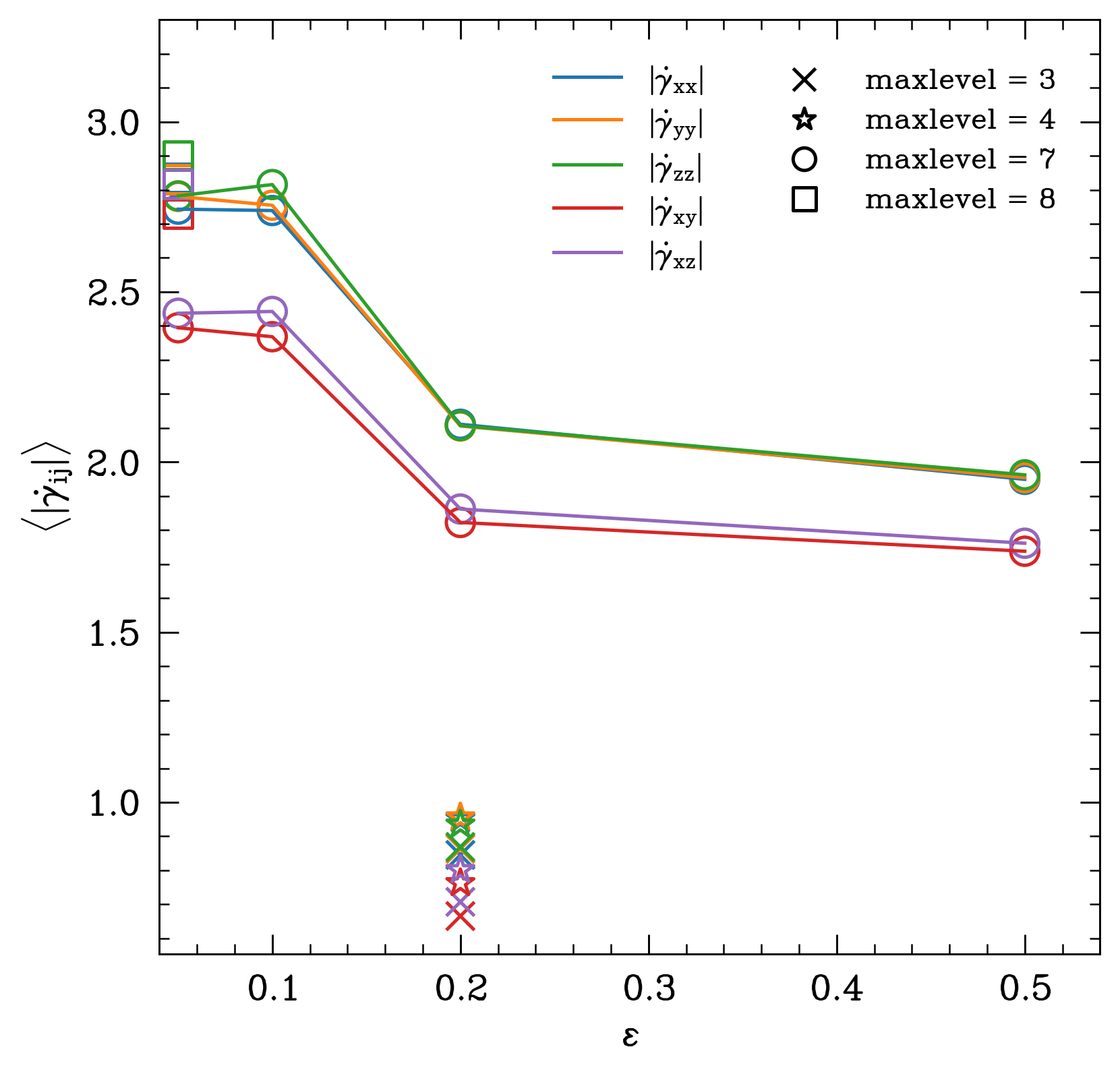}
   \caption{Volume- and time-averaged absolute shear rate within a cube of side length $2\ab$ centered on the binary's center of mass, averaged over the last 50 orbital periods for simulations with $q=1$ and fixed orbital separation, as a function of the softening radius $\epsilon$.}
\label{fig:abShear}
\end{figure}

In the context of ideal magnetohydrodynamics, the magnetic energy density evolution equation reads 
\begin{equation}
    \frac{\partial E_B}{\partial t} = \left(\bB \otimes \bB \right) {:} \bnabla \buu - E_B \bnabla \cdot \buu - \bnabla \cdot \left( E_B \buu \right) \ ,
\end{equation}
where $E_B = \bB \cdot \bB /2$ is the magnetic energy density, and the terms on the right hand side correspond to the change in magnetic energy from the stretching of the magnetic field lines by the fluid elements ($\left(\bB \otimes \bB \right) {:} \bnabla \buu$), the changes in magnetic energy due to expansion or compression effects ($E_B \bnabla \cdot \buu$), and changes in local magnetic energy from field advection ($\bnabla \cdot \left( E_B \buu \right)$). The first two terms, associated with magnetic energy amplification, are proportional to the local shear rate,
\begin{equation}
     \left(\bB \otimes \bB \right) {:} \bnabla \buu  = \frac{1}{2}B_iB_j \dot{\gamma}_{ij} \quad {\rm and} \quad E_B \bnabla \cdot \buu  = \frac{1}{4}B_iB_i \dot{\gamma}_{jj} \ ,
\end{equation}
where
\begin{equation}
    \dot{\gamma}_{ij} = \frac{\partial u_i}{\partial x_j} + \frac{\partial u_j}{\partial x_i} 
\end{equation}
is the local shear rate.

In Figs.~\ref{fig:Shearq1} and \ref{fig:Shearq033}, we show snapshots of the absolute horizontal shear rate $|\dot{\gamma}_{xy}|$ on the orbital plane respectively after 300 orbital periods for our fixed $q=1$ runs, and $\sim 650$ orbital periods for our ``live'' $q=1/3$ runs. In Fig.~\ref{fig:abShear}, we show the absolute shear rate components, averaged within a cubic region of edge length $2$ encompassing the binary system, over last 100 orbital periods for our fixed $q=1$ runs, as a function of the softening radius $\epsilon$. We find an increase in local shear rate with higher grid resolution and smaller softening radius, each effect observed independently while the other parameter is held constant. This shows the importance of sufficiently high resolution and small softening radi to understand angular momentum transport and magnetic field amplification in common envelopes. Increased resolution indeed allows the capture of more detailed turbulence features and velocity gradients, while smaller softening radii renders the binary's gravitational body force more accurate. The often lower resolution and larger softening radii commonly used in \emph{ab initio} CEE simulations may not accurately capture these effects and may thus lead to an incomplete and potentially erroneous depiction of the dynamics, evolution, and outcome of common envelopes. Unfortunately, achieving such high resolution and small softening radii is often computationally prohibitive for \emph{ab initio} simulations. 
Of course, the presence of magnetic fields and the associated Lorentz force in our simulation would act back on the fluid, damping the local shear rate, and promoting a more ordered flow. 

 \section{Summary and discussions}\label{sec:discussion}

In this work, we conducted a detailed investigation of the dynamics of two point-mass cores orbiting inside a shared envelope with application to the post-dynamical inspiral phase of common envelope evolution. Special attention was given to the effects of the gravitational potential softening and the numerical resolution of the central region. We found that the spline-softened potential offers the best overall performance and we used it for most of our simulations (Table~\ref{tab:runs}). In all our simulations we observed that the two cores become surrounded by a shared envelope, which is to a high degree hydrostatic and corotating with the orbit (Figs.~\ref{fig:rhoPhiq1}, \ref{fig:rhoPhi}, and \ref{fig:HE}). We investigated torques on the central binary both when the semi-major axis was held fixed and when it was allowed to respond to simulation events. We found that for sufficiently resolved simulations the orbital evolution timescales converges to $\sim 10^5\, P_\text{orb}^i$ for $q=1/3$ and $\sim 10^6\, P_\text{orb}^i$ for $q=1$  (Figs.~\ref{fig:jgrav_fix} and \ref{fig:liveb}). We observed that the softening length and resolution affect the efficiency of mass ejection in complex ways (Fig.~\ref{fig:massspline}), but that higher binary mass ratios typically result in more flattened ejecta (Fig.~\ref{fig:iso_aniso}). In certain situations, our simulations develop clear polar outflows escaping through the low-density polar funnel even in the absence of the collimating effect of a magnetic field. These outflows are pressure-driven, and result from the intense shock-heating of the gas in the binary's vicinity. Using kinetic helicity as an indicator of $\alpha$-dynamo, we found that large-scale magnetic fields were unlikely to develop through the usual $\alpha$-effect in our simulations (Fig.~\ref{fig:h}). Finally, we used the local shear rate as another indicator of turbulence-driven magnetic field amplification and observed that it increases with higher grid resolution and smaller softening radius. Each of these effects was observed independently, while the other parameter was held constant. We found that quantities such as the volume-averaged shear rate --measured within a small central region surrounding the binary-- and the orbital evolution timescale appear to converge for $\epsilon \le 0.1\,a_\text{b}^i$ and a grid spacing inside the binary's orbit $\delta \le 6 \times 10^{-3}\,a_\text{b}^i$. Smaller values of $\epsilon$ require correspondingly smaller sizes of $\delta$ to ensure the resolution of pressure gradients within the softening spheres. We now discuss the broader implications of our findings for CEE.

\subsection{Implications for current simulations}

One of the open questions in CEE theory is what drives the transition from the fast dynamical inspiral to the much slower post-dynamical inspiral phase. The key physical effects influencing this transition are typically considered to be a reduction in gas density or a decrease of the Mach number. Each of these factors would decrease the drag felt by the two cores \citep[e.g.,][]{Ostriker1999,Kim2007,Roepke2023}. 
For instance, 3D hydrodynamical simulations of \citet{Reichardt2019} suggest that the end of the dynamical inspiral coincides with the corotation of the gas (i.e. to zero relative Mach number), consistent with the idea of Bondi-Hoyle-Lyttleton accretion \citep[e.g.,][]{Edgar2004}, where accreted mass transfers momentum to the cores. This theory assumes a steady, stable, and homogeneous medium and a low mass perturber. These assumptions are generally not met in the context of CEE. Our simulations reveal a different mechanism. We find that it is not the corotation of the gas that stalls the inspiral, but rather the fact that the gas surrounding the cores reaches a quasihydrostatic equilibrium. The associated symmetry properties of the gas distribution implies (almost) no net gravitational torque and lead to the end of the dynamical inspiral phase. This is in line with the results of \citet{Kim2010}. 

Building upon the findings presented here and in \citet{Gagnier2023,Gagnier2024}, we propose a coherent picture for the post-dynamical inspiral phase, applicable to systems with high to moderate mass ratios and cores experiencing weak accretion feedback. The dynamical inspiral ends when the cores become enveloped by a corotating, nearly-hydrostatic structure resembling a contact binary, which is itself eventually embedded in a low-density cavity. Unless the outer envelope is fully ejected during the dynamical phase, the subsequent evolution is driven by interactions between the central binary and the outer envelope layers. These outer layers cannot maintain corotation with the central binary, giving rise to spiral waves, shocks, and turbulence. Angular momentum transport is primarily driven by local advective torques from the mean flow and Reynolds stresses from turbulence. This transport occurs outward in a disk-like structure around the orbital plane and inward along the polar axis. The variability in mass accretion within the envelope is remarkably similar to that of circumbinary disks. Apart from brief phases of orbital expansion, the binary's orbit shrinks over a timescale of $\sim 10^5$--$10^6\,P_\text{orb}$, though significant core accretion can shorten this timescale to about $10^3\,P_\text{orb}$. Depending on the properties of the remaining envelope, the contraction timescale of the central binary can be either longer or shorter than the envelope's thermal timescale. Consequently, different processes, such as energy input from the contracting binary, dust- and recombination-driven winds, and feedback from core accretion,  may more or less contribute to the ultimate ejection of the envelope across different stars. With current 3D simulations, often limited by both their short duration and range of included physics, it remains challenging to assess the relative importance of these processes.

Magnetic field amplification and potential launching of jets or similar collimated outflows is another topic of importance for theory and observations. Our results on kinetic helicity, shear rate, and our earlier ideal MHD simulations with excised central region tend to suggest that the envelope with a central binary does not efficiently produce large-scale organized magnetic fields. However, this statement has a limited validity due to us not simulating earlier CEE stages, where magnetic field amplification can occur around the plunging companion \citep{Ohlmann2016b,Ondratschek2022,Vetter2024}, and because we have not yet performed MHD simulations explicitly including dynamics around the two cores. At the same time, we showed here that thermally-launched polar outflow  collimated by the centrifugally-evacuated polar funnel can occur at least some amount of time without the presence of magnetic fields. Our earlier simulations indicate that the existence and properties of the polar funnel can vary considerably due to the turbulence in the envelope \citep[Fig.~3 in][]{Gagnier2024}. This implies that magnetic fields are not a necessary condition for launching at least intermittent time-variable jets, but it is not yet clear whether this is also the case for jets that are stronger and sustained over longer times.

\subsection{Relation to contact binaries}

The corotating hydrostatic structure surrounding the two cores bears a remarkable similarity to the shared envelope of ordinary contact binaries. Despite decades of observations and theoretical efforts, the internal structure and evolution of contact binaries remain unsolved. The fundamental problem is the \citet{Kuiper1941} paradox arising from the observational fact that the two cores of very different mass and luminosity are able to maintain nearly constant effective temperature across the shared surface. Proposals to solve this problem typically involve strong subsurface heat-redistributing flows and absence of global thermal equilibrium \citep[e.g.,][]{Lucy1968,Shu1981,Stepien2006}. 

Based on the analogy with contact binaries, we can now speculate about two potential effects that this corotating hydrostatic shared envelope can have for the two cores inside CEE. First, the shared envelope can develop internal flows, which can redistribute heat from one core to the other one. If this transport is sufficiently efficient and long-lasting, it might be incorrect to treat the cores as single stars, for example, by emulating the effect of CEE as a simple rapid envelope removal. Second, after the dynamical inspiral it is unlikely that the spin of the individual cores would corotate with the new binary orbit. As a result, shear will necessarily develop between the corotating hydrostatic envelope and the two cores, potentially leading to generation and amplification of magnetic fields. Indeed, contact binaries typically show highest levels of magnetic activity when compared to single stars of the same type \citep[e.g.,][]{Vilhu1987}. Neither of these two effects would be captured by existing 3D simulations due to insufficient resolution.

It is also important to discuss how long can the corotating hydrostatic structure survive around the two cores. If one of the cores is capable of accreting gas, the mass of the envelope will decrease. However, the accretion rate might be very small, for example when one of the cores is a stellar-mass black hole accreting at the Eddington limit, giving long accretion timescales. More disruptive is the potential feedback from the accreting core in the form of disk winds or jets. Even if neither of the cores accrete, the shared envelope should eventually disperse due to ordinary stellar winds once the outer envelope is cleared. 

Finally, it is interesting to note the corotating hydrostatic envelope formed very rapidly within the first few tens of orbits. This suggests that similar structures might be formed in events such as rapid mass transfer preceding CEE to make binaries appear in contact similarly to what was observed in the pre-explosion orbital variability in V1309~Sco and other similar transients \citep{Tylenda2011,Pejcha2014,Pejcha2017,Blagorodnova2020}.

\subsection{Future work}\label{sec:disc}

 There are several areas where our work can be improved in the future. First, we did not account for accretion onto the cores. This process, often implemented as numerical mass sinks, prevents potentially unrealistic mass accumulation around the point masses by removing excess mass at a set rate. The absence of such sinks hinders the formation of asymmetric structures around the point masses. Their implementation could lead to an increase in total gravitational torque, potentially resulting in smaller final orbital separations, as well as differential accretion towards $q=1$. Additionally, accretion would complexify the gas dynamics and thus affect magnetic energy amplification.
However, whether accretion occurs on the primary, secondary, or both cores, and at what rate, remains highly uncertain. These factors heavily depend on the nature of the cores and the thermodynamic and radiative properties at their interface with the envelope. Furthermore, feedback from the accretion could similarly fundamentally change the gas dynamics, however, resolving these effects would require much higher resolution. Second, our calculations could be repeated in the limit of ideal magnetohydrodynamics, shedding further light on the amplification of magnetic fields. 
Finally, implementing self-gravity of the gas would improve the accuracy of the shared envelope's internal dynamics and its interaction with the outer envelope. This may, in turn, influence the eccentricity of the binary orbit, which could be excited or damped by the effects of the surrounding gas.

\begin{acknowledgements}
   The research of D.G. and O.P. has been supported by Horizon 2020 ERC Starting Grant ‘Cat-In-hAT’ (grant agreement no. 803158). D.G. acknowledges support by the Klaus-Tschira Foundation. D.G. acknowledges funding by the European Union (ERC, ExCEED, project number 101096243). Views and opinions expressed are, however, those of the authors only and do not necessarily reflect those of the European Union or the European Research Council Executive Agency. Neither the European Union nor the granting authority can be held responsible for them. This work was supported by the Ministry of Education, Youth and Sports of the Czech Republic through the e-INFRA CZ (ID:90254).
\end{acknowledgements}

\bibliographystyle{aa}
\bibliography{bibnew}
\onecolumn
\begin{appendix}

\section{Initial density and pressure profiles}\label{app:ic}

\begin{figure} 
\centering
      \includegraphics[ width=0.5\textwidth]{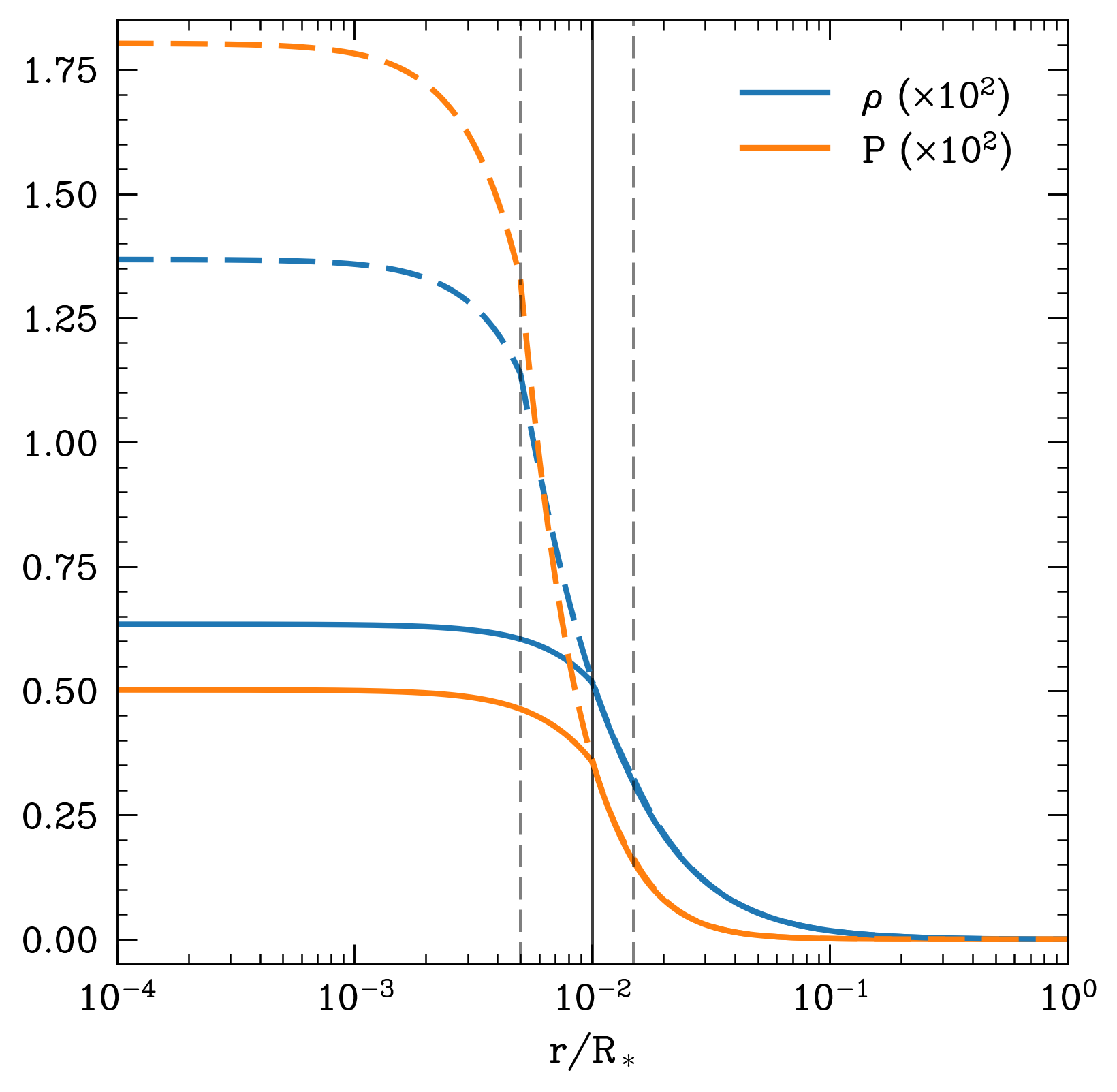}
   \caption{ Initial density and  pressure profiles of the envelope  for $q=1$ (full lines) and $q=1/3$ (dashed lines), obtained from Eqs.~(\ref{eq:rho_sh}), (\ref{eq:rho_in}), and (\ref{eq:rho_inb}). The vertical lines represent the values of $r_< = \min(r_1,r_2)$ and $r_>= \max(r_1,r_2)$ for $q=1$ (full lines) and for $q=1/3$ (dashed lines). When $q=1$, $r_1 = r_2$, and thus $r_> = r_<$}. See Eq.~(\ref{eq:units} for an example how to scale plotted quantities to physical systems.
\label{fig:IC}
\end{figure}

The radial density and pressure profiles outside of the binary orbit, i.e. at $ r \ge \max(r_1,r_2)= r_>$, are
\begin{align}\label{eq:rho_sh}
    \frac{\rho(r \ge r_>)}{\rho(r_>)} &=  \left( 1 + C \left(\frac{B}{3} \left(\frac{1}{r^3} - \frac{1}{r_>^3} \right) - A \left(\frac{1}{r} - \frac{1}{r_>} \right) \right) \right)^n,\\
    P(r \ge r_>) &= K \rho^\Gamma \ ,
    \end{align}    
where $n \equiv 1/(\Gamma - 1) = 3/2$ is the polytropic index and
\begin{equation}
\begin{aligned}
   A & = GM \ ,  &  B &= \frac{3A\ab^2 q}{8(1+q)^2} \ , \\
   A^\prime &= A \left( \frac{1}{R_\ast} - \frac{1}{r_>} \right) \ ,    &   B^\prime &= \frac{B}{3} \left ( \frac{1}{R_\ast^3} -  \frac{1}{r_>^3} \right) \ , \\
  C &=   \frac{\kappa  - 1}{B^\prime - A^\prime} \ , &  K &= \frac{(1-\Gamma)(B^\prime - A^\prime)}{\Gamma (\kappa -1) \rho(r_>)^{1/n}} \ ,\\ \kappa^n &= \rho(R_\ast)/\rho(r_>) \ .
\end{aligned}
\end{equation}
$\rho(r_>)$ is calculated from the prescribed mass of the envelope outside of the orbit $M_{\rm env}$.

Unlike \citet{Gagnier2023,Gagnier2024}, we do not excise an inner sphere containing the binary. We therefore need to derive the radial density and pressure profiles inside of the binary orbit.
The nonsoftened gravitational potential of the two point-mass objects at $r \leq \min(r_1,r_2) = r_<$ reads
\begin{equation}\label{eq:fullphi1}
     \Phi(r \leq r_<)=  -4 \pi \sum_{i=1}^2  \sum_{k = 0}^\infty \sum_{m=-2k}^{2k} \frac{G M_i}{4k +1} \frac{r^{2k}}{r_i^{2k +1}} (Y_{2k}^m (\theta_i,\phi_i))^\ast Y_{2k}^m (\theta,\phi)    \ ,
\end{equation}
where $Y_\ell^m$ are the usual normalised scalar spherical harmonic functions. The latitude- and time-averaged binary potential for spherical harmonics degrees $\ell \leq 2$ (or equivalently $k \leq 1$) within the inner binary orbit ($r \leq r_<$) reads
\begin{equation}\label{eq:phi_in}
    \langle \Phi(r \leq r_<)\rangle_\theta =  -A \left( \frac{1+q^2}{\ab q}-  B^{\prime\prime} r^2\right) \ ,
\end{equation}
where $\langle \cdot \rangle_\theta$ indicates a time and latitude average, and 
\begin{equation}
     B^{\prime\prime} =\frac{\left(1+q\right)^2}{8\ab^3} \left(\frac{1+ q^4 }{q^3} \right) \ .
\end{equation}
Injecting Eq.~(\ref{eq:phi_in}) into the equation of hydrostatic equilibrium, 
\begin{equation}\label{eq:HSE}
    \bnabla P = -\rho \bnabla \Phi \ ,
\end{equation}
yields
\begin{align}\label{eq:rho_in}
    \frac{\rho(r \leq r_<)}{\rho( r_<)} &=  \left( 1 + B^{\prime\prime\prime} \left( r^2 -  r_<^2\right)\right)^n,\\
    P(r \leq r_<)&= K \rho^\Gamma \ ,
\end{align}    
where
\begin{equation}
     B^{\prime\prime\prime} = \frac{B^{\prime \prime} (\kappa-1)}{B^\prime - A^\prime} \left(\frac{\rho(r_>)}{\rho(r_<)} \right)^{1/n} \ .
\end{equation}
In simulations initialized with $q=1$ (and $e = 0$), $r_< = r_> = \ab^i/2$. However, when we set $q < 1$, we further need to compute the density and pressure profiles in the spherical shell located between the orbits of the two point masses. The $\ell \leq 2$ latitude and time averaged binary potential at $r_< \leq r \leq r_>$ reads
 \begin{align}\label{eq:phi_inb}
   &\langle \Phi(r_< \leq r \leq r_>)\rangle_\theta = \\ &-A \left( \frac{q}{\ab} + \frac{1}{(1+q)r} - \frac{qB}{3(1+q)Ar^3} - \frac{B^{\prime \prime} q^4 r^2}{2(1+q^4)} \right) \nonumber \ .
\end{align}
Injecting Eq.~(\ref{eq:phi_inb}) into the equation of hydrostatic equilibrium Eq.~(\ref{eq:HSE}) finally yields
\begin{align}\label{eq:rho_inb}
    \frac{\rho(r_< \leq r \leq r_>)}{\rho( r_>)} &= \left( 1 + \frac{\kappa - 1}{B^\prime - A^\prime} \left( \frac{qB}{3(1+q)} \left( \frac{1}{r^3} - \frac{1}{r_>^3} \right)  -\right. \right.  \\ & \left. \left. \frac{A}{1+q}  \left( \frac{1}{r} - \frac{1}{r_>} \right) + \frac{A B^{\prime \prime} q^4}{2(1+q^4)} \left(r^2 - r_>^2 \right)         \right) \right),\nonumber \\
    P(r_< \leq r \leq r_>)&= K \rho^\Gamma \ .
\end{align}   
 
In Fig.~\ref{fig:IC}, we show the initial density and pressure profiles for $q=1$, $q=2/3$, and $q = 1/3$. We note that while we have ensured the continuity of $\rho$ and $P$ throughout the envelope, the first derivative of Green's function, hence the gravitational potential gradient, is discontinuous at $r= r_>$ and $r = r_<$. This naturally leads to discontinuities in the initial density and pressure gradients at these radii, which slightly disrupt hydrostatic equilibrium. We find such perturbations to be sufficiently weak to not require special attention.

\section{Plummer vs Ruffert vs Spline}\label{app:A}

\begin{figure*} 
\centering
      \includegraphics[ width=.33\textwidth]{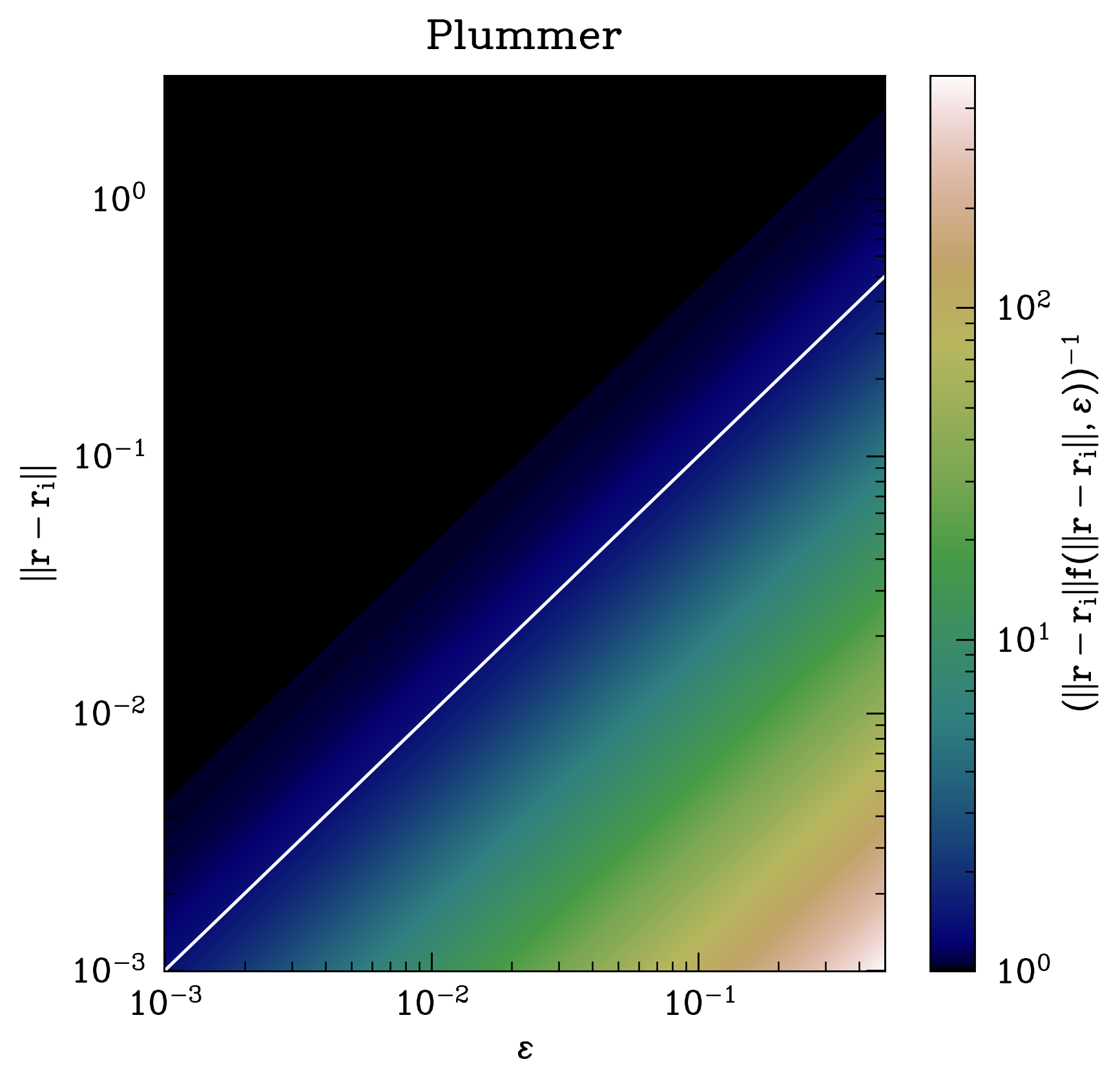}
            \includegraphics[ width=.33\textwidth]{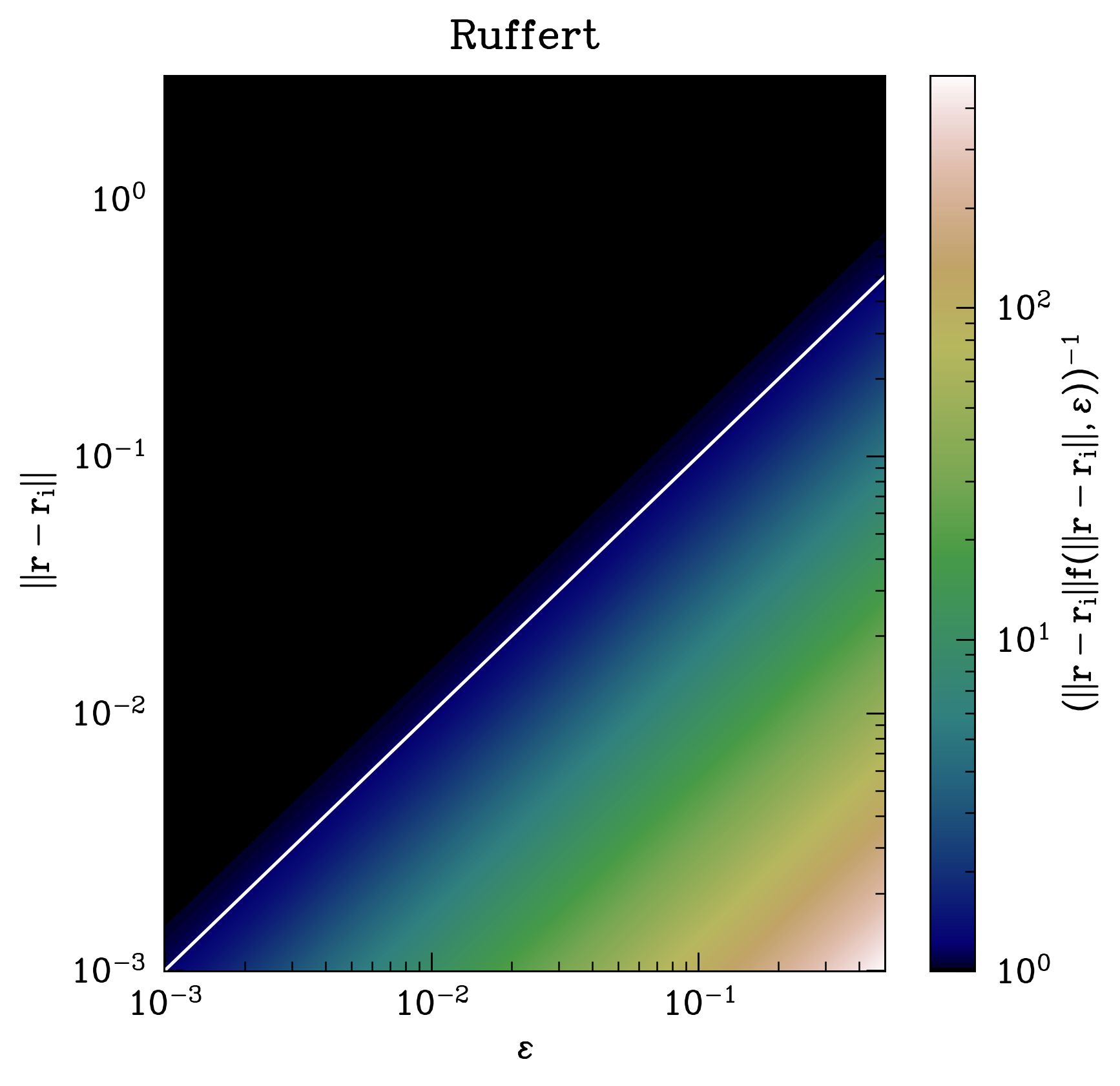}
            \includegraphics[ width=.33\textwidth]{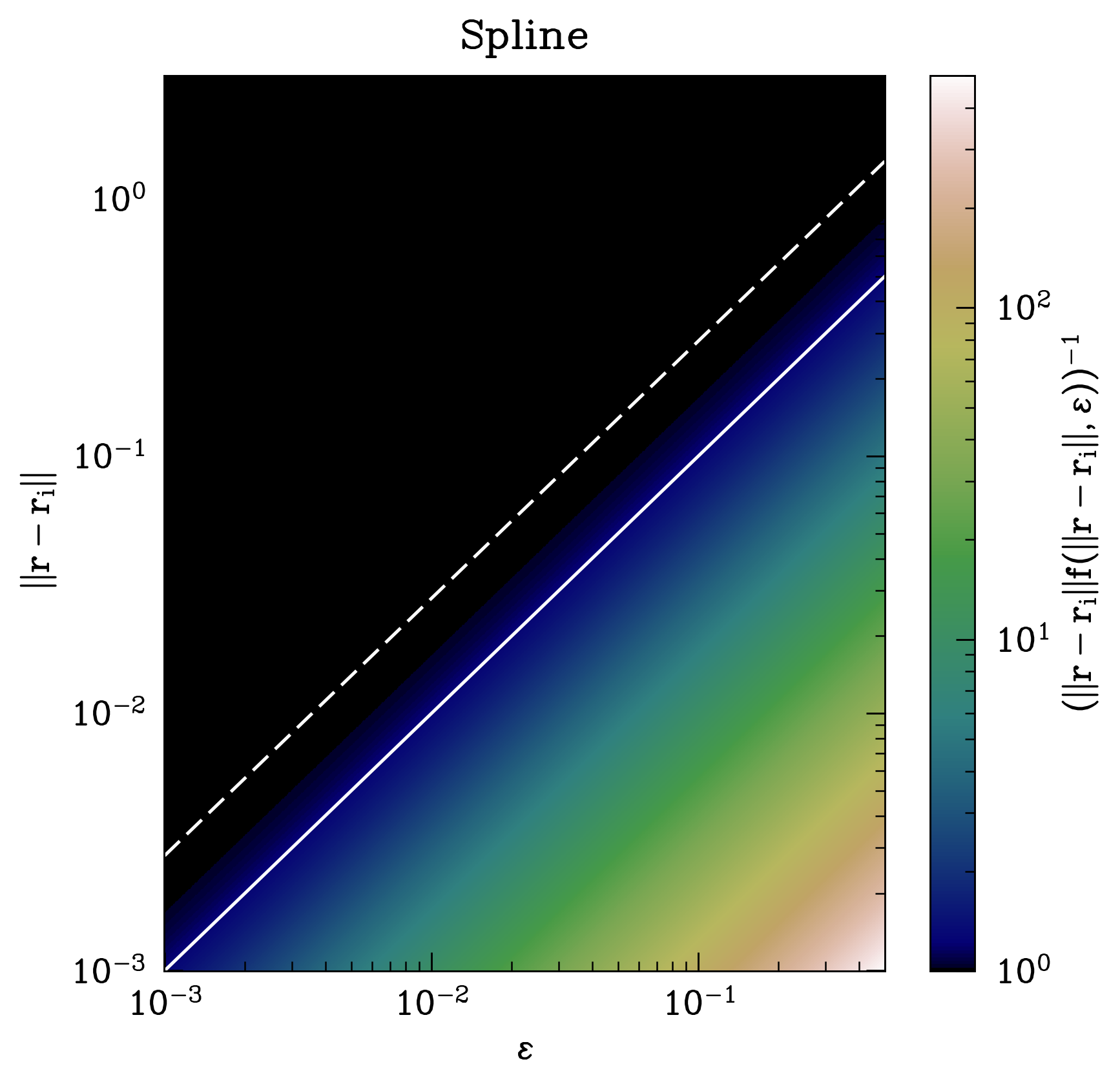}\hfill
            \includegraphics[ width=.33\textwidth]{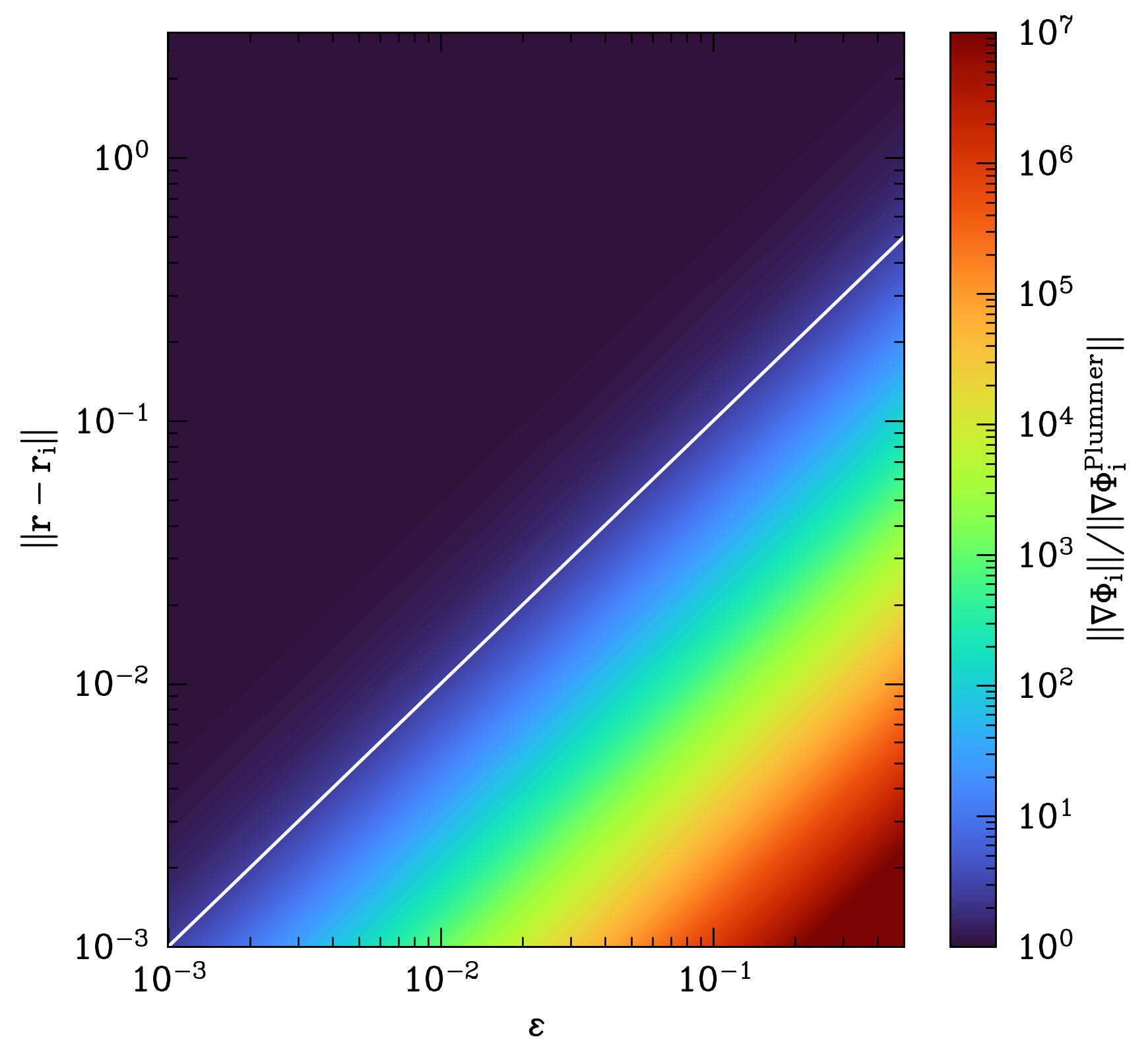}
                        \includegraphics[ width=.33\textwidth]{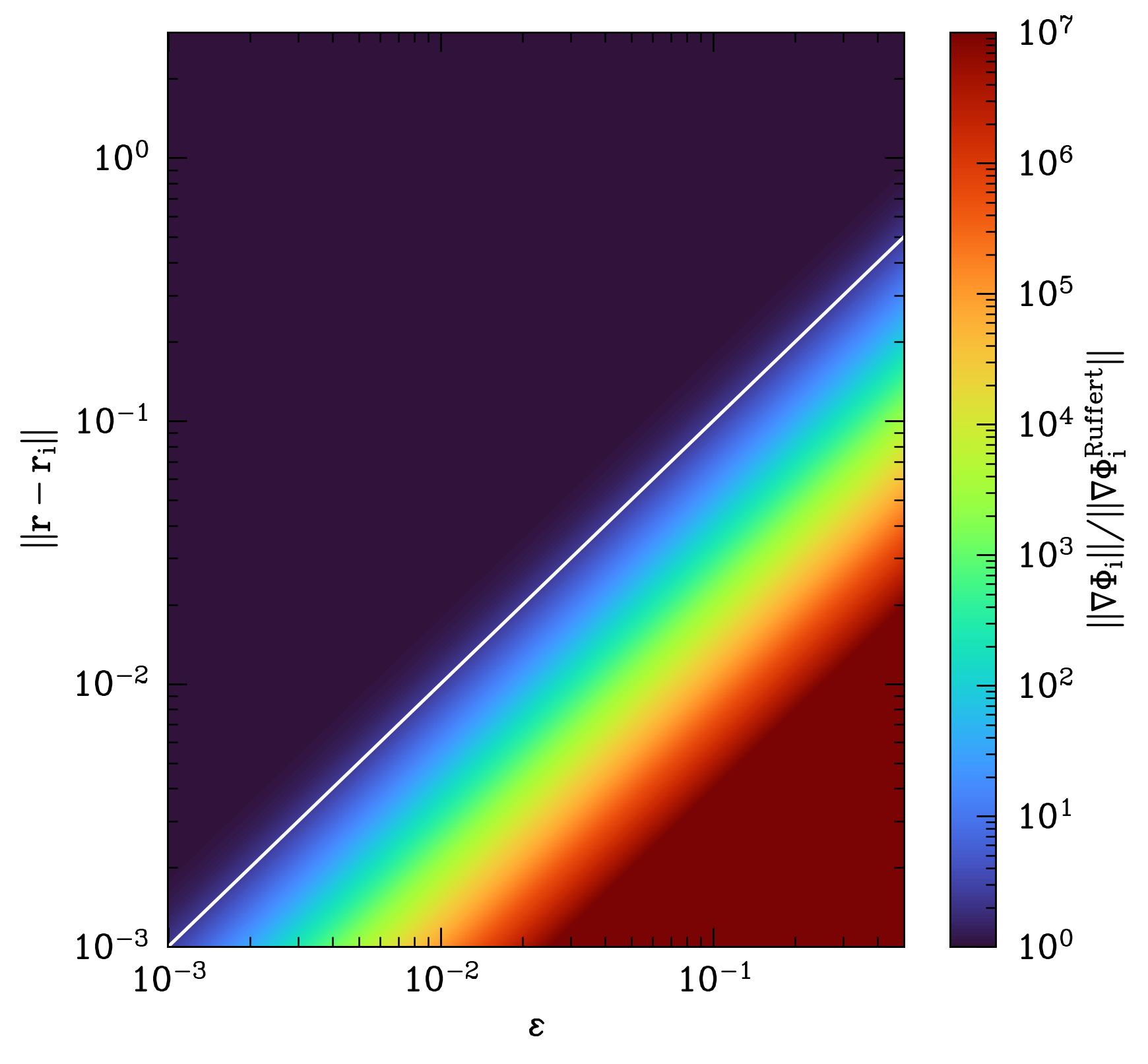}
            \includegraphics[ width=.33\textwidth]{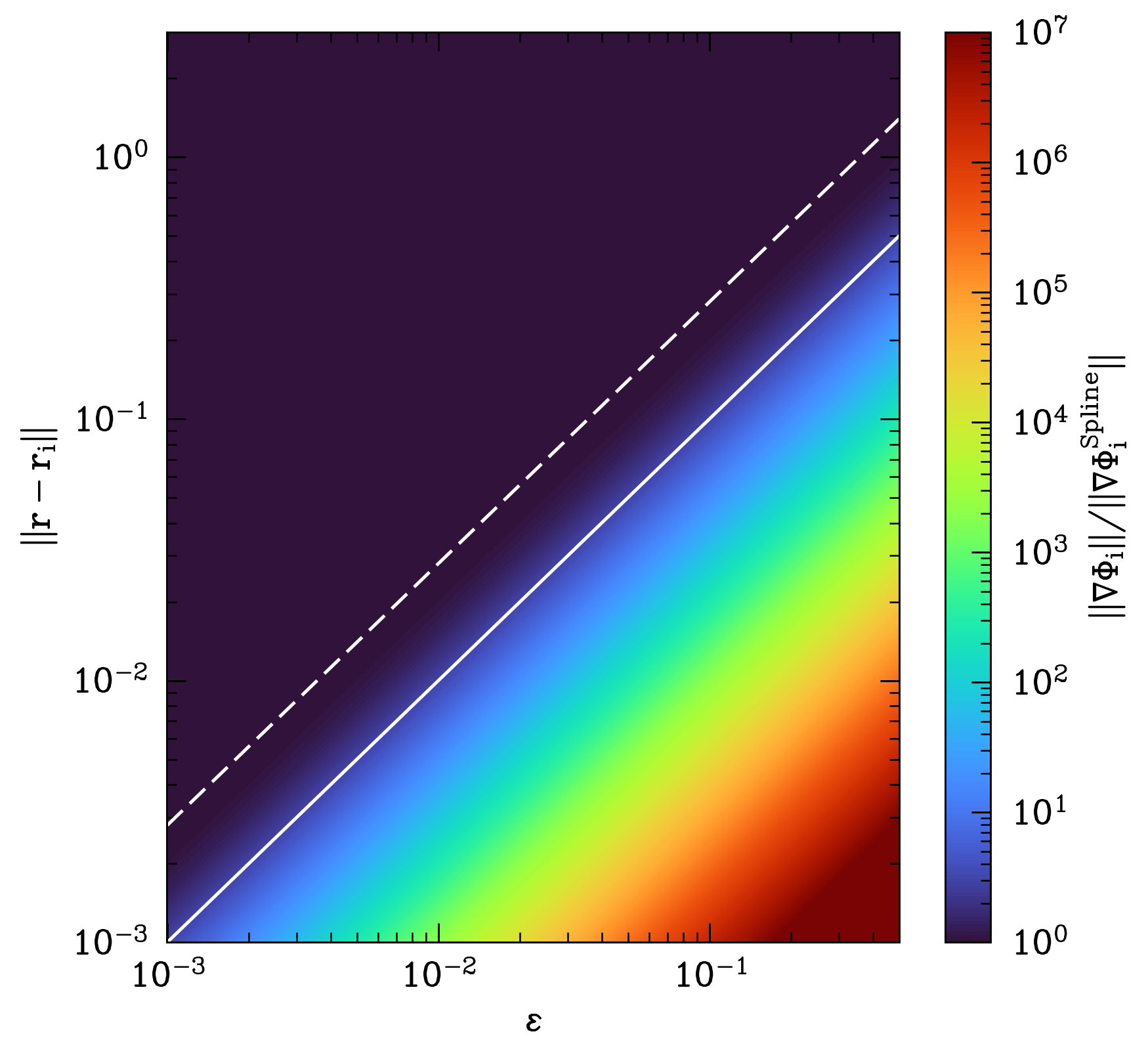}
            
   \caption{Top row: Ratio between the nonsoftened gravitational potential and the Plummer-, Ruffert-, and spline-softened gravitational potentials, as a function of the softening radius and of the distance to the core. Bottom row: same as top row, but for the gravitational torque density associated with one of the two binary components. The white full lines indicates $\left.\lVert \br - \br_i \right.\rVert  = \epsilon$ and the white dashed line indicates $\left.\lVert \br - \br_i \right.\rVert  = h = 14\epsilon/5$. }
\label{fig:ratio}

\end{figure*}

We show the ratio between the non-softened gravitational potential and the Plummer (Eq.~\ref{eq:plum}), Ruffert (Eq.~\ref{eq:ruf}), and spline-softened (Eq.~\ref{eq:spline}) gravitational potentials as a function of the softening radius $\epsilon$ and the distance to the core in Fig.~\ref{fig:ratio}. This figure also shows the ratio between the non-softened and softened gravitational potential gradients as a function of the softening radius and the distance to the core for the three softening methods. The same quantities for a fixed softening radius ($\epsilon = 0.2$) are displayed in Fig.~\ref{fig:ratioeps02}.
Using the definition $h \equiv  14 \epsilon/5$, the three softening methods yield very similar gravitational potentials with a difference of less than or equal to 1\% up to a distance of $\epsilon / 5$. Beyond this distance, the Plummer-softened potential converges to Newton's law more slowly than the Ruffert- and spline-softened potentials, with the spline-softened potential becoming exactly Newtonian at a distance greater than or equal to $h$.
\begin{figure} 
\centering
      \includegraphics[ width=0.5\textwidth]{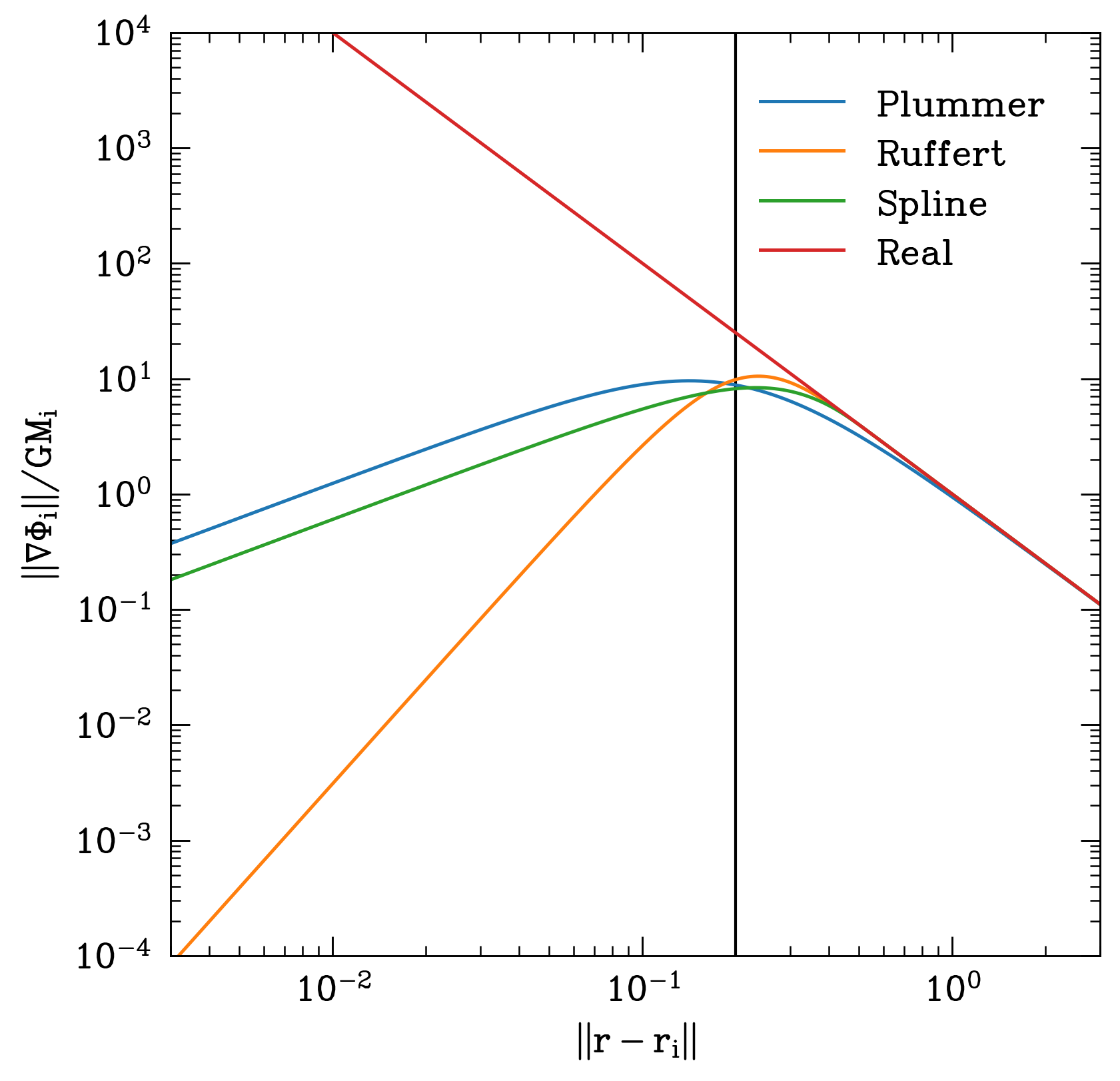}
   \caption{Comparison between the Euclidian norm of the gradient of the different potentials as a function of the distance to the core, for a softening radius $\epsilon = 0.2$ indicated by the vertical line.}
\label{fig:grad}

\end{figure}
\begin{figure} 
\centering
      \includegraphics[ width=.49\textwidth]{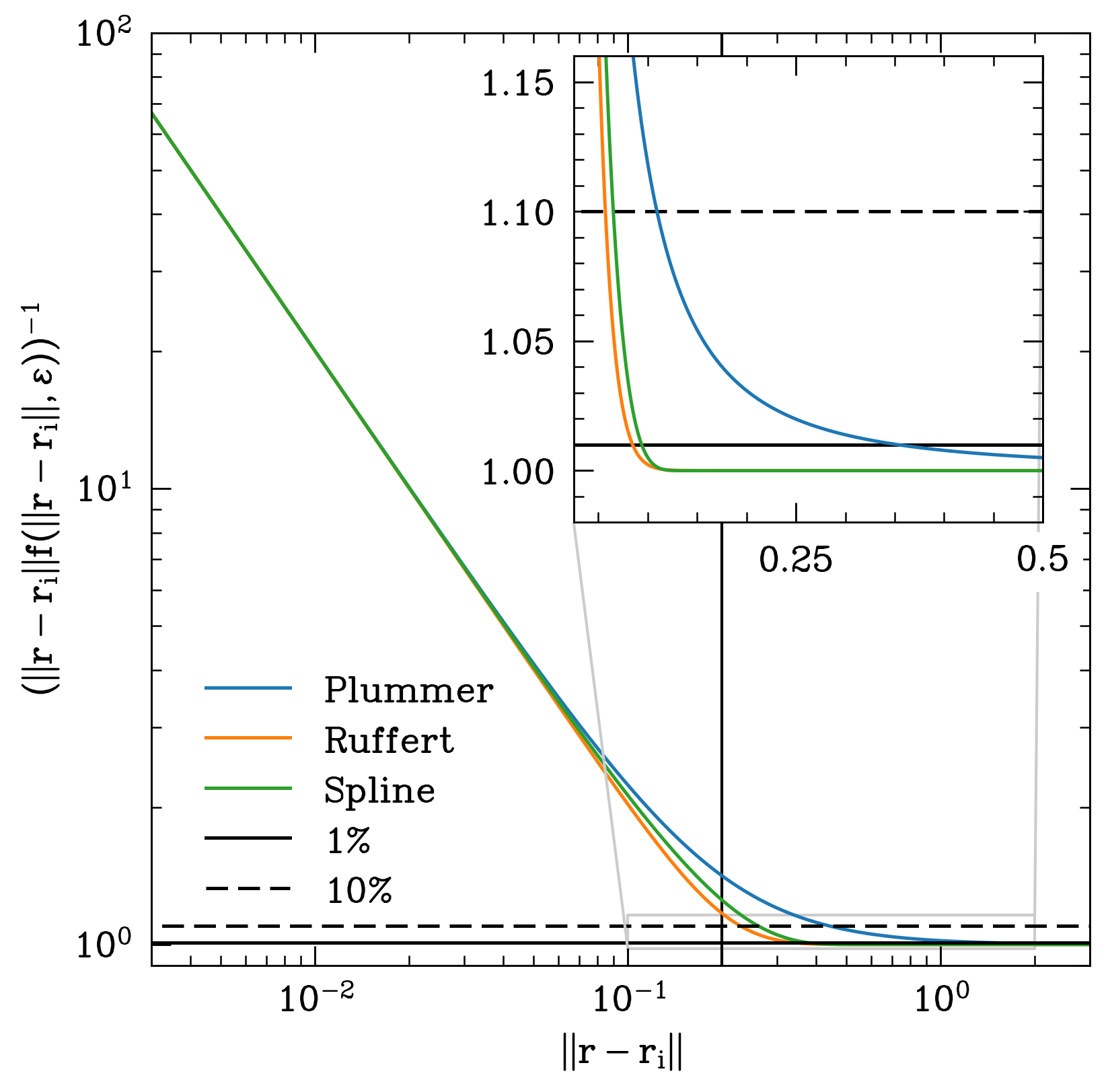}
      \includegraphics[ width=.49\textwidth]{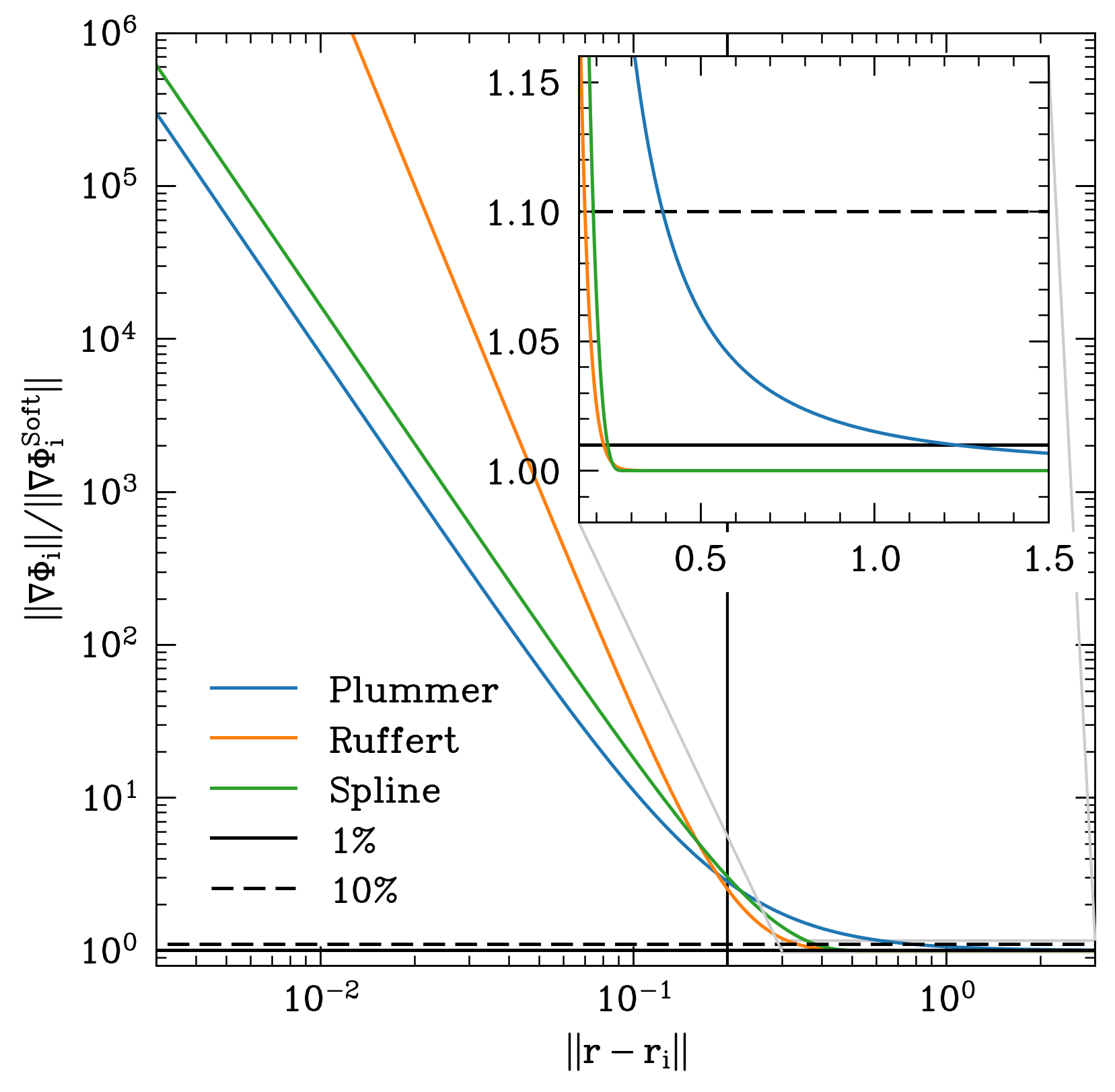}
   \caption{Left: Ratio between the nonsoftened gravitational potential and the Plummer- and spline-softened gravitational potentials, as a function of the distance to the core, for $\epsilon = 0.2$. Right: Same, but for the gravitational torque density associated with one of the two binary components. Horizontal black lines indicate at relative difference of $1\%$ (full line) and $10\%$ (dashed line) with the nonsoftened value.}
\label{fig:ratioeps02}
\end{figure}

Comparing the softened potentials is however meaningless as the equations of hydrodynamics only involve their gradient. The softened gradients are underestimated by orders of magnitude for $\left.\lVert \br - \br_i \right.\rVert  \lesssim \epsilon$, with the three softening formulations (see Figs.~\ref{fig:ratio}, \ref{fig:grad}, and \ref{fig:ratioeps02}). Hence, as intended, the binary's body force density is dramatically underestimated within the softening regions. We further note that
\begin{equation}
    \frac{\left\lVert \bnabla \Phi_i\right\lVert}{\left\lVert\bnabla \Phi_i^{\rm Soft}\right\lVert} = \frac{\dot{\ell}_{z,i}}{\dot{\ell}_{z,i}^{\rm Soft}} \ ,
\end{equation}
that is the gravitational torque density is equally underestimated. While dramatically underestimated, the Plummer and spline-softened gravitational potential gradients are better at reproducing the true  gravitational torque and gravitational force densities than the Ruffert alternative, provided $\left.\lVert \br - \br_i \right.\rVert  \lesssim \epsilon$. 
As for the softened potential, the Plummer-softened potential gradient converges slower to Newton's law than the Ruffert and spline-softened potential gradients, with the spline-softenend potential gradient becoming exactly Newtonian at  $\left.\lVert \br - \br_i \right.\rVert  \geq h$. In particular, one can derive that, using the spline softening formulation, the relative difference\begin{align}\label{eq:spline_rel}
  \frac{\left\lVert \bnabla \Phi_i\right\lVert - \left\lVert\bnabla \Phi_i^{\rm Spline}\right\lVert}{\left\lVert\bnabla \Phi_i^{\rm Spline}\right\lVert}  \begin{cases}
      \leq    0.8 & \text{if}\ \left.\lVert \br - \br_i \right.\rVert \gtrsim 1.295 \epsilon.\\[2mm]
    \leq    0.5 & \text{if}\ \left.\lVert \br - \br_i \right.\rVert \gtrsim 1.448 \epsilon.\\[2mm]
  \leq    0.1 & \text{if}\ \left.\lVert \br - \br_i \right.\rVert \gtrsim 1.911 \epsilon.\\[2mm]
  \leq    0.01, & \text{if}\  \left.\lVert \br - \br_i \right.\rVert \gtrsim 2.325 \epsilon.\\[2mm]
  =  0, & \text{if}\  \left.\lVert \br - \br_i \right.\rVert \geq h = 14\epsilon/5.
  \end{cases}
  \end{align}
  Equivalently, using the Plummer-softened potential
\begin{align}\label{eq:plummer_rel}
  \frac{\left\lVert \bnabla \Phi_i\right\lVert - \left\lVert\bnabla \Phi_i^{\rm Plummer}\right\lVert}{\left\lVert\bnabla \Phi_i^{\rm Plummer}\right\lVert}  \begin{cases}
  \leq    0.8 & \text{if}\ \left.\lVert \br - \br_i \right.\rVert \gtrsim 1.444\epsilon.\\[2mm]  
  \leq    0.5 & \text{if}\ \left.\lVert \br - \br_i \right.\rVert \gtrsim 1.795\epsilon.\\[2mm]  
  \leq    0.1 & \text{if}\ \left.\lVert \br - \br_i \right.\rVert \gtrsim 3.904\epsilon.\\[2mm]
  \leq    0.01, & \text{if}\  \left.\lVert \br - \br_i \right.\rVert \gtrsim 12.258 \epsilon.\\[2mm]
  \to  0, & \text{as}\  \left.\lVert \br - \br_i \right.\rVert \to \infty.
  \end{cases}
  \end{align}  
  And the Ruffert-softened potential
\begin{align}\label{eq:Ruffert_rel}
  \frac{\left\lVert \bnabla \Phi_i\right\lVert - \left\lVert\bnabla \Phi_i^{\rm Ruffert}\right\lVert}{\left\lVert\bnabla \Phi_i^{\rm Ruffert}\right\lVert}  \begin{cases}
  \leq    0.8 & \text{if}\ \left.\lVert \br - \br_i \right.\rVert \gtrsim 1.149\epsilon.\\[2mm]  
  \leq    0.5 & \text{if}\ \left.\lVert \br - \br_i \right.\rVert \gtrsim 1.264\epsilon.\\[2mm]  
  \leq    0.1 & \text{if}\ \left.\lVert \br - \br_i \right.\rVert \gtrsim 1.672\epsilon.\\[2mm]
  \leq    0.01, & \text{if}\  \left.\lVert \br - \br_i \right.\rVert \gtrsim 2.209 \epsilon.\\[2mm]
  \to  0, & \text{as}\  \left.\lVert \br - \br_i \right.\rVert \to \infty.
  \end{cases}
  \end{align}  
  Hence, for a given softening radius $\epsilon$, the Ruffert and spline-softened potential gradients converge similarly to the real potential with distance to the core. In particular, Ruffert softened potential gradient is more accurate than
the spline-softened one up to $ \left.\lVert \br - \br_i \right.\rVert \simeq 2.459\epsilon$. At $  \left.\lVert \br - \br_i \right.\rVert = h$, Ruffert's softened potential gradient's deviation from the real potential is $\sim 0.047 \%$.

\begin{figure} 
\centering
      \includegraphics[ width=0.7\textwidth]{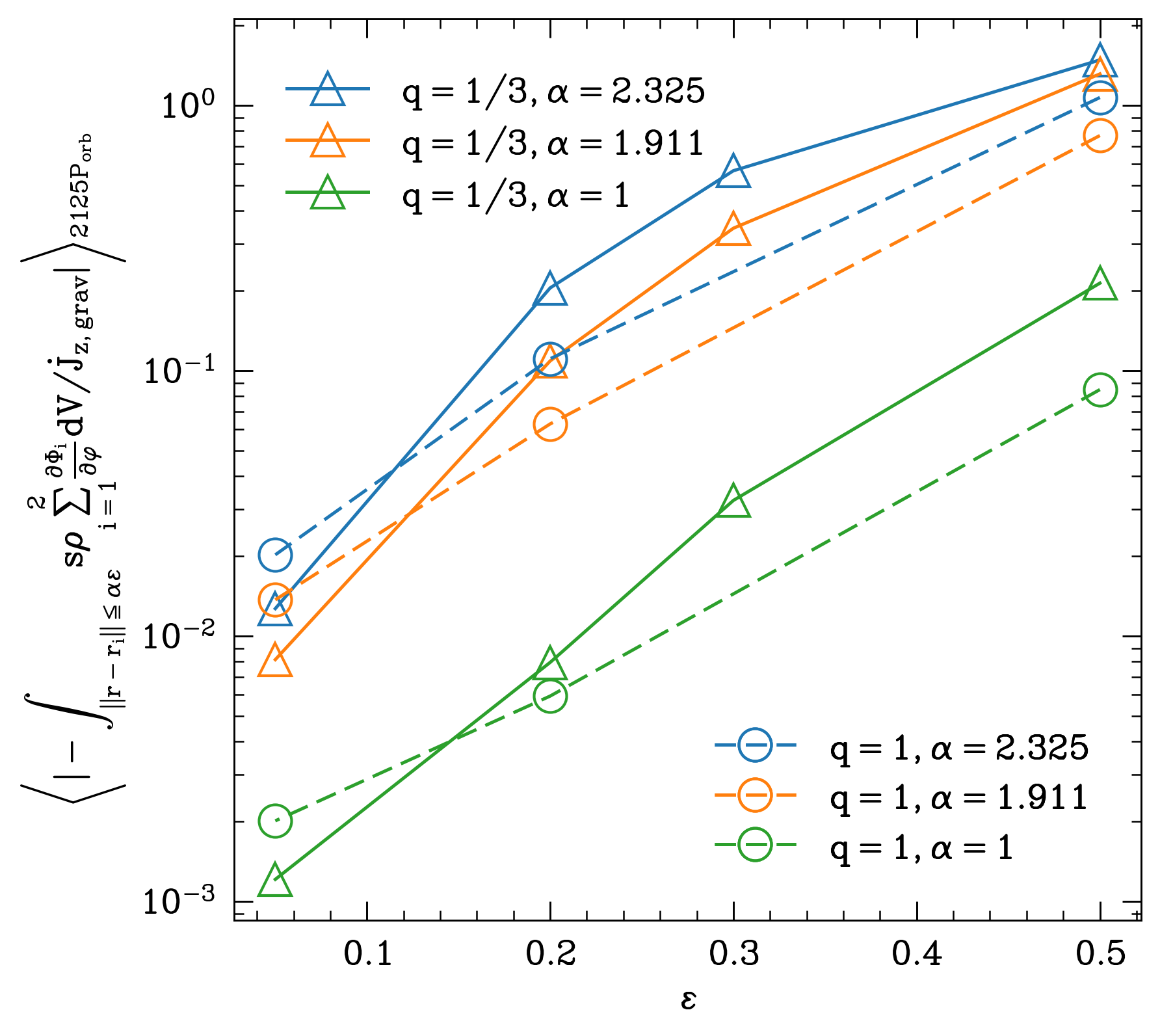}
   \caption{Simulation average of the ratio between the gravitational torque contribution from the softened region and the total gravitational torque, for various values of $\epsilon$ and $\alpha$, for our live binary simulations,  using the spline formulation.}
\label{fig:condition}
\end{figure}

Comparing Eqs.~(\ref{eq:spline_rel}) and (\ref{eq:plummer_rel}), it is clear that the adopted definition $h \equiv 14\epsilon/5$ is far from implying any form of relevant equivalence between the different softening methods. For example, taking $\epsilon = 0.2$ implies that a spline-softened potential gradient will be underestimated by less than $1\%$ at a distance $\left.\lVert \br - \br_i \right.\rVert \gtrsim 0.465$ from the core. On the other hand, for the same value of $\epsilon$, a Plummer-softened potential gradient will be underestimated by less than $1\%$ at a distance $\left.\lVert \br - \br_i \right.\rVert \gtrsim 2.452  $ from the core, that is far beyond the companion core. Similarly, while the body force density and gravitational torque density is underestimated by more than $10\%$ within  $\left.\lVert \br - \br_i \right.\rVert \lesssim  0.3822  $ with a spline softening, is goes up to a distance $\sim 0.781  $ with a Plummer softening.
One may instead define $h$ by imposing 
\begin{equation}
    \frac{\left\lVert \bnabla \Phi_i\right\lVert - \left\lVert\bnabla \Phi_i^{\rm Spline}\right\lVert}{\left\lVert\bnabla \Phi_i^{\rm Spline}\right\lVert} = \frac{\left\lVert \bnabla \Phi_i\right\lVert - \left\lVert\bnabla \Phi_i^{\rm Plummer}\right\lVert}{\left\lVert\bnabla \Phi_i^{\rm Plummer}\right\lVert} = 0.01
\end{equation}
for instance, that is $h^\prime \simeq 14.7612 \epsilon$. This alternative definition becomes a physically relevant equivalence between the two formulations provided any nonaxysymmetric structure (e.g. minidisks in circumbinary disks) significantly contributing to the total torque and body force, are properly resolved outside of the softened regions.
In other words,  an asymptotic regime of small $\epsilon$ can be considered reached for softening radii $\sim$ 2--5 times larger using a spline-softened potential compared to using a Plummer-softened potential. A necessary but insufficient condition for convergence to an asymptotic regime of small $\epsilon$ is, for example,
\begin{equation}\label{eq:condition}
  - \int_{\left.\lVert \br - \br_i \right.\rVert 
 \leq \alpha \epsilon} s \rho \sum_{i=1}^2 \frac{\partial \Phi_i}{\partial \varphi} \dd V \ll \dot{J}_{z, \rm grav} \ .
\end{equation}
where $\alpha$ can be chosen from Eqs.~(\ref{eq:spline_rel}), \ref{eq:plummer_rel}), and  (\ref{eq:Ruffert_rel}), depending on desired inaccuracy tolerance.
We note that if such condition is not satisfied, the Plummer and spline softening methods may provide a less inaccurate representation of the binary's body force than the Ruffert's potential. However, such simulations may not be used to obtain any physical insight nor predictions. We check criterion (\ref{eq:condition}) for different runs in Fig.~\ref{fig:condition}.
Finally, we note that azimuthal grid asymmetry relative to the plane orthogonal to the orbital plane and parallel to the $\br_i - \br_j$ vector induces torque artificially, independent of gas density distribution and coordinate system. This anomalous gravitational torque is primarily driven by the gravitational torque density near the cores, hence specifically within softening spheres. This again suggests the importance of satisfying condition (\ref{eq:condition}).

\end{appendix}
\end{document}